%% file: main.tex
\documentclass[fleqn,usenatbib]{mnras}

\usepackage{newtxtext,newtxmath}

\usepackage[T1]{fontenc}

\DeclareRobustCommand{\VAN}[3]{#2}
\let\VANthebibliography\thebibliography
\def\thebibliography{\DeclareRobustCommand{\VAN}[3]{##3}\VANthebibliography}

\usepackage{graphicx}	
\usepackage{amsmath}	
\usepackage{bm}
\usepackage{cleveref}
\usepackage{comment}
\usepackage{subcaption}
\usepackage{cprotect}

\usepackage[normalem]{ulem}

\usepackage{lipsum}

\renewcommand{\d}{{\mathrm{d}}}

\newcommand{\fix}[1]{\textcolor{violet}{#1}}

\title[Galaxy Populations in the IllustrisTNG Caustic Skeleton]{Galaxy Populations in the IllustrisTNG Caustic Skeleton}

\author[B. Hertzsch et al.]{
Benjamin Hertzsch,$^{1, 2}$\thanks{E-mail: benjamin.hertzsch@ed.ac.uk}
Job Feldbrugge,$^{1}$ and
Rien van de Weygaert$^{2}$
\\
$^{1}$School of Physics and Astronomy, University of Edinburgh, United Kingdom\\
$^{2}$Kapteyn Astronomical Institute,  University of Groningen, Netherlands
}

\date{Accepted XXX. Received YYY; in original form ZZZ}

\pubyear{\the\year{}}

\begin{document}
\label{firstpage}
\pagerange{\pageref{firstpage}--\pageref{lastpage}}
\maketitle

\begin{abstract}
The caustic skeleton is a parameter-free and mathematically rigorous formalism for tracing the hierarchical formation history of the multiscale cosmic web from the singularities in the underlying dark matter flow. In the present study, we explicitly use the multistreaming nature of the cosmic mass distribution to address the influence of the weblike embedding on the galaxy populations and discern their properties in different web environments. To this end, we construct the multiscale caustic skeleton of the dark mass distribution in the state-of-the-art suite of the large-scale IllustrisTNG simulations. In addition to the multistreaming dark matter density field, we assess the characteristic properties of the intergalactic baryonic gas in the vicinity of the caustics. Next, we associate the galaxies with the voids, walls, filaments and cluster nodes, and investigate their colours and star formation activities. A unique feature of the analysis is that it explicitly addresses the multiscale aspects with respect to the galaxy population, assessing issues such as the fraction of (blue) galaxies as a function of the scale of the cosmic web pattern and its caustic features. We find that the galaxy properties form a continuum in the scale-space cosmic web. Intimately coupled to the hierarchical build-up of the cosmic structure, it also allows us to systematically assess the impact of the formation time of the various structural components of the cosmic web on the galaxy properties. This furthers insight into the establishment of the observed colour-density relation of galaxies. 
\end{abstract}




\input{contents.tex}

\section*{Acknowledgements}
We thank Bram Alferink for useful discussions and willingness to share results. 
This work relied heavily on the computational facilities of the \textit{Hábrók} supercomputer of the University of Groningen. BH is supported by a Science and Technology Funding Council (STFC) PhD studentship, JF is supported by the STFC Consolidated Grant ‘Particle Physics at the Higgs Centre,’ and by a Higgs Fellowship, RvdW acknowledges funding from EU Horizon Europe (EXCOSM, grant nr. 101159513). For the purpose of open access, the authors have applied a Creative Commons Attribution (CC BY) license to any Author Accepted Manuscript version arising from this submission.

\section*{Data Availability}

This work made use of the publicly available data catalogue of the IllustrisTNG cosmological simulation suite, hosted at \verb|https://www.tng-project.org/|.



\bibliographystyle{mnras}
\bibliography{bibliography} 








\bsp	
\label{lastpage}
\end{document}

%% file: contents.tex
\section{Introduction}

\begin{figure*}
    \centering
    \includegraphics[width=\linewidth]{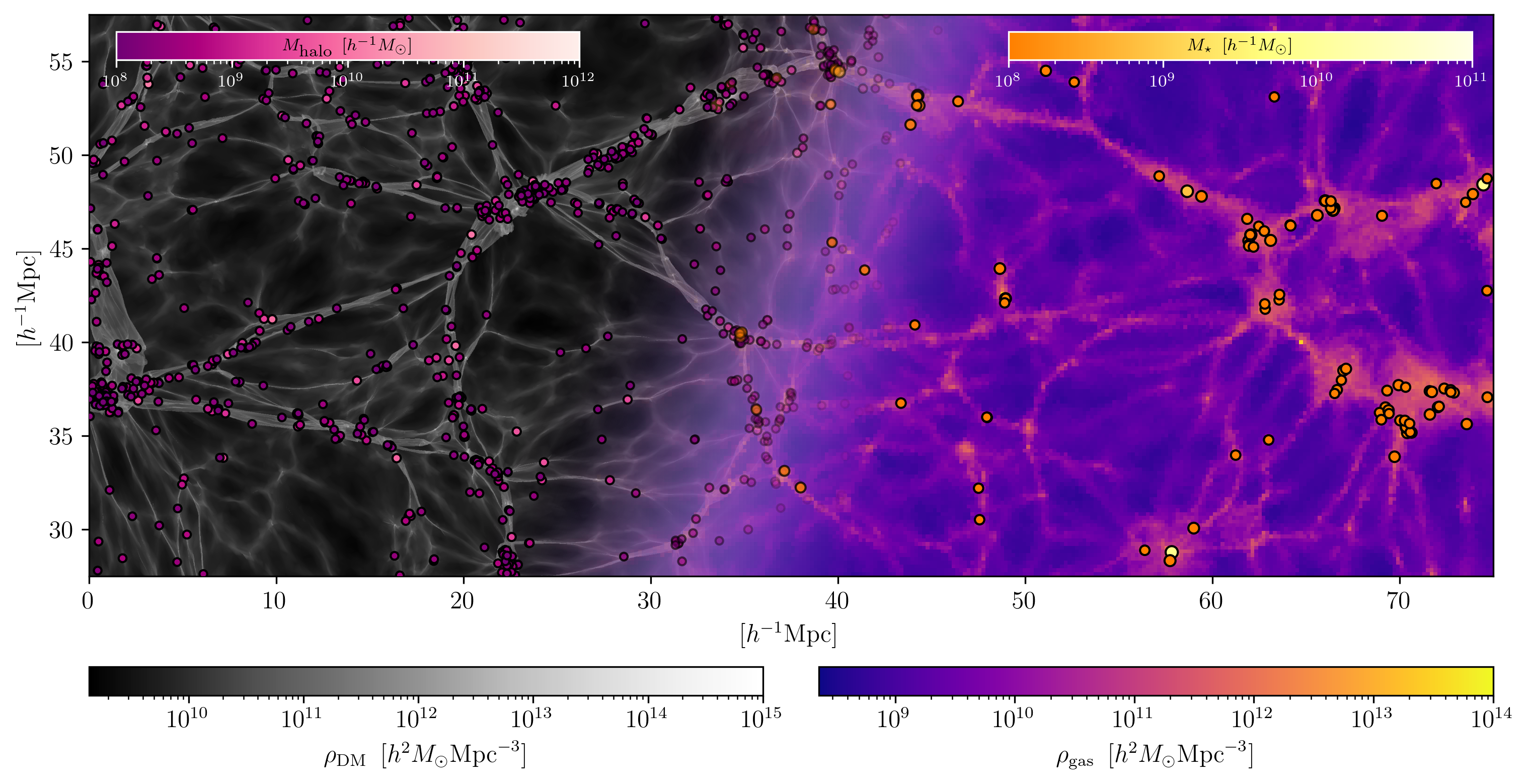}
    \caption{The dark and baryonic cosmic web in the TNG100 simulation of the IllustrisTNG suite. Shown is the density of dark matter (left) and baryonic gas (right) in a slice of width $70\,h^{-1}\textrm{Mpc}$ and height $30\,h^{-1}\textrm{Mpc}$. The left and right hand also display the haloes and galaxies, respectively, contained within a slab of thickness $\epsilon = 0.1\,h^{-1}\textrm{Mpc}$ around the shown slice, with their total or stellar masses indicated by the corresponding colour bars.}
    \label{fig:dm_gas_density}
\end{figure*}

The \textit{Cosmic Web} is the intricate multiscale network \citep{Zeldovich1970, JoeveerEinastoTago1977, BondKofmanPogosyan1996, WeygaertBond2008, Cautun+2014} defining and representing the fundamental spatial organisation of galaxies and matter on scales of a few up to a hundred megaparsecs. Galaxies, intergalactic gas and dark matter arrange themselves in a prominent wispy pattern of dense compact clusters, elongated filaments, and tenuous, sheet-like walls surrounding near-empty void regions. The filaments are the most visually outstanding features of the Megaparsec Universe, in which around $50\%$ of the mass and galaxies in the Universe reside. On the other hand, almost $80\%$ of the cosmic volume belongs to the interior of voids \citep[see e.g.][]{Cautun+2014, Ganeshaiah+2018}. Together, they define a complex spatial pattern of intricately connected structures, displaying a rich geometry of various morphologies and shapes. In addition to the rich pattern comprising a diversity of morphologies, the cosmic web is marked by at least three additional key aspects: the distinct multiscale nature, the complex connectivity of its structural constituents, and the large asymmetry between overdense and underdense volumes. This intrinsic multiscale nature of the cosmic web, including objects over a considerable range of spatial scales and densities, considerably enhances its complexity. For example, its filamentary backbone contains large and dense arteries that define the spine of the web-like pattern and form the major transport channels on megaparsec scales, while these branch out into a network of as tenuous, low-density tendrils that populate the underdense void regions. Likewise, the void population includes dominant expansion basins, as well as a large fraction of minor voids, most of which are subvoids within their larger peers, often near their boundaries with surrounding walls and filaments \citep[see e.g.][]{ShethWeygaert2006}. The connectivity of this rich assembly of features, the nature of how the various structures connect to establish the pervasive network, has only recently been recognised as an important defining -- topological -- aspect of the cosmic mass distribution \citep{AragonCalvo+2010, CodisPogosyanPichon2018, Wilding2022, FeldbruggeYanWeygaert2023}. It may even offer one of the most outstanding probes for inferring cosmological parameters \citep{Wang2026}. 

In the observational reality, the existence and structure of the Cosmic Web have been revealed in the most detail by maps of the nearby cosmos produced by large galaxy redshift surveys. Starting from the first revelation of the web-like arrangement of galaxies by the CfA2 survey \citep[e.g.][]{LapparentGellerHuchra1986}, subsequent surveys such as 2dFGRS, the SDSS, the 2MASS and GAMA redshift surveys~\citep{Colless+2003,SDSS2004,Huchra+2012,Liske:2015} established the web-like arrangement of galaxies as a fundamental characteristic of cosmic structure. Maps of the galaxy distribution at larger cosmic depths, such as VIPERS~\citep{Vipers+2013}, showed its existence over a sizeable fraction of cosmic time. Undoubtedly, at the moment, the most-detailed visualisation of the Cosmic Web is provided by the stunning 3D map of the intricate web-like galaxy distribution in the Local Universe recently obtained by the DESI (Dark Energy Spectroscopic Instrument) galaxy survey, comprising no less than 47 million galaxies \citep[see e.g][]{Zapata-Zuluaga:2026}. Current and upcoming Stage-IV surveys, such as the DESI survey, the Euclid surveys, and the Legacy Survey of Space and Time (LSST), unveil in ever-increasing detail the distribution of galaxies in the local Universe, and thus provide a traced view of the cosmic mass distribution out to moderate redshifts. The web-like nature of the galaxy network will further be highlighted by broader and deeper surveys such as the Roman Space Telescope \citep{Roman2022}, and spectroscopic observations by 4MOST \citep{4MOST2019} and WEAVE \citep{WEAVE2024} will not only provide unprecedented detailed maps of the web-like galaxy distribution, but also allow for detailed studies of the physical properties of galaxies as a function of their web-like environment. 

The intergalactic gaseous medium closely follows the web-like structure defined by the dark matter, the principal component of the cosmic web. A range of observational probes have detected the web-like structure over which intergalactic
gas \citep[for a review, see][]{Meiksin2009} has diffused itself. While baryons represent no more than a sixth of the Universe's matter content, the vast majority of these baryons are  not locked up in galaxies, but in the diffuse, largely ionised, intergalactic medium. A range of studies have addressed the question of the influence on the thermal and spatial gas properties as a function of its spatial web-like distribution, closely related but spatially more diffuse than the dark matter cosmic web \citep[e.g.][]{Sembach:2004, Dolag:2006, Tuominen:2021, Cui:2018, Bonjean:2018}. The baryonic gas traces the dark matter distribution, and some fraction of it constitutes the reservoir out of which the luminous galaxies are assembled and accrete gas. Despite its notoriously elusive nature in cosmic surveys, recent observational efforts by \cite{Connor+2025} demonstrate that the baryonic gas is to become an observationally relevant probe in the near future. Ly$\alpha$ absorption lines in the spectra of bright background sources such as QSOs are piercing through the web-like assembly of neutral hydrogen gas in the cosmic web at high redshifts \citep{Ostriker:1996,Cen:1997}. The combination of sufficiently close linear probes even allows a reconstruction of the full three-dimensional intergalactic hydrogen lines \citep{Pichon:2001}. It has already led the CLAMATO survey \citep{Khee-Gan+2018} to successfully produce fascinating maps of the full three-dimensional gaseous cosmic web at high redshifts. Recent observations by the MUSE integral field unit on the Very Large Telescope even managed to see the Ly$\alpha$ emission from the filamentary gaseous extensions around clusters directly \citep{MUSE}. At lower redshifts, most of the intergalactic gas has heated up as it settled in the deepening potential wells of the dark matter cosmic web. This warm gas, the so-called WHIM, is assumed to represent the major share of baryons in the current Universe. As such is a prime target for detection and mapping \citep{Nicastro:2018,Macquart:2020}, although it has proven to be notoriously hard to detect. It has been more straightforward to detect the hot gas residing in the strongest filamentary features in the cosmic web at high redshifts, filling the short dense bridges between  adjacent clusters. The hot gas reveals itself through the Sunyaev-Zeldovich upscattering of CMB photons. It has even allowed the detection of a few individual high-redshift gaseous filaments, where their ubiquitous presence has been revealed by stacking Sunyaev-Zeldovich observations of numerous cluster pairs \citep{Bonjean:2018,deGraaff:2019}.

Beyond the luminous tracers, recent observational advances have enabled first, yet significant glimpses into the web-like nature of the dark matter from weak lensing surveys, see \cite{Massey:2007,Dietrich:2012,Xia:2020,Scognamiglio:2026, Cha:2026}. Accounting for about 86 per cent of the Universe's matter, it is the dark matter that shapes the cosmic web and defines its underlying geometry that is traced by the galaxies. Augmented by the data catalogues of peculiar velocities from the CosmicFlows4 project \citep{Tully:2023}, the DESI survey \citep{DESI1, DESI2}, the upcoming 4MOST HS survey \citep{Colless:2024} and the LSST supernovae measurements \citep{LSST}, it will be possible to assess the dynamics -- the forces and tides -- shaping the cosmic web as well as get a more detailed inventory of the dark matter distribution that ultimately defines its gravitationally directing component.

\subsection*{Galaxies and the cosmic web}
Noting that the galaxies do still represent the most informative probe of the structure and dynamics of the cosmic web,
it is of crucial importance to understand the nature, morphology and properties of the galaxies that form and evolve in the cosmic
matter distribution, and in how far they constitute a fair discrete sample of the underlying (dark) matter distribution. To this end,
we need to assess the fundamental characteristics of galaxies as a function of cosmological environment. This involves
systematic studies of environmental influences on galaxy formation and assembly, and relating these to the morphological location
of observed galaxies. 

It has been known since the 1980s that the large-scale structure environments impact the properties of the embedded galaxies. In a seminal study, \cite{Dressler1980} observed the galaxy populations in 55 clusters and discovered a pronounced correlation between galaxies' morphologies and their environmental densities. His observations revealed that the densest clusters are populated largely by the \textit{early-type galaxies} (ellipticals and lenticulars), whereas the fraction of \textit{late-type galaxies} (spirals and irregulars) increases for decreasing environmental density. Commonly known nowadays as the \textit{morphology-density relation}, this correlation has since been extended to other web environments and confirmed in observational probes over six orders of cosmic density magnitudes, see e.g. \cite{PostmanGeller1984, Dressler+1997, Goto+2003}.

A direct measure of the influence of the cosmic web environment on galaxy properties is provided by the masses and luminosities of galaxies in clusters, filaments, walls, and voids. Voids and walls are mainly populated by low-mass objects, while regular Milky-Way mass haloes reside in filaments, and the most massive ones in clusters (for a telling illustration, see e.g. fig. 17 in \cite{Cautun+2014}). This leads to the overall question of the bias of galaxies, i.e. the spatial variations of galaxy number density with respect to the underlying dark matter distribution, or with respect to other galaxy populations. In particular, the influence of the cosmic web on the growth and assembly history of haloes and galaxies has been recognised as a key factor in determining their present-day properties. The relation between halo growth and cosmic web environment is partially a reflection of the tidally directed dynamical evolution of and anisotropic inflow onto the haloes \citep{AragonCalvo+2007a, Dalal+2008,Hahn+2009,Porciani+2017,Musso+2018,Paranjape+2018b,Zhang+2021}. The result is a halo assembly history and time that is modulated by the cosmic web, and translates into a halo and galaxy bias known as \textit{assembly bias} \citep{Gao+2005,Wechsler+2006,Dalal+2008,Mao+2018}. \citet{Salcedo+2020} found observational evidence for such bias in the SDSS survey. It may also be reflected in the relation between the brightness of galaxies and their level of clustering, as has been established for the GAMA survey~\citep{Jarrett+2017} and SDSS survey~\citep{Paranjape+2018}.

The most outstanding and immediate dynamical impact of the web-like environment on a major characteristic of galaxies is that of the spin of galaxies. The angular momentum of galaxies is a major factor in determining their morphologies. The same tidal field that shapes the cosmic web \citep[see eg.][]{Kugel+2026} is also the source of angular momentum build-up in collapsing haloes and galaxies. This is neatly encapsulated by Tidal Torque Theory (TTT), which explain how in the linear stages of evolution the tidal field torques the non-spherical collapsing proto-haloes to generate a net rotation \citep{Hoyle+1949,Peebles+1969,EfstathiouJones+1980,White+1984,LeePen+2000,Porciani+2002,Schafer+2009,Lopez+2024}. More recent theoretical and simulation-based studies have revealed the mass-dependence of the spin orientation, with high-mass galaxy haloes having a preferential perpendicular alignment, while low-mass haloes tend to parallelise their spins to the filament in which they reside \citep{AragonCalvo+2007a,Hahn+2007,Codis+2012,Tempel+2013,Ganeshaiah+2018,Ganeshaiah+2019,Ganeshaiah+2021,Lopez+2021,Lopez+2025}. The reality of the effect has also been confirmed by observational studies \citep{Welker+2020}. The transition mass from halo spins preferentially perpendicular to preferentially parallel is known as the spin-flip mass. There exists a diverse set of explanations for this phenomenon, with most of the responsible processes having to do with the nature of halo late-time mass accretion \citep{Bertschinger+1985}, in which the induced inflow of mass along a filament may be of decisive importance \citep{Ganeshaiah+2021}.

Another important environmental impact is that of the mass and gas flows along the components of the cosmic web, their impact on the accretion of gas in galaxies and the resulting star formation rate and history of galaxies. While not largely recognised, it may be that the filamentary nature of the cold gas accretion is a determining factor \citep{DekelBirnboim+2006, Keres+2005, Keres+2009}. The recent recognition by \cite{AragonNeyrinck+2016} of the central role of the filamentary web in determining galaxy star formation histories -- via its regulation of cold gas inflow -- highlights the importance of assessing how the cosmic web environment governs galaxy properties.

In terms of specific large-scale environments, voids have as yet allowed the most systematic observational appraisal of their low density  on the properties of galaxies residing in their interior. Several observational programs have been exploring the characteristics of void galaxies. The Void Galaxy Survey (VGS) has been investigating the gas content, star formation activity and structural parameters of galaxies living in voids \citep{Kreckel+2011,Kreckel+2012,Beygu+2016,Beygu+2017}. The currently ongoing CAVITY survey \citep{Conrado+2024,Perez+2024,GarciaBenito+2024} has been following this up with a more detailed program, including that of IFU spectroscopic observations. Nearly all void galaxies are small, low-mass galaxies, often somewhat irregular, with blue colours indicating active star formation activity, confirming the earlier results obtained by Vogeley, Hoyle and collaborators on the basis of the SDSS survey \citep{Rojas+2004,Hoyle+2005,Rojas+2005,Hoyle+2012}. Recently, the CAVITY survey even managed to obtain significant information on the detailed star formation history of void galaxies \citep{DominguezGomez+2023}.

Observations over the past 25 years, most notably by the Sloan Digital Sky Survey (SDSS) and Hubble Space Telescope, revealed that population of galaxies in our neighbourhood consists of a bimodal distribution of blue, star-forming galaxies on the one hand and red, quiescent galaxies on the other hand, see \cite{Strateva+2001, Baldry+2004, Balogh+2004, Menanteau+2005} to name but a few relevant papers. Comprehensive analyses by e.g. \cite{Kaufman+2004, Hogg+2004, Balogh+2004, Blanton+2005, Rojas+2005, Kraljic+2017, Kuutma+2017, Kuchner+2022, GalarragaEspinosa+2022} and numerous subsequent works demonstrated that the galaxies' colours and star formation activities are highly dependent on the environmental densities, as denser environments host redder, less active galaxies, whereas blue and active galaxies are typically found in the less dense regions of the large-scale structure (see also \cite{Rojas+2005,Kreckel+2011,Tempel+2013,Alpaslan:2014,Kraljic+2017,GalarragaEspinosa+2022}). In analogy to Dressler's morphology-density relation, these results are nowadays succinctly known under the name of the \textit{colour-density relation}, as first termed by \cite{Cooper+2007}. While being in approximate mutual correspondence, \cite{Bamford+2009} showed that the colour-density relation is more robust over the full range of dark matter densities across the cosmic web than the galaxies' morphologies. Concretely, the authors demonstrated that the relationship between the colour and morphology is more nuanced than a simple one-to-one mapping. With transformations in the galaxy colours occurring on shorter timescales than those of their morphologies, the galaxy colour is generally more sensitive to the galaxy mass function, and ultimately the environmental dark matter density. The more inert nature of the galaxy morphologies therefore explains the existence of atypical galaxy populations\footnote{Concretely, the study showed that although early-type galaxies always exhibit higher red fractions, blue ellipticals do nonetheless exist. Conversely, a non-negligible fraction of spiral galaxies are red, underscoring that colour and morphology are not in strict one-to-one correspondence.}, rendering the density-morphology relation a less rigorous correlation when seen over the whole of the cosmic web.

In summary, a systematic survey of galaxy properties and their interpretation in terms of their formation histories and evolution is required. However, such an analysis is much more complicated for the majority of galaxies, as they reside in walls, filaments and clusters. Their application to the Cosmic Web in the Local Universe will allow us to construct systematic samples of galaxies populating filaments and relate their morphology, colour, star formation activity, and a range of structural galaxy parameters, and statistical correlations of these properties, to the nature of the environment in which they live. The analyses by a range of studies \citep[e.g.][]{HaynesGiovanelli:1986,Alpaslan:2014, Metuki+2015,Kraljic+2017, Kuutma+2017,Xu+2020, GalarragaEspinosa+2022, Kuchner+2022, Hasan+2023, Hasan+2025,Yu+2025, NandiPandeySarkar2026} have highlighted quantifiable trends of the dependence of galaxy properties on the large-scale web environments. 

\subsection*{Galaxy formation simulations \& cosmic web classification}
Augmenting observational probes, the availability of large-scale cosmological simulations has driven significant advances in modelling the cosmic web and its embedded galaxies. In recent years, this has enabled a series of detailed analyses of galaxy properties across different environments of the simulated cosmic web, e.g. in the SIMBA \citep{Metuki+2015}, EAGLE \citep{Xu+2020}, the IllustrisTNG \citep{Hasan+2023, Yu+2025, NandiPandeySarkar2026} suites, and most recently the Flamgino and Colibre simulation suites \citep{Schaye:2023, Kugel:2023, Schaye:2025}. Specifically tuned simulation explorations of weblike environmental effects on halo and galaxy formation involve constrained setups, such as the MIP formalism of \cite{AragonCalvoMIP+2016} and the Splicing technique by \cite{Cadiou+2021} (see \cite{Storck+2025} for a telling illustration wrt. filament influence on halos). The simulation-based analyses support the observational evidence of the cosmic surveys and have confirmed in greater detail that the populations of galaxies in different web environments exhibit a pronounced bimodality in their characteristic properties. Intermediate- and high-density environments, namely the cosmic filaments and clusters, tend to accommodate predominantly ``quenched'' or ``quiescent'' galaxies, that is, galaxies with low star formation rates. These typically host cooler, redder stellar populations whose haloes have been efficiently depleted of their baryonic content. In contrast, the galaxy populations in lower-density environments, namely the cosmic voids and walls, are typically bluer and actively star-forming. The active galaxies contain younger stars with low metallicities, and their haloes have retained a considerable baryon fraction up to the current time. The simulation-based analyses therefore support the observational evidence of cosmic surveys, and give a more nuanced view of the results of \cite{Conselice2006}: while the galaxy mass is a key factor in determining the galaxies' physical properties, it is the environmental density embedded in the large-scale web that implicitly defines the characteristic galaxy masses. 

However, while the colour-related properties are resolved by simulations in greater detail than from the observations presently available, it is at present challenging to robustly infer optical morphologies from the simulation's kinematical particle data; for a discussion of the shortcomings of the IllustrisTNG case, see \cite{RodriguesGomez+2018}. We suspect that this is why, to our knowledge, simulation-based analysis of galaxy morphologies across different web environments are yet absent from the literature. We likewise defer a detailed treatment of galaxy morphology to future studies, and instead focus in the present article on the bimodality of the galaxies' physical properties in the large-scale structure.

To some extent, the environmental influences yielded by these studies depend on the characterisation and classification of the web-like environment, which in most situations are based on heuristic criteria underlying the various available schemes. A major share of these use the dark matter distribution in simulations as the tracer sample, while a range of these have also been able to resort to the observed galaxy distribution to analyse the cosmic web in the observational reality. The review by \cite{Libeskind+2018} discusses and compares a wide range of these formalisms. They involve topological methods, such as the \verb|DisPerSE| method \citep{Sousbie2011} and \verb|SpineWeb| formalism \citep{AragonCalvo+2010}, stochastic methods such as the Bisous method \citep{Tempel+2016}, Graph and percolation methods \citep{Alpaslan:2014} and the Monte Carlo Physarum Machine (MCPM) that exploits the structural resemblance of slime moulds to the cosmic web \citep{Elek:2021, Elek:2022}. Turning to more physical criteria, dynamical and kinematical considerations,  underlie the T-web and V-web formalisms \citep{Hahn+2007, Forero-Romero+2009, Hoffman+2012, Libeskind:2012} which look at the local tidal and velocity shear signature to characterise the cosmic environment. Most analyses address the cosmic matter distribution or velocity flows at one particular physical (filter) scale. The MMF-\verb|NEXUS(+)| formalism \citep{AragonCalvo+2007, Cautun+2013} explicitly addresses the fundamental aspect of the multiscale nature of the cosmic web in considering morphological filters -- resorting to either the mass density, tidal field or velocity flows -- over a range of scales, within the context of its scale-space formalism, yielding at each location within a simulation or survey volume the web-like morphology at one locally dominant scale. A major step in the attempt to physically underpin the cosmic web classification has been made in several formalisms that exploit the phase-space structure of the -- mostly simulated -- mass distribution \citep{Shandarin2011, ShandarinSalmanHeitmann2012, AbelHahnKaehler2012, Falck:2012}. 

\subsection*{Galaxies and the caustic skeleton}
The present study takes the fundamental step of basing its cosmic web characterisation on a firm physical footing that involves the key aspects of its multiscale and structural character with that of a true physical, theoretical framework for its dynamical formation and development. In other words, it not only seeks to base the cosmic web characterisation on the web-like pattern in the mass and/or galaxy distribution at one -- often the current -- cosmic epoch, but intends to characterise the cosmic web on the basis of the formation history and time of its structural constituents. In this, it follows the realisation that the dynamics and formation histories of voids, walls, filaments and clusters are integral to the present-day morphology of the cosmic web. The \textit{caustic skeleton model} \citep{Feldbrugge+2018} is the expression of a phase-space dynamics-based identification of the structural singularities that emerge in the evolving cosmic mass distribution as gravity moulds the primordial Gaussian density field into an ever more complex pattern \citep{Peebles:1980}. It follows up on the seminal work of Zel'dovich and collaborators \citep{Zeldovich1970, ArnoldShandarinZeldovich1982, Arnold1986}.

Zeldovich \citep{Zeldovich1970} showed how the present-day (dark) matter distribution arises as an intricate multistreaming network that emerged through the foldings of the initially smooth dark matter sheet into a `creased', `overlapping', web-like configuration. Formalised into a rigorous description of cosmic structure formation, this phase-space perspective of the dark matter dynamics has given rise to the caustic skeleton model  \cite{ArnoldShandarinZeldovich1982,Hidding+2013, Feldbrugge+2018}, in which the emergence of the web elements is identified through the occurrence of caustics in the dark matter flow. A mathematically rigorous and parameter-free formalism, the caustic skeleton traces the formation history of the cosmic web and associates the morphologies and dynamical origin of the present-day walls, filaments and clusters to an exhaustive set of \textit{elementary catastrophes (caustics)} \cite{Thom1972}. Moreover, the caustic skeleton offers a natural framework to study the multiscale nature of the cosmic mass distribution, imprinted already in the morphology of the primordial density perturbation. In recent years, a series of papers \citep{FeldbruggeWeygaert2023, FeldbruggeYanWeygaert2023, FeldbruggeWeygaert2024, Hertzsch+2026} have provided significant progress in our understanding of the statistical properties and morphology of the large-scale structure elements traced by the dark matter caustics. Notably, by means of constraint simulations \citep{Bertschinger:1987,HoffmanRibak:1991,WeygaertBertschinger1996,FeldbruggeWeygaert2024}, the caustic skeleton model has provided detailed insights into the nature of the cosmic voids, walls, filaments and cluster nodes, and how these relate to primordial conditions out of which the Universe originated. 

To assess the properties of the population of galaxies within the mass distribution, we resort to the use of a state-of-the-art cosmological simulation that includes both the dark matter and gas distribution as well as the emerging galaxy population. This allows us to study the properties of the galaxies with respect to the emerging web-like pattern that signals the transition from linear to non-linear development. To this end, we employ the IllustrisTNG simulations \cite{Pillepich+2017a, Nelson+2019}, which may be rightfully seen as a community standard structure and galaxy formation suite. The evolving cosmic web in the simulation is followed by applying the caustic skeleton formalism. Using the formalisms' unprecedented phase-space insights, we trace the scale-cosmic web through the multiscale network of caustics. With this rigorous definition of the voids, walls, filaments and cluster nodes, we provide a comprehensive analysis of the galaxy properties in the different web environments. As such, we for the first time take into account the multistreaming nature of the cosmic mass distribution to discern the characteristic colours and star formation activities of the embedded galaxy populations, and will directly confirm the observationally supported colour-density relation. As a major step beyond traditional analyses of the present-day density field, we investigate the formation time of the cosmic web elements and study their impact on the galaxy properties.

\subsection*{Overview of the present study}
The structure of this paper is as follows. In \cref{sec:caustics}, we review the multistreaming nature of the cosmic web and the caustic skeleton formalism tracing its geometric backbone. We here place particular emphasis on the multiscale nature of the cosmic mass distribution and how it is recovered by the scale-space caustic skeleton. In \cref{sec:illustris}, we discuss the IllustrisTNG simulation suite and introduce two computational methods that enable us to trace the multistreaming dark matter and investigate the Lagrangian flow provided by $N$-body simulations. We then use the latter to construct the caustic skeleton of the TNG100 and TNG300 simulations. In \cref{sec:field}, we investigate the multistreaming dark matter density field and use the web identification provided by the caustic skeleton to study the properties of the baryonic gas in the different web environments. Our discussion leads up to \cref{sec:galaxies}, the main part of this article, in which we investigate the galaxy properties in the different web environments. We study the colour, star formation and related properties and provide a comprehensive analysis of galaxies in the multiscale caustic network. Further, we discuss the overall population of galaxies across the different scale-space web environments. In \cref{sec:formation_time}, we investigate the impact of the web formation time and quantify its correlation with the embedded galaxies' colours and star formation activities. Finally, we conclude our results and give an outlook on further work.


\section{The caustic skeleton of cosmological structure formation}
\label{sec:caustics}

We commence our study with a discussion of the caustic skeleton, delineating the geometric backbone of cosmological structure formation.

\subsection{The multistreaming cosmic web}
\label{subsec:caustics-multistream}

\begin{figure*}
    \centering
    \includegraphics[width=\linewidth]{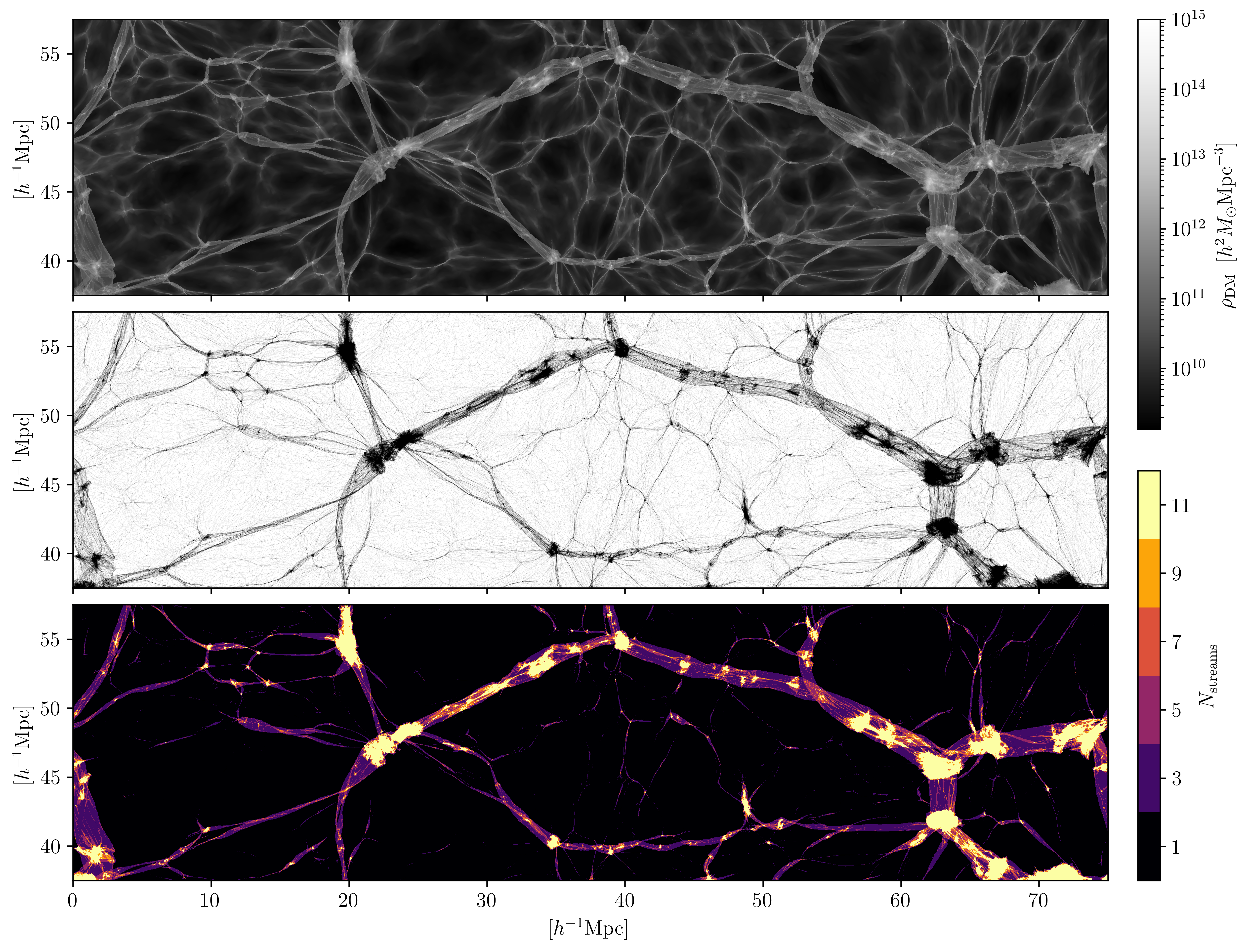}
    \caption{Different views of the multistreaming cosmic web in the TNG100 simulation. The upper panel shows a slice through the multistream dark matter density field, evaluated from \cref{eq:Eulerian_density} using the PS-DTFE method (see \cref{subsec:illustris-comp_methods}). The middle panel shows the corresponding slice through the dark matter particle mesh. The lower panel counts the number of incoming dark matter streams, again evaluated using the PS-DTFE method.}
    \label{fig:triple_plot}
\end{figure*}

The cosmic web is not merely the result of smooth density accretion, but rather the geometric manifestation of an intricate folding of the dark matter sheet over cosmic time. Ever since the seminal work by \cite{Zeldovich1970} and collaborators, it has been appreciated that the web-like morphology of the dark matter density distribution as observed today may be inferred from the properties of the dark matter flow and the corresponding formation history of the large-scale structure. This \textit{phase-space perspective} of cosmological structure formation is suitably understood in the \textit{Lagrangian fluid formalism}, where the dark matter particles stream from initial coordinates $\bm{q}$ at high redshift to the present-day configuration $\bm{x}$ through a displacement field $\bm{s}_t({\bm{q}})$ as
\begin{equation}
    \bm{x}_t(\bm{q}) = \bm{q} + \bm{s}_t({\bm{q}})\,.  \label{eq:Lagrangian_fluid}
\end{equation}
It is the properties of the displacement field $ \bm{s}_t({\bm{q}})$ that give rise to the plethora of structures making up the present-day cosmic web. Starting from a smooth distribution of dark matter in the early Universe, the primordial density perturbations grow as dark matter starts to accumulate. At early times, the displacements are small and the Universe remains in a \textit{single-streaming} configuration, in which the mapping between the coordinates $\bm{q}$ and $\bm{x}_t$ is one-to-one. That is, for each $\bm{x}_t$, there is a single corresponding initial coordinate $\bm{q}$ out of which the final coordinate has emerged.

As time progresses, this ceases to be true. With the overdensities becoming considerable, the displacements grow accordingly and particle trajectories eventually start to cross. The qualitative properties of $\bm{s}_t({\bm{q}})$ therefore change as the Universe enters a \textit{multistreaming} configuration, in which several initial coordinates  $\bm{q}$ stream into the same final coordinate $\bm{x}_t$. These  \textit{shell-crossing} events are at the heart of the seminal Zel'dovich approximation \citep{Zeldovich1970}. After shell-crossing, the mapping \cref{eq:Lagrangian_fluid} is no longer one-to-one, and the growing overdensities are traced out by the cosmic multistreaming volume in which several incoming dark matter streams overlap. For a more detailed discussion, we refer to \cite{Hertzsch+2026}.

The present-day cosmic web is the result of these shell-crossing events and constitutes an inherently multistreaming structure, the geometry of which is traced out by the dynamics of the dark matter sheet over cosmic time. See \cref{fig:triple_plot} for an analysis of the multistream regions in a slice of the IllustrisTNG simulation. The web-like morphology of the density field in the upper panel follows the structure of the overlap of the particle mesh in the middle panel. The lower panel reveals that the dominant overdensities correspond to regions with a large number of incoming streams, whereas the cosmic voids and minor overdensities remain in single-stream regions.

From the phase-space perspective of the dark matter flow, the density field at the time $t$ is given by the sum over all incoming streams,
\begin{equation}
    \rho \left( \bm{x}, t \right) = \sum_{\bm{q} \in \bm{x}_t^{-1}(\bm{x})} \frac{\rho(\bm{q}) }{|\det \nabla_{\bm{q}} \bm{x}_t(\bm{q})|} \,.
    \label{eq:Eulerian_density}
\end{equation}
Here, the determinant of the deformation tensor weighs the different contributions. It is immediately apparent that density diverges when the determinant vanishes. This is the physical manifestation of the shell-crossing events, at which the dark matter sheet folds over to give rise to a higher-streaming configuration, thus delineating the overdense cosmic web elements. In mathematical terminology, these events are also known as \textit{catastrophes} or \textit{caustics} (see the foundational work and reviews by \cite{Thom1972, Arnold1972, Zeeman1977, Saunders1980, Arnold1992, ArnoldGuseinZadeVarchenko2012}).

\subsection{Caustic skeleton theory}
\label{subsec:caustics-theory}

Multistreaming is central to the formation of the cosmic web. Caustics mark the points where the folding of the particle mesh occurs and the density formula, \cref{eq:Eulerian_density}, diverges. These insights were first appreciated in the early work of \cite{ArnoldShandarinZeldovich1982}. However, it was only recently that \cite{Feldbrugge+2018} formalised these concepts into the \textit{caustic skeleton model}, in which the morphology and formation history of the walls, filaments and cluster nodes are rigorously identified with an exhaustive list of geometrically distinct caustics arising in the large-scale dark matter flow. These are known as the \textit{fold}, the \textit{cusp}, the \textit{swallowtail}, the \textit{butterfly} and the \textit{umbilics}, which have different formation histories, distinct geometries and local environments in the cosmic mass distribution. It is this identification that will be the central tool in our analysis of the cosmic web and the galaxy properties in the IllustrisTNG simulations.

\begin{figure}
    \centering
    \includegraphics[width=\linewidth, trim={17cm 4.5cm 17cm 3.5cm},clip]{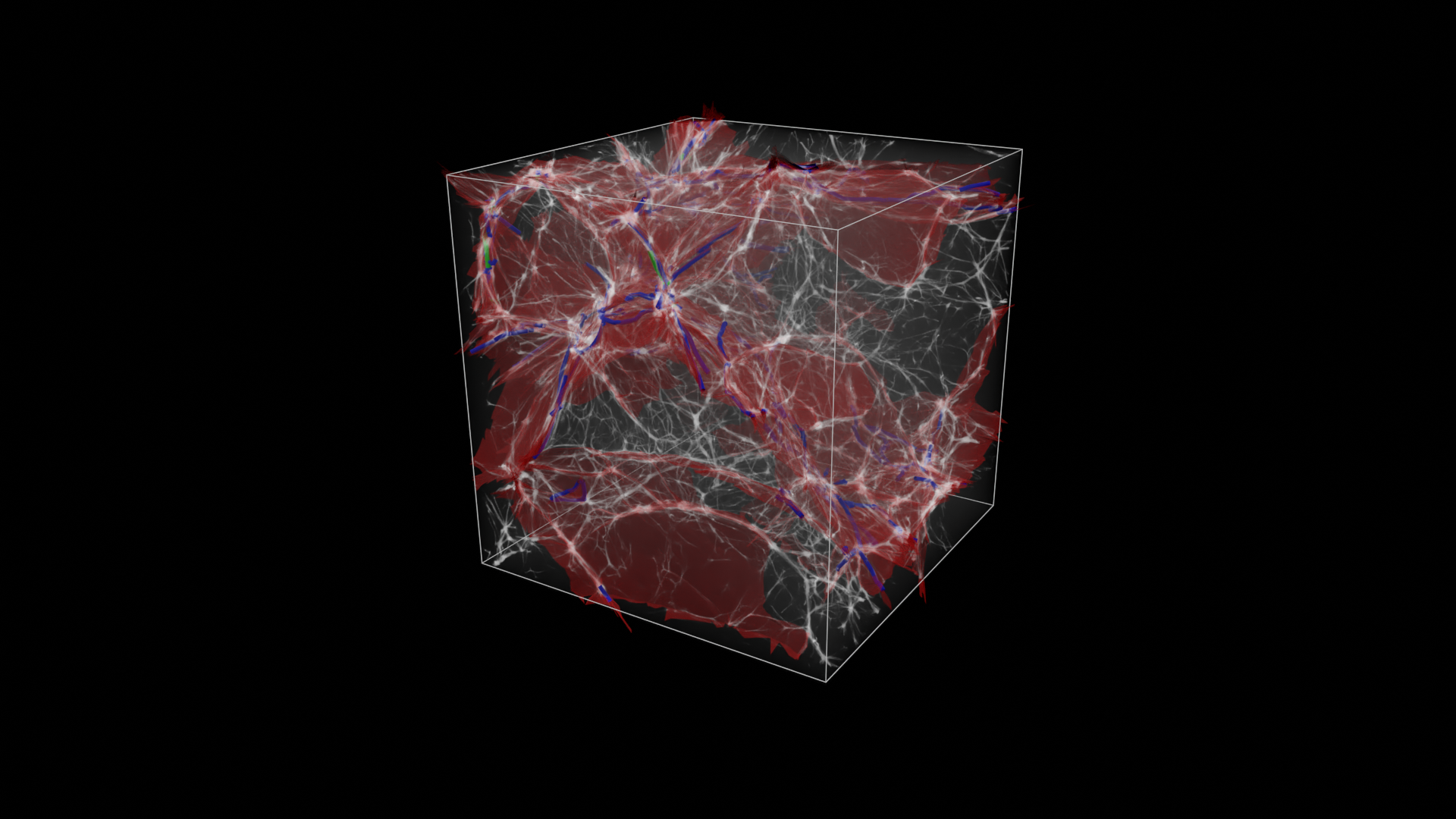}
      \cprotect\caption{Three-dimensional view of the cosmic web and the caustic skeleton in the TNG100 simulation. Shown is a volume of side length $50 \,h^{-1}\textrm{Mpc}$, with the large-scale cusp sheets (walls) and the swallowtail and umbilic filaments in red, blue and green, respectively. A rotating-view animation of the scale-space caustic skeleton, tracing the entirety of the multistreaming web through the network of small- to large-scale caustics (see \cref{subsec:caustics-scale_space}), is provided at \verb|benhertzsch.github.io/papers/2026_IllustrisTNG_Caustics|.}
    \label{fig:caustics_3D}
\end{figure}

\begin{figure*}
    \centering
    \includegraphics[width=\linewidth]{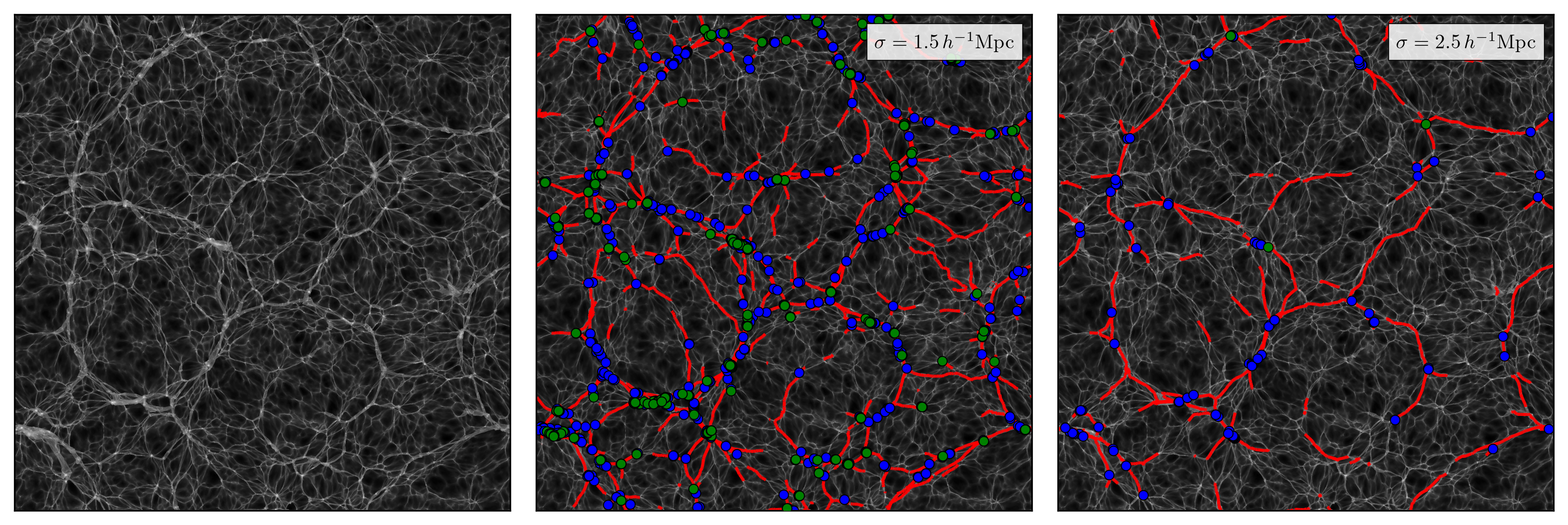}
    \caption{Density field and caustic skeleton of the TNG300 simulation. Shown is a slice through the entire simulation box of side length $205 \,h^{-1} \textrm{Mpc}$. The middle panel shows the web traced by the intermediate-scale caustic skeleton, evaluated at $\sigma=1.5 \,h^{-1} \textrm{Mpc}$. The slices through the walls and swallowtail and umbilic filaments are displayed in red, blue and green, respectively (see the 3D visualisation of \cref{fig:caustics_3D}). The right panel shows the large-scale web features traced by the caustic skeleton evaluated at $\sigma=2.5 \,h^{-1} \textrm{Mpc}$.}
    \label{fig:caustics_tng300}
\end{figure*}

In \Cref{fig:caustics_3D}, we show a three-dimensional render of the cosmic density field in the TNG100 simulation along with the large-scale caustic skeleton tracing its geometrical backbone. Remarkably, the caustics identify the entirety of the large-scale filamentary morphology of the overdense multistream regions with stunning accuracy. Moreover, the accompanying animation demonstrates that the smaller-scale web features are equally traced by the caustics at smaller length scales, as we will discuss in more detail in \cref{subsec:caustics-scale_space}.
The accuracy of the caustic skeleton model is even better appreciated in the two-dimensional slices of the TNG300 density field shown in \cref{fig:caustics_tng300}. Here, the middle and right panels show the caustic skeleton at intermediate and large length scales, respectively. The slices through the sheet-like walls (red) reveal the web-like nature of the caustics and the present-day dark matter distributions, while the slices through the line-like filaments (blue and green) reside exactly in the pronounced overdensities seen in the left panel. These illustrations demonstrate that the caustic skeleton, as a dynamical theory of the formation history of the cosmic matter distribution, is an unprecedented and powerful identification tool of the present-day cosmic web. Having illustrated its practical merit, we now summarise the theory of the caustic skeleton model. We refer to \cite{Hertzsch+2026} for a more detailed review.

\begin{table*}
	\centering
	\begin{tabular}{lccr}
		\hline
		name & symbol & caustic condition & cosmic web element\\
		\hline
		fold & $A_2$ & $\lambda_1 = b_c^{-1}$ & multistream region\\
        cusp & $A_3$ & $\bm{v}_1 \cdot \nabla \lambda_1 = 0$ & wall\\
        swallowtail & $A_4$ & $\bm{v}_1 \cdot \nabla \left( \bm{v}_1 \cdot \nabla \lambda_1 \right) = 0$ &  filament\\
        butterfly & $A_5$ & $ \bm{v}_1 \cdot \nabla\left(\bm{v}_1 \cdot \nabla \left( \bm{v}_1 \cdot \nabla \lambda_1 \right) \right) = 0$ &  cluster (node) \\
        umbilic & $D_4$ & $\lambda_1 = \lambda_2 = b_c^{-1}$ & filament\\
        parabolic umbilic & $D_5$ & $ \det (\mathcal{H}_{\bm{v}_1\bm{v}_2}(\det \nabla \bm{x}_t))  = 0$ & cluster (node)\\
		\hline
	\end{tabular}
    \caption{Summary of the caustic conditions in the Zel'dovich approximation with the correspondence to the structural elements of the cosmic web in the three-dimensional Universe. Here, $\mathcal{H}_{\bm{v}_1\bm{v}_2}$ represents the Hessian in the $\bm{v}_1\bm{v}_2$-plane.}
    \label{tab:caustic_skeleton}
\end{table*}

First conceptualised by \cite{ArnoldShandarinZeldovich1982, Arnold1986}, and systematically formalised for the three-dimensional Universe by \cite{Feldbrugge+2018}, the caustic skeleton identifies the cosmic web elements by the singularities in the dark matter flow marking the formation of multistream regions. 
For visualisations in 2D, see \cite{Hidding+2013}; for detailed illustrations of the 3D case, we refer to the \cite{Hertzsch+2026}. Applying catastrophe theory (see e.g. \cite{Thom1972, ArnoldGuseinZadeVarchenko2012}), \cite{Feldbrugge+2018} showed that the caustics are identified by an exhaustive set of \textit{caustic conditions} on the deformation tensor $\nabla_{\bm{q}} \bm{s}_t$. These take the form of non-linear algebraic conditions determining special configurations of time- and space-dependent eigenvalue and eigenvector fields $\mu_i(\bm{q}, t)$ and  $\bm{v}_i(\bm{q}, t)$, defined by the eigenequation
\begin{equation}
    (\nabla_{\bm{q}} \bm{s}_t) \bm{v}_i(\bm{q}, t) = \mu_i(\bm{q}, t) \bm{v}_i(\bm{q}, t)  \,. 
\end{equation}

The importance of the eigenvalue fields has already been recognised in the seminal work by \cite{Zeldovich1970}, and can in fact be directly linked to the multistream density formula, \cref{eq:Eulerian_density}: by substituting  \cref{eq:Lagrangian_fluid}, one can write the determinant in terms of the eigenvalues as $|\det \nabla_{\bm{q}} \bm{x}_t(\bm{q})| = |\det \left( I +\nabla_{\bm{q}} \bm{s}_t(\bm{q}) \right)| = |1 +  \mu_1(\bm{q}, t)||1 +  \mu_2(\bm{q}, t)||1 +  \mu_3(\bm{q}, t)|$. The density $\rho(\bm{x},t)$ therefore diverges when any of the eigenvalues attains the value $  \mu_1(\bm{q}, t) = -1$, corresponding to the particle mesh foldings we described earlier. The importance of the eigenvalue fields has given rise to the widely used \textit{T-web formalism} and variations thereof \citep{Hahn+2007, AragonCalvo+2007, Forero-Romero+2009, Hoffman+2012}, in which the different cosmic web elements are identified with different numbers of eigenvalues attaining some threshold value. However, the community has, up to this point, remained largely oblivious to the role of the corresponding eigenvector fields. The caustic skeleton reveals that these play an equally central role, and that it is the configuration of the eigenvalue and eigenvector at a given time that determines the formation of wall sheets, filaments and cluster nodes, thus tracing out the cosmic web over time, see \cite{Feldbrugge+2018, Hertzsch+2026}. This is seen in \cref{tab:caustic_skeleton}, where we list the caustic conditions in the Zel'dovich approximation (see details below).

While we leave a detailed discussion of the caustic conditions aside for the present article, we now summarise the morphology of the different caustics, and refer the reader to \cite{Feldbrugge+2018, Hertzsch+2026} for details. The lowest in the hierarchy of caustics is the $A_2$ fold caustic, which determines the formation of a multistream region by the condition $\mu_1(\bm{q}, t)=-1$. This is the condition that we have seen above as the cause of the divergences in the density field; the caustic skeleton shows that these correspond to the dark matter particle mesh folding over into a multistream configuration. Each multistream region is bisected by an $A_3$ cusp sheet, corresponding to the progenitors of the late-time cosmic walls. When the cusp sheet folds onto itself, an $A_4$ swallowtail filament is born, which may in turn fold up into an $A_5$ butterfly cluster (node)\footnote{the Caustic Skeleton formalism covers the entire spectrum of scales, in which the nodes of the cosmic web may be, dependent on scale, be identified with galaxy scale halos at earlier cosmic epochs while at the current epoch they correspond to massive clusters situated at or near the nodes of the cosmic web. This will be elaborated upon in a forthcoming study. In the present study, the name \textit{clusters} when strictly speaking it concerns the \textit{nodes} of the weblike network.}. Catastrophe theory shows that there exists a second family of catastrophes, with the $D_4$ elliptic and hyperbolic umbilics corresponding to filaments and the $D_5$ parabolic umbilics corresponding to cluster nodes. The $A$ and $D$ families are qualitatively different, as for the latter collapse occurs due to two eigenvalue fields. The $A_4$ and $D_4$ filaments are therefore expected to exhibit distinct geometric properties in the cosmic web, the first evidence of which was presented in \cite{Feldbrugge+2018, FeldbruggeWeygaert2023, Hertzsch+2026}. Here, it is found that the swallowtail filaments are more elongated and less overdense, corresponding to the boundaries of or the creases in the extended wall sheets. The umbilic filaments, in turn, are found to be shorter, thicker and more overdense. Moreover, they have a three-fold symmetry and form the junctions of three incoming wall sheets. A detailed study of the dynamical origin and morphology of the distinct filament families will be given in \cite{HertzschFeldbruggeWeygaert2026}. For this article, we focus on the general identification of the walls, filaments and cluster nodes by the caustic skeleton, and leave these detailed considerations aside.

Written in the form presented above, the caustic skeleton is a mathematically rigorous formalism that is applicable to the non-linear dynamics of gravitationally driven structure formation (as given e.g. by $N$-body simulations). However, the caustic theory fully reveals its practical power when combined with an analytical approximation for the dynamics. Notably, by adopting the seminal \textit{Zel'dovich approximation} (ZA) \citep{Zeldovich1970}, one can already identify the caustics marking the nascent and evolving cosmic web in the initial conditions imprinted on the primordial potential perturbation.

In the ZA, the fluid elements follow linear trajectories from their initial positions and the displacement field is given by the gradient map
\begin{equation}
    \bm{x}_t(\bm{q}) = \bm{q} - b(t) \nabla \Psi(\bm{q}) \,.
    \label{eq:ZA}
\end{equation}
Here, the \textit{primordial displacement potential} $\Psi(\bm{q})$ is related to the primordial potential perturbation $\phi(\bm{q})$ by \cite{Hidding+2013, FeldbruggeWeygaert2024}
\begin{equation}
    \Psi(\bm{q}) = \frac{2}{3 H_0^2 \Omega_m} \phi(\bm{q}) = \frac{2}{3 b(t) a(t)^2 H_0^2 \Omega_m} \phi_{\textrm{lin}}(\bm{q}) \,,
    \label{eq:displacement_potential}
\end{equation}
where in we used  the Hubble function $H(t)$, the scale factor $a(t)$, the matter density parameter $\Omega_m$ and the linearly extrapolated potential perturbation $\phi_{\textrm{lin}}$ at time $t$.
The temporal dependence is absorbed by the linear growth factor  $b(t)$, which gives the solution for the linear growth of perturbations through the defining equation \citep{Peebles1994}
\begin{equation}
    \ddot{b}(t) + 2 H(t) \dot{b}(t) - 4 \pi G \rho_0 b(t)=0 \,,
\end{equation}
with the boundary conditions $b(0)=0$ and $b(t_0)=1$, the primordial mean density $\rho_0$ and the gravitational constant  $G$.

In the ZA, the deformation tensor is proportional to the Hessian of the primordial displacement potential,  $\nabla_{\bm{q}} \bm{s}_t = -b(t)\mathcal{H}\Psi$. The caustic conditions reduce to functions of the eigenvector and eigenvalue fields $\lambda_i(\bm{q})$ and $\bm{v}_i(\bm{q})$ of the Hessian $\mathcal{H}\Psi$, defined by the eigenequation
\begin{equation}
    \left[\mathcal{H} \Psi(\bm{q})\right] \bm{v}_i(\bm{q}) = \lambda_i(\bm{q}) \bm{v}_i(\bm{q})  \,. 
\end{equation}
The caustic conditions in the ZA, summarised in \cref{tab:caustic_skeleton}, are functions of space only \fix{(i.e. in the space of initial conditions)}. This means that the primordial density fluctuations directly govern the emerging cosmic web through the associated tidal eigenvalue and eigenvector fields.

At this point, it is instructive to comment on the applicability of the caustic skeleton from the ZA; we refer the reader to \cite{Hertzsch+2026} for a more detailed discussion. It is well known that as structure formation progresses into the mildly non-linear regime, the ZA ceases to accurately describe the cosmic density field, as it assumes particles to follow linear trajectories and ignores turn-around, phase-mixing and virialisation into haloes after shell-crossing. Nevertheless, the linear trajectories of the ZA suitably predict the locations and formation times of the dominant shell-crossing events that make up the large-scale cosmic web elements and mark the onset of strongly non-linear gravitational collapse. Consequently, by evaluating the caustic conditions on  $\Psi(\bm{q})$ according to the ZA, and by subsequently evolving the identified objects into the final-configuration space with the non-linear displacement field $\bm{s}_t(\bm{q})$ (modelled by the $N$-body simulation), one aptly traces the cosmic web as emerging out of the fully non-linear structure formation dynamics (see \cref{fig:caustics_3D,fig:caustics_tng300} for a demonstration). 
Moreover, \cref{fig:caustics_scale_space} highlights the fundamental role of the tidal eigenvalue field: not only do the caustic conditions identify the relevant shell-crossing events, but the filamentary morphology of the late-time universe is already imprinted in the non-Gaussian nature of the primordial eigenvalue field (see \cite{FeldbruggeYanWeygaert2023, FeldbruggeWeygaert2023, Hertzsch+2026} for more details). The visual correspondence between the density distribution in the upper panel and the eigenvalue field in the lower panel is striking.  Furthermore, \cref{fig:caustics_scale_space} illustrates another important aspect of the caustic skeleton model, as it shows the caustics evaluated at different length scales. We now turn our attention to this aspect, and discuss in more detail the multiscale nature of the caustic skeleton outlining the physical cosmic web.

\subsection{Primordial conditions and the scale-space cosmic web}
\label{subsec:caustics-scale_space}

\begin{figure*}
    \centering
    \includegraphics[width=\linewidth]{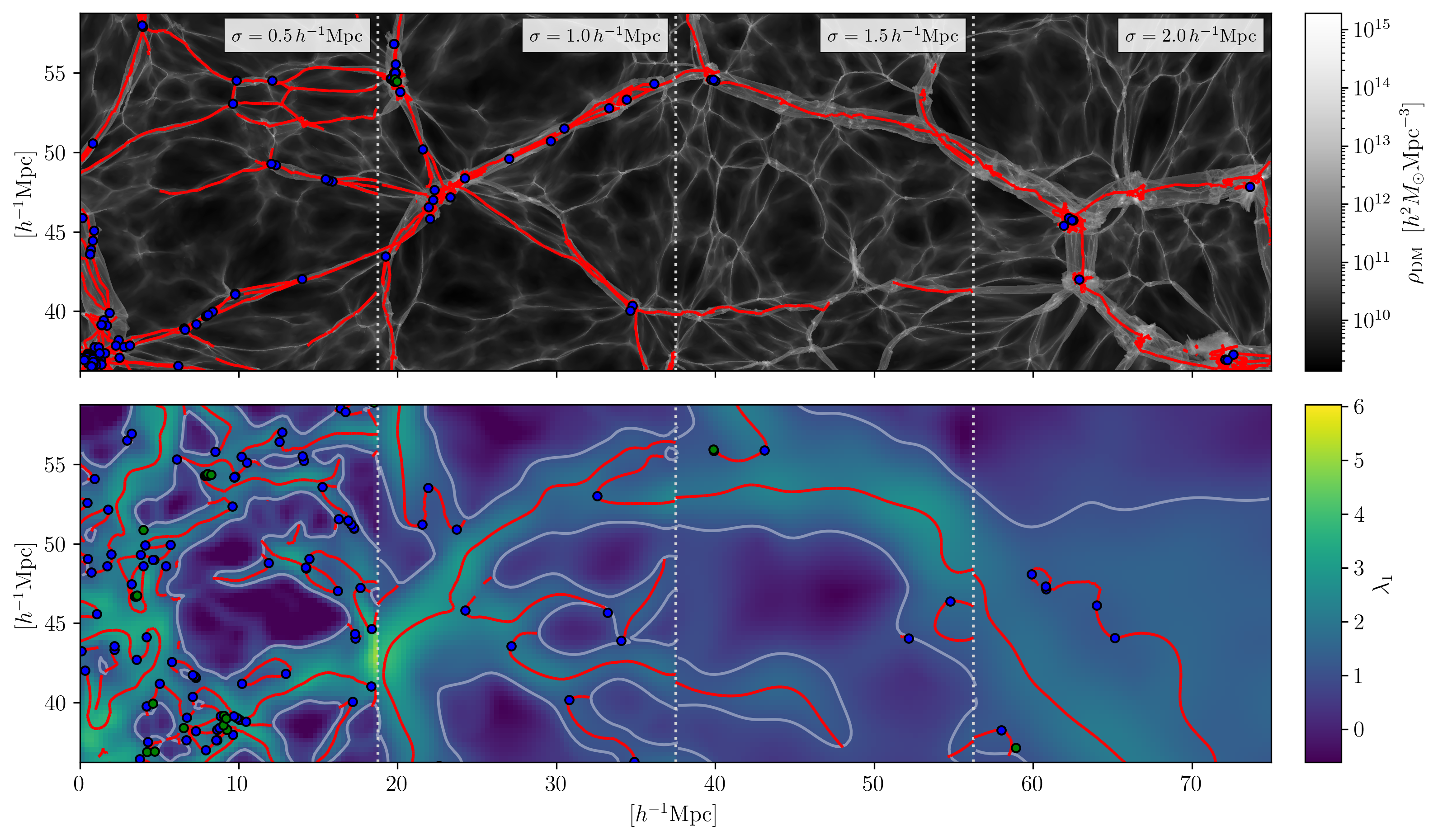}
    \caption{The scale-space caustic skeleton and its origin in the primordial tidal fields. The upper panel shows a slice through the present-day density field of the TNG100 simulation (the same as in \cref{fig:triple_plot}), along with the caustics tracing the cosmic web elements. The cusp sheets (walls) and swallowtail and umbilic filaments are shown in red, blue and green, respectively. The caustics are evaluated at different smoothing scales to trace the small- to large-scale features of the multiscale cosmic mass distribution. The lower panel shows the (approximately) corresponding slice through the initial conditions. Shown is the first eigenvalue field $\lambda_1$ of the Hessian $\mathcal{H} \Psi$, with $\Psi$ smoothed at the respective smoothing scales according to \cref{eq:smoothing_real_space} The caustics are evaluated through the caustic conditions listed in \cref{tab:caustic_skeleton}, and evolved into the final-configuration space to trace the present-day web elements (upper panel). The correspondence between the upper and lower panels is striking, with the filamentary nature of the present-day mass distribution clearly imprinted in the non-Gaussian nature of the first eigenvalue field.}
    \label{fig:caustics_scale_space}
\end{figure*}

Observations of the cosmic microwave background suggest that the late-time cosmic web originated from tiny fluctuations with respect to an overall homogeneous and isotropic background. As several analyses before, the Planck 2018 best-fit results \citep{Planck2018} indicate that the primordial potential perturbation $\phi(\bm{q})$ is close to a Gaussian random field with a nearly scale-invariant power spectrum.

The caustic skeleton model is sensitive to the multiscale nature of the primordial perturbations, and the caustics identify the entirety of the small-scale to the large-scale features of the cosmic web from the primordial yields. Yet, it is well known that, at the current time, the ZA is accurate only at large scales. One might therefore expect that the formalism, as outlined in the preceding section, is only applicable to the largest scales of the present-day cosmic web. Fortunately, 
\cite{Hertzsch+2026} recently demonstrated that the caustics obtained from the ZA accurately describe the cosmic web elements down to even below-megaparsec scales (see left panel of \cref{fig:caustics_scale_space}), thus making the entirety of the multistreaming network accessible to the caustic conditions on the primordial fields.

However, in practice, one is typically not interested in the entirety of the multiscale cosmic web, but rather in the features occurring at a given length scale. Following up on early work by \cite{Zeldovich1970, Doroshkevich1970, PressSchechter1974, DoroshkevichShandarinSaar1978} and collaborators, \cite{BondKofmanPogosyan1996} solidified this notion in the \textit{scale-space} picture of the cosmic web. The multiscale description of the cosmic web has since been firmly established in the community, and has led to numerous dedicated studies; for a review, see \cite{Hertzsch+2026}.

In the context of the caustic skeleton model, we follow \cite{Feldbrugge+2018, FeldbruggeWeygaert2023, FeldbruggeYanWeygaert2023, FeldbruggeWeygaert2024, Hertzsch+2026} and identify the cosmic web elements at a given length scale $\sigma$ by the caustics evaluated on the smoothed displacement potential $\Psi_{\sigma}(\bm{q})$. The latter is obtained by convolving the displacement potential with a smoothing window function $W_{\sigma}(\bm{q})$ as

\begin{equation}
    \Psi_{\sigma}(\bm{q}) = \int \d \bm{q}^{\prime} \Psi(\bm{q} - \bm{q}^{\prime} ) W_{\sigma}(\bm{q}^{\prime} ) \,.
    \label{eq:smoothing_real_space}
\end{equation}
In this work, we choose $W_{\sigma}(\bm{q} )$ to be given by the Gaussian filter
\begin{equation}
    W_{\sigma}(\bm{q}) = \frac{1}{2 \pi \sigma^2} e^{-\frac{\bm{q}^2}{2 \sigma^2}}
    \label{eq:window_function}
\end{equation}
with the smoothing scale $\sigma$.

Note that \cref{eq:smoothing_real_space} is a non-local operation averaging out the potential at a Lagrangian position $\bm{q}$ with that of its nearby vicinity. The caustics identified at smoothing scale $\sigma$ are those that persist in the smoothing process, and therefore have an extent of the order of or greater than $\sigma$ in the initial condition space (i.e. on the primordial potential perturbation). The entirety of the scale-space caustic skeleton is a fractal-like structure, with small networks of caustics occurring within the large-scale caustics and reflecting the fractal-like nature of the cosmic web (see e.g. \cite{Einasto+2020, Einasto2025}) over the range of cosmologically relevant length scales. This is suitably illustrated in \cref{fig:caustics_scale_space}, in which the small-scale caustics are sensitive to minor multistreaming overdensities, embedded in the dominant, large-scale structural components identified by the large-$\sigma$ caustics. First explorations of statistical properties of the scale-space caustic skeleton are found in \cite{FeldbruggeWeygaert2023, FeldbruggeYanWeygaert2023, FeldbruggeWeygaert2024} for the 2D case, and in \cite{Hertzsch+2026} for the 3D case. However, the statistics of the primordial fields are beyond the scope of the present study. Instead, we investigate the phenomenology of the cosmic web and its embedded galaxies, and place particular emphasis on its scale-space nature as identified in the caustic skeleton formalism.


\section[]{The I{\sevensize\bf llustris}TNG Simulations}
\label{sec:illustris}

In this section, we outline the IllustrisTNG simulation suite and describe the construction of the caustic skeleton for the TNG100 and TNG300 simulations.

\subsection{Overview}
\label{subsec:illustris-overview}

Cosmological simulations are the cornerstone of numerical structure formation research. Ever since the early simulation efforts by \cite{KlypinShandarin1983}, the worth of and need for high-resolution, large-scale cosmic simulations has been firmly established. After decades of ever-increasingly resolved and sophisticated simulation suites (for reviews, see e.g. \cite{SomervilleDave2015, AnguloHahn2022, CrainVoort2023}), the \textit{IllustrisTNG} suite \citep{Nelson+2019} represents one of the most developed and large-scale cosmic simulations, and has become a modern community standard.

The IllustrisTNG project is a suite of large-volume, gravity and magnetohydrodynamical cosmological simulations that solve for the coupled evolution of dark matter, gas, luminous stars, and supermassive black holes from the starting redshift $z = 127$ into the current Universe, $z=0$. This is achieved by the massively parallel moving-mesh code \verb|AREPO| \citep{WeinbergerSpringelPakmor2020} on a $\Lambda $CDM background with fixed cosmology $\{\Omega_{\Lambda,0} = 0.6911, \Omega_{m,0}=0.3089, \Omega_{b,0}=0.0486, \sigma_8=0.8159, n_s=0.9667, h = 0.6774\}$ \citep{Nelson+2019}, which we accordingly adopt throughout this article. The galaxy formation model \citep{Weinberger+2017, Pillepich+2017a} used in the IllustrisTNG simulations is an empirically calibrated subgrid description that supplements gravity and ideal magnetohydrodynamics with prescriptions for gas cooling and heating, star formation, stellar evolution and chemical enrichment, stellar winds, black-hole seeding and growth, and multi-mode AGN feedback, tuned to reproduce the key low-redshift observables such as the stellar mass function and cosmic star formation history. The (sub-)halo catalogues are computed using the \verb|Subfind| algorithm \citep{Springel+2001}, which extends the traditional friends-of-friends (FoF) method to identify gravitationally bound (sub-)haloes within the FoF groups.

The IllustrisTNG project contains simulations of varying resolutions and cosmological volumes. In this work, we are interested in the TNG100 and TNG300 simulations, which were presented in a series of five presentation papers by \cite{Marinacci+2018, Naiman+2018, Nelson+2017, Pillepich+2017b, Springel+2017}. Both simulations are available at three resolutions, either including the full baryonic physics model or as the dark-matter-only counterpart. Crucially, for both the TNG100 and TNG300 simulations, the different runs simulate the same realisation of the chosen volume, respectively. That is, while the galaxy properties are suitably studied from the high-resolution baryonic runs, the large-scale dark matter fields can be evaluated from the low-resolution dark-matter-only runs. We will make use of this in the following. Concretely, the simulations have the following specifications:
\begin{itemize}
    \item \textbf{TNG100}: volume $L=75^3 \, h^{-3} \textrm{Mpc}^3$ with low-resolution and high-resolution dark matter particle numbers $455^3$ and $1820^3$ respectively. The high-resolution dark matter and gas particle masses are $m_{\textrm{DM}}=7.5 \cdot 10^6 M_{\odot}$ an $m_{\textrm{gas}}=1.4 \cdot 10^6  M_{\odot}$ respectively, with $\approx 4.4 \cdot 10^6$ subhaloes identified at $z=0$.
     \item \textbf{TNG300}: volume $L=205^3 \, h^{-3} \textrm{Mpc}^3$ with low-resolution and high-resolution dark matter particle numbers $625^3$ and $2500^3$ respectively.  The high-resolution dark matter and gas particle masses are $m_{\textrm{DM}}=5.9 \cdot 10^7 M_{\odot}$ an $m_{\textrm{gas}}=1.1 \cdot 10^7  M_{\odot}$ respectively, with $\approx 1.4 \cdot 10^7$ subhaloes identified at $z=0$.
\end{itemize}

Consequently, with the given particle numbers and comoving volumes, the TNG100 and TNG300 simulations aptly resolve the below-megaparsec-scale physics that we analyse in this article, and at the same time contain a representative number of large-scale cosmic web elements. Moreover, with more than a million subhaloes being identified at the current time, both simulations contain a substantial sample of galaxies to study the properties across the different cosmic web environments. 

We note that the particle mass resolutions are different for the two simulations. While the TNG300 simulation covers a larger volume and hence contains more large-scale cosmic web elements, as well as substantially more galaxies, the individual galaxy properties in the TNG100 suite are more accurately computed due to its finer resolution. The issue of numerical convergence is addressed in detail in \cite{Pillepich+2017a} and further raised in \cite{Nelson+2019}. The authors state that care must be taken when combining analyses of the TNG100 and TNG300 data, as some galactic properties may be quantified differently by the different mass resolutions, as is shown for the galactic stellar mass at fixed subhalo mass in \cite{Pillepich+2017a}. At this point, the resolution dependence of the IllustrisTNG results represents an inherent shortcoming of the simulation suite. For the present article, however, we find these quantitative effects to be subdominant to the web dependence of the galaxy properties of interest.

\subsection{Computational methods}
\label{subsec:illustris-comp_methods}

Throughout this article, we study the properties of smooth and continuous physical fields and their embedded galaxies from the output snapshot files of the IllustrisTNG simulations. These contain the relevant information at the discrete particle positions in their configuration at a given time, and therefore provide the smooth fields as samples on an unstructured grid. We here summarise the two computational methods we employ to obtain smooth and continuous field estimates of physical fields from the simulations.

\paragraph*{Phase-space DTFE.} Of primary importance to our study is the present-day multistreaming dark matter density field, obtained from the density formula \cref{eq:Eulerian_density}. Despite its solid theoretical underpinning, it is challenging to numerically evaluate for $N$-body simulations. Consequently, the multistreaming nature of the cosmic web has rarely been exploited in analysis of the dark matter density, see e.g. \cite{Libeskind+2018}. Recently, \cite{Feldbrugge2024} proposed the \textit{phase-space DTFE} (PS-DTFE) method, which extends the traditional \textit{Delaunay tessellation field estimator} (DTFE) method (\cite{SchaapWeygaert2000, WeygaertSchaap2009}, also see \cite{BernardeauWeygaert1996}) for the smooth and adaptive numerical evaluation of \cref{eq:Eulerian_density} from a tessellation of the coordinates at high redshift and their subsequent evolution into the present-day configuration. In essence, the PS-DTFE tracks the deformation of the primordial simplex mesh and identifies the present-day multistream regions through the overlap of the simplex cells. As such, the method extends previous phase-space field estimators proposed by \cite{Shandarin2011, ShandarinSalmanHeitmann2012, AbelHahnKaehler2012, Hahn+2015} and significantly improves upon the DTFE evaluations in the multistream regions, while reducing to the same in the single-streaming volume, see also \cite{Hertzsch+2026}. Throughout this article, we evaluate all dark matter density fields and the number-of-streams fields with the PS-DTFE method. To this end, we make use of a stable and efficient implementation provided by the publicly available \verb|julia| package \verb|PhaseSpaceDTFE.jl| \cprotect\footnote{Hosted at  \verb|github.com/jfeldbrugge/PhaseSpaceDTFE.jl| and available for installation from the \verb|julia| package manager.}, written by \cite{FeldbruggeHertzsch2025}.

\paragraph*{Discrete Natural Neighbours.} As we will highlight in \cref{subsec:illustris-caustics}, a major challenge in our analysis is posed by the glass pre-initial conditions of the IllustrisTNG suite (see \cite{Pillepich+2017a}). To enable a Lagrangian analysis from simulation data, one requires a robust method to infer the fields of interest on a regular grid in the primordial (i.e. initial) conditions. Multistreaming is not a concern here, and the reconstruction of fields from discrete samples at particle positions reduces to the standard -- yet numerically challenging -- problem of interpolation on an unstructured grid. 

A suitable method for this task is the \textit{natural neighbour interpolation} (also known as \textit{Sibson interpolation}) \citep{Sibson1980}, which applies Voronoi tessellations of the unstructured grid to give a smooth and adaptive, ``natural'' estimate of the field to be interpolated. In simple terms, the natural neighbour interpolation evaluates the field at a given point by ``naturally'' weighing the contributions of the nearest data points based on their distance (or, more precisely, volume in a tessellation simplex). Unfortunately, Voronoi tessellations are computationally costly and not feasible when the number of data points becomes large. This makes it impossible to naively apply the method to the IllustrisTNG data with several hundred particles per box side. Fortunately, \cite{Park+2006} proposed a clever method to approximate the volume calculations of the Voronoi simplices by discretised point-in-cell counts. This method, which authors refer to as \textit{discrete Sibson interpolation}, significantly reduces the computational cost of natural neighbour interpolation, while preserving its accuracy when the number of points is large. For this work, we have developed a stable and efficient implementation of the algorithm in \verb|julia| language, available as the registered package \verb|DiscreteNaturalNeighbors.jl|\cprotect\footnote{Hosted at \verb|github.com/benhertzsch/DiscreteNaturalNeighbors.jl| and available for installation from the \verb|julia| package manager.}. A key ingredient in our study, we use this implementation for the interpolation of the dark matter displacements, as well as the present-day gas density, temperature and metallicity fields.

\subsection{Evaluating the caustic skeleton of the IllustrisTNG simulations}
\label{subsec:illustris-caustics}

We now come to the central theoretical development of our study and discuss the construction of the caustic skeleton delineating the geometric backbone of structure formation in the IllustrisTNG simulations.

\subsubsection{Method}

For a given simulation, we obtain the caustic skeleton by evaluating the caustic conditions of \cref{tab:caustic_skeleton} on a regular grid of the smoothed primordial displacement potential $\Psi_{\sigma}$ out of which the particles originated. Often in cosmological simulations, this information is available, see e.g. \cite{Hertzsch+2026}. For the IllustrisTNG simulations, however, this information is not directly accessible, as the suite is set up with glass pre-initial conditions (see \cite{Pillepich+2017a}) rather than primordial particle displacements on a regular grid. This marks a major computational challenge and necessitates the use of an accurate and efficient interpolation routine, as we introduced in the preceding section. Making use of the same, we now outline in detail the numerical construction of the simulations' caustic skeleton.

Given a simulation, we infer the primordial displacement potential $\Psi$ from the snapshot files. These contain information about the particle positions $\bm{x}_i$ at time $t_i$, where $i$ here denotes the $i$th snapshot file and $\bm{x}_i$ is the set of all dark matter particle positions. The IllustrisTNG data do not contain information about the initial configuration of the particles and their velocities at time $t_0$ (i.e. at the time of recombination). Instead, the earliest available snapshot files are at given times $t_1, t_2$ shortly after the time of recombination. We use these to infer the initial-configuration coordinates $\bm{q}$, along with the respective velocities, out of which the simulation originated. 

In the early universe, the Zel'dovich approximation is a suitable description of the $N$-body dynamics that has not yet entered its non-linear stage. We therefore assume that the linear mapping of \cref{eq:ZA} accurately captures the displacement of the particles to both time $t_1$ and $t_2$. Using the linear displacement, we hence construct the initial-space particle positions $\bm{q}$ (i.e. the particle positions at time $t_0$) as
\begin{equation}
    \bm{q} = \bm{x}_1 - (\bm{x}_2-\bm{x}_1) \frac{b(t_2) - b(t_1)}{b(t_1)} \,.
    \label{eq:q_calculation}
\end{equation}
Using the initial-space particle positions, we then define the displacement field from $t_0$ to $t_1$ as $\bm{s}_1 = \bm{x}_1 - \bm{q}$ and identify the primordial displacement potential according to the ZA by writing
\begin{equation}
    \bm{s}_1 = b(t_1) \nabla \Psi \,.
    \label{eq:s1_ZA}
\end{equation}
That is, we take the displacement field  $\bm{s}_1$ to be given by the gradient of the potential we wish to construct.  

At this point, we have constructed $\bm{s}_1$ on the irregular grid $\bm{q}$ set by the glass pre-initial conditions of the IllustrisTNG simulations \citep{Pillepich+2017a}. Crucially, this is where the discrete natural neighbour interpolation comes in, as we can now construct $\bm{s}_{1, \textrm{grid}}$ on the regular grid $\bm{q}_{\textrm{grid}}$ by employing the method we motivated in \cref{subsec:illustris-comp_methods}.

From the interpolated displacement field $\bm{s}_{1, \textrm{grid}}$, we subsequently evaluate $\Psi$ by solving \cref{eq:s1_ZA}. For numerical stability, we do not directly invert the gradient operation, but instead employ standard Fourier methods to calculate the density perturbation $\delta_{\textrm{grid}} = t_1^{-1} \nabla \cdot \bm{s}_1$ and hence obtain $\Psi$ as the inverse Laplacian $\Psi = \nabla^{-2} \delta_{\textrm{grid}}$, where we omit the grid subscript for clarity here and in the following. Finally, the smoothed potential $\Psi_{\sigma}$ is obtained by convolving $\Psi$ with the Gaussian window function according to \cref{eq:smoothing_real_space}.

The Lagrangian-space caustic skeleton is numerically calculated on $\Psi_{\sigma}$, and evolved into the final-configuration space with the displacement field
\begin{equation}
    \bm{s}_E = \bm{x}_{E} - \bm{q} \,,
\end{equation}
which we read off from the particle positions $\bm{x}_{E}$ contained in the final snapshot file at $z=0$. Again, we use discrete natural neighbour interpolation to evaluate $\bm{s}_E$ on the regular grid $\bm{q}_{\textrm{grid}}$. This allows us to evolve the caustic skeleton from the space of initial conditions into its current-time configuration.

\subsubsection{Numerical specification}

We end this section with a brief discussion of the numerical specification of the IllustrisTNG caustic skeleton. As was outlined in \cref{subsec:illustris-overview}, the TNG100 and TNG300 simulations are available at different resolutions in both the full baryonic version and its dark-matter-only counterpart. For the determination of the caustic skeleton, we only use the dark matter particle positions. Moreover, it suffices to consider the particle positions from the low-resolution runs, as these suitably resolve the large-scale cosmic web elements, and hence the caustics, down to the relevant below-megaparsec scales.

Using the low-resolution dark matter particle positions, we construct the smoothed potential $\Psi_{\sigma}$ and the caustic conditions on a grid resolution such that fluctuation of length scale $\sigma$ can be appropriately resolved. Concretely, we use the following specifications:
\begin{itemize}
    \item \textbf{TNG100:} The low-resolution run contains $455^3$ particles in a box of side length $L=75\, h^{-1} \textrm{Mpc}$. We use these to interpolate the primordial potential on a grid of size $256^3$, thus giving an effective resolution of about $0.3\,h^{-1}\textrm{Mpc}$. We calculate the caustic skeleton for the full box at smoothing scales $\sigma= 0.5, \, \ldots,\, 2.5 \, h^{-1} \textrm{Mpc} $ in steps of $0.1\, h^{-1} \textrm{Mpc}$.
    \item \textbf{TNG300:} The low-resolution run contains $655^3$  in a box of side length $L=205\, h^{-1} \textrm{Mpc}$. We use these to interpolate the primordial potential on a grid of size $352^3$ particles, thus giving an effective resolution of about $0.8\,h^{-1}\textrm{Mpc}$. We calculate the caustic skeleton for the full box at scales $\sigma= 1.0,\, 1.5,\, 2.0,\, 2.5 \, h^{-1} \textrm{Mpc}$.
\end{itemize}

Using these specifications, we construct the caustics to trace the TNG100 and TNG300 cosmic web at different length scales. We omit the $D_5$ caustics here, as these are numerically challenging to identify and require higher grid resolution. This omission does not affect our present study, as we expect that the $D_5$ caustics, representing one family of cluster nodes, are always surrounded in close vicinity by the more prominent $A_5$ cluster nodes\footnote{Moreover, we do not expect their distinction to be as relevant to the properties of the embedded galaxies as that of the swallowtail and umbilic filaments, see \cref{sec:galaxies}.}. We therefore do not distinguish between these and identify the full cluster node population through the $A_5$ family.

\begin{figure*}
    \centering
    \includegraphics[width=\linewidth]{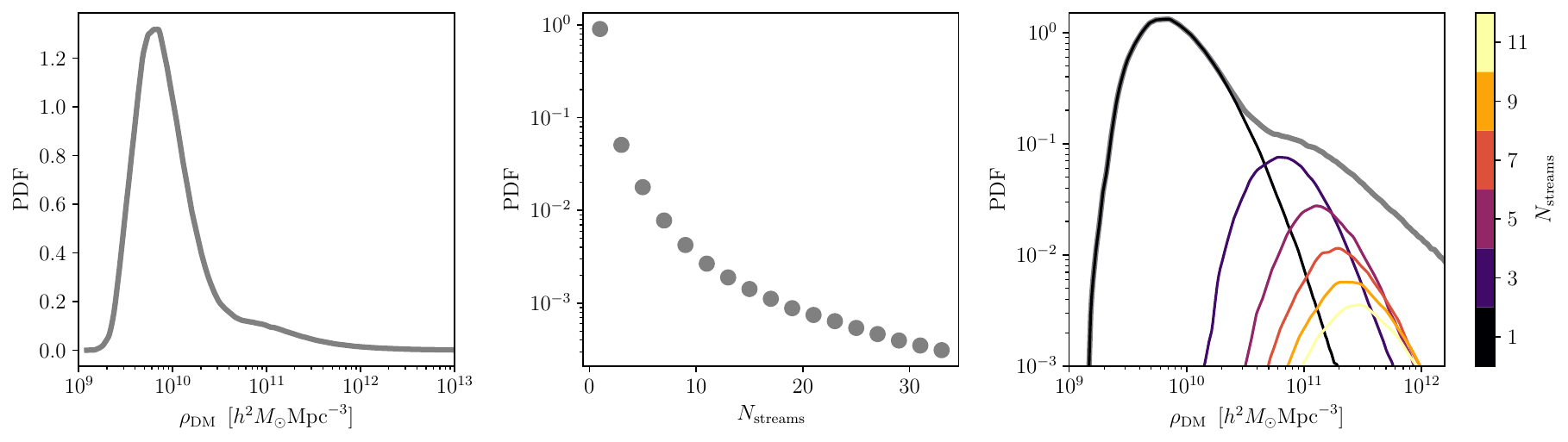}
    \cprotect\caption{Probability density functions (PDFs) of the multistream dark matter density (left) and number-of-streams field (middle) of the TNG100 simulation. The PDFs are evaluated as histograms of the field values over a grid of spacing $0.2 \,h^{-1} \textrm{Mpc}$ through the entire simulation box. The right panel shows the density field from the left panel on a log-log scale, displaying the contributions from the different streaming configurations (see colour bar). The overall log-normal nature of the density field (shown in grey) arises through the summation of the different multistream contributions. This figure was first conceptualised by the authors of \cite{Alferink+2026}. }
    \label{fig:density_nstreams_plot}
\end{figure*}


\section{Properties of dark matter and gas in the cosmic web}
\label{sec:field}

The present-day cosmic web is an intricate conglomerate of dark and baryonic matter woven together through the dynamics of structure formation over cosmic time. While it is the dark matter and its associated tidal field that are responsible for shaping the assembly of complex web-like geometry, it is baryonic gas embedded in the same that eventually clumps into the luminous galaxies we observe in cosmic surveys. Leading up to the detailed analysis of the galaxy properties, we discuss in the following a multistream dark matter field and some properties of the baryonic gas.

\subsection{The multistreaming dark matter field}

The dark matter density field has been a key object in the development of the modern cosmological standard model. Originating from near-perfect Gaussian initial conditions, it is well known that the current-time dark matter density field follows an approximately log-normal distribution, with a significant tail of high overdensities associated with the dense filaments and clusters in the cosmic web. \cite{Shandarin2011} and \cite{AbelHahnKaehler2012} addressed several aspects of the multistream character of the cosmic mass distribution, and its spatial relation with the existence and nature of structural features. The upcoming paper by \cite{Alferink+2026} extends this towards a systematic quantitative and statistical analysis, whose findings are confirmed by the related results outlined below. This concerns in particular the conclusions by \cite{Alferink+2026} with respect to the lognormal nature of both the density field and multistream field.

The left panel of \cref{fig:density_nstreams_plot} confirms that the multistreaming dark matter density field inferred from the TNG100 simulation confirms the well-known lognormal shape. The corresponding number-of-streams distribution shows that while the majority of the cosmic volume at present is still in a single-stream configuration, higher numbers of incoming streams are significant. Indeed, due to the phase mixing in the collapsing web structures, the probability density function (PDF) of the number-of-streams field has an extended tail to configurations with $N_{\mathrm{streams}} \approx 10^2-10^3$ incoming streams. It should be noted, however, that these configurations typically appear in superdense clusters and virialised regions, in which baryonic gas and dark matter are mixed such that the multistreaming perspective is no longer applicable. Nevertheless, the significant proportion of multistreaming volume with moderate streaming numbers suggests that the density field and, therefore, the cosmic web from its multistream perspective requires a careful treatment. The right panel of  \cref{fig:density_nstreams_plot} demonstrates that the log-normal distribution of the cosmic dark matter density field is due to the summation of the contributions from higher-streaming configurations, each of which contributes a higher mean density. 

The dark matter distribution is at the heart of the physical cosmic web. The histograms of \cref{fig:fields} show that cosmic web elements, as identified by the caustic skeleton at scale $\sigma=1.0\,h^{-1}\textrm{Mpc}$, constitute environments of significantly different dark matter densities. With the mean cosmic density being  $\approx 8.6 \cdot 10^{10}\,h^2 M_{\odot} \textrm{Mpc}^{-3}$, we find that the voids have a characteristic density contrast of $\rho / \bar{\rho} \sim 0.1-1$, therefore constituting extended cosmic underdensities. The walls, made up of the extended cusp sheets, are moderately overdense with $\rho / \bar{\rho} \sim 1-10$. The filaments cover a higher and significantly more extended mass range, with the typical overdensities being around $\rho / \bar{\rho} \sim 10-100$. It is also apparent that the umbilic filaments are typically more dense than the swallowtail filaments; this will be investigated in more detail in \cite{HertzschFeldbruggeWeygaert2026}.  Finally, the clusters constitute the densest environments, with density contrasts reaching up to $\rho / \bar{\rho} \sim 100 - 2000$. These results are consistent with numerous previous analyses of the dark matter density in the $\Lambda$CDM cosmic web, see \cite{Forero-Romero+2009, AragonCalvo+2010a,  ShandarinSalmanHeitmann2012, Hoffman+2012, Cautun+2014, Libeskind+2018}. Note, however, that with the exception of \cite{ShandarinSalmanHeitmann2012} these studies did not take into account the multistream nature of the cosmic web, and care must be taken when comparing to our results obtained with the PS-DTFE method. 

Comparing the densities of \cref{fig:fields} to the multistream densities in \cref{fig:density_nstreams_plot}, it is evident that the overdensities in the cosmic web elements arise as higher-streaming configurations, in accordance with the prediction of the caustic skeleton model, see \cite{Hertzsch+2026}. While we omit a further analysis of the number-of-streams field here, it can be readily concluded that voids consist of single-stream regions, while the moderately overdense walls can be associated with moderately low multistream regions. As for the filaments and cluster nodes, phase-mixing into highly multistreaming configurations is key in their evolution into significant overdensities.

\begin{figure*}
    \centering
    \includegraphics[width=\linewidth]{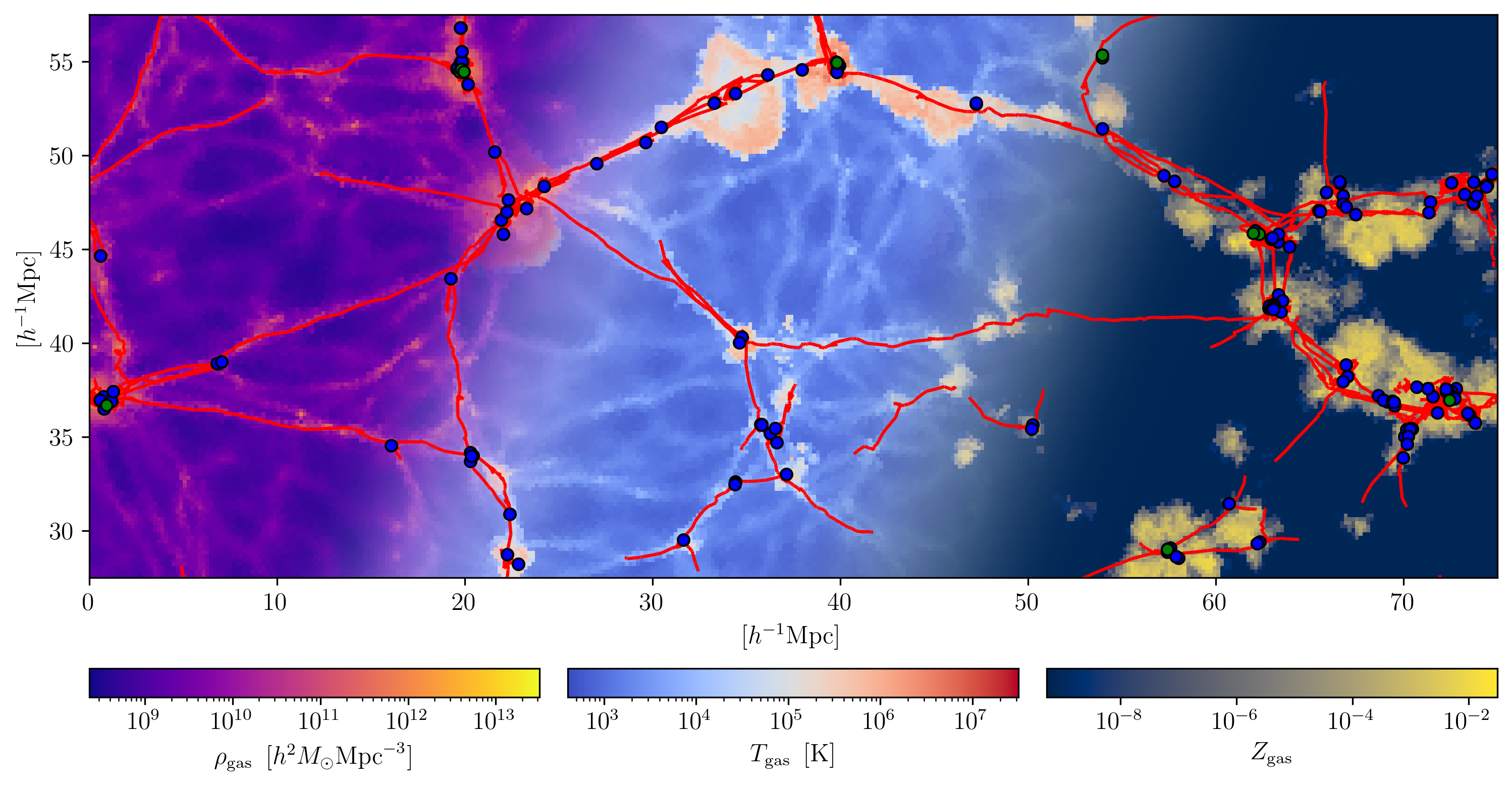}
    \caption{Different views of the baryonic gas web in the TNG100 simulation. Shown in the slice (same as in \cref{fig:dm_gas_density}) are the gas density (left), the gas temperature (middle) and the gas metallicity (right). Superimposed is the corresponding slice through the intermediate-scale caustic skeleton, evaluated at $\sigma=1.5\,h^{-1}\textrm{Mpc}$. The cusp walls and swallowtail and umbilic filaments are shown in red, blue and green, respectively, as above in  \cref{fig:caustics_tng300,fig:caustics_scale_space}.}
    \label{fig:baryonic_web}
\end{figure*}

\subsection{Properties of the baryonic gas}
\label{subsec:fields-gas}

The dark matter fluid is the fundamental building block of the cosmic web. Accounting for about 86 percent of the Universe's matter content, it is the collisionless and multistreaming dynamics of the dark matter that traced out the filamentary nature of the cosmic mass distribution through the caustics of its Lagrangian flow. Nevertheless, the baryons are as key an ingredient in the assembly of the visible web through the formation of stars. It is nowadays known that the majority of baryons is not locked up in the luminous galaxies; instead, latest measurements from Fast Radio Bursts, \cite{Connor+2025}, suggest that as much as 76 percent of the baryons after the Big Bang reside in the diffuse and ionised \textit{intergalactic medium}. Thus, while the dark matter digs the potential wells of cosmic large-scale structure, the gas falls into the same and henceforth constitutes the dominant baryon reservoir out of which the luminous galaxies are assembled. At the same time, the dynamics of the baryons in the cosmic web are significantly more complicated than the non-linear gravity-driven structure formation of the collisionless dark matter. This is because the gas is not only subject to gravity but also to hydrodynamical processes such as pressure gradients, radiative cooling and heating, shocks, turbulence, and feedback from galaxies and AGN. For an excellent review, we refer to \cite{Meiksin2009}. A detailed investigation of these processes in the vicinity of the dark matter caustics is beyond the scope of the present article; instead, we here focus on a few characterising continuum fields of the baryonic gas tracing the dark matter web.

\subsubsection{Overview of baryonic gas properties}
\label{subsubsec:fields-gas_overview}

We investigate the following gas properties:

\begin{enumerate}
    \item \textbf{Gas density}$\quad$ The density of the baryonic gas is directly obtained from the respective values stored in the simulation's gas cells.
    
    \item \textbf{Gas temperature}$\quad$ We follow \cite{GalarragaEspinosa+2021} in calculating the gas temperature field $T$ from the gas cells' internal energies $u$ and the electron abundances $x_e$. Under the assumption of a perfect monatomic gas, the equation of state reads
    \begin{equation}
        T = (\gamma - 1)\frac{\mu}{k_B} u
    \end{equation}
    with the adiabatic index $g=5/3$, the Boltzmann constant $k_{B}$ and the molecular weight $\mu$, which can be estimated as
    \begin{equation}
        \mu = \frac{4 m_p}{1 + 3 X_{H} + 4 X_{H} x_e} \,.
    \end{equation}
    Here, $m_p$ is the proton mass and $X_H \approx 0.76$. Within the scope of the present article, we do not consider gas pressure and do not differentiate between different gas phases. We thus leave further investigations of the results of, e.g., \cite{Martizzi+2019, GalarragaEspinosa+2021} in the context of the caustic skeleton to future work.
    
    \item \textbf{Gas metallicity}$\quad$ The dimensionless gas metallicity $Z_{\rm{gas}}$ is defined as the mass fraction $Z_{\rm{gas}} = M_{\rm{metals}}/M_{\rm{gas}}$ of metals (i.e. elements heavier than helium) in a given gas mass element. The gas metallicity is directly obtained from the respective values stored in the simulation's gas cells.
\end{enumerate}

\begin{figure*}
    \centering
    \includegraphics[width=\textwidth]{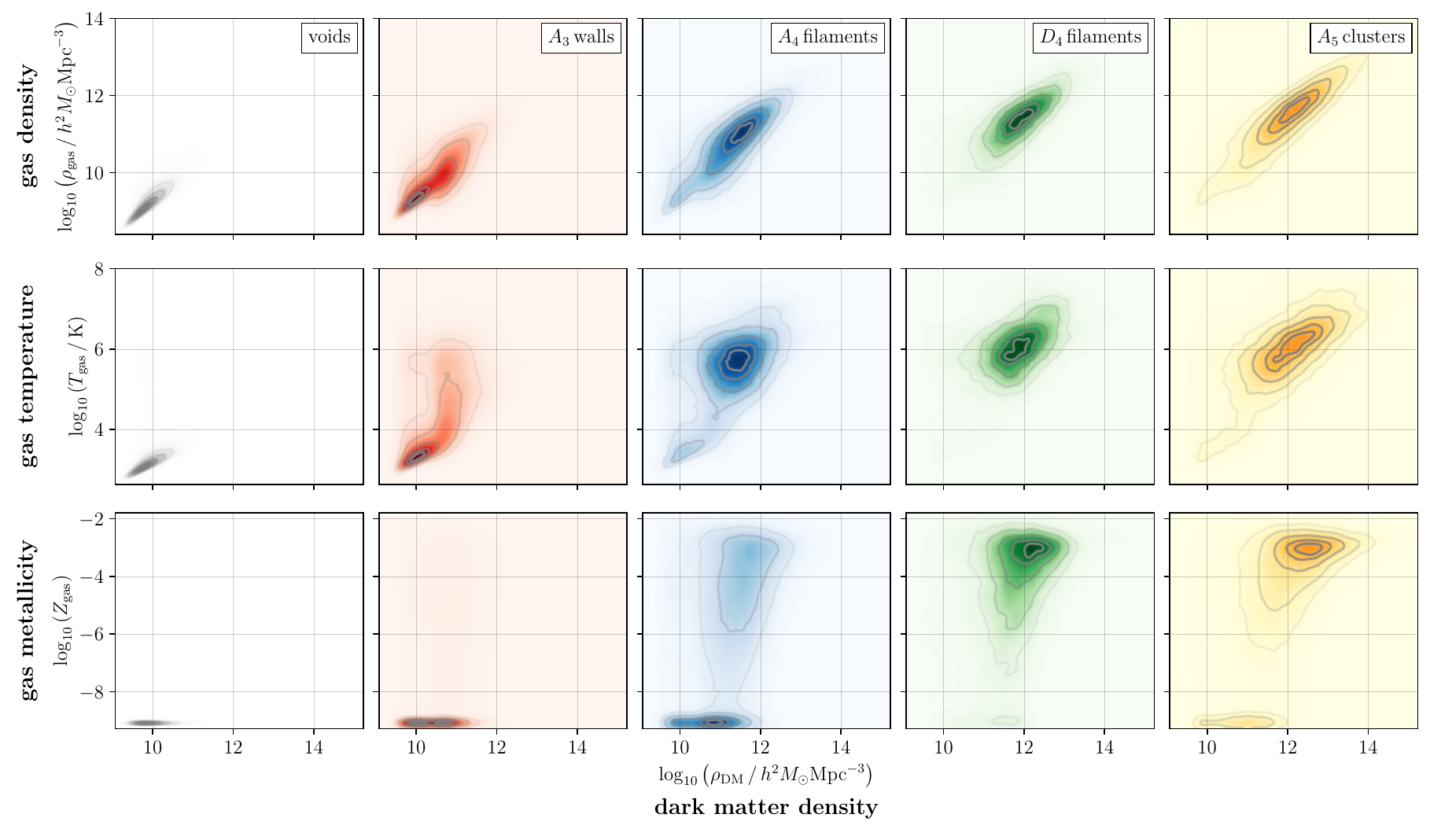}
    \caption{Properties of the baryonic gas across different web environments in the TNG100 simulation. The web is identified by the intermediate-scale caustic skeleton, evaluated at $\sigma=1.5\,h^{-1}\textrm{Mpc}$. Shown are the PDFs of the gas density (upper row), gas temperature (middle row) and gas metallicity (lower row) against the environmental dark matter density, evaluated from kernel density estimation of the histograms of the field values over a grid of spacing $0.2 \,h^{-1}\textrm{Mpc}$ over the full simulation box. The columns show the PDFs in the different caustic environments. The contours of the PDFs are displayed at levels $\{0.01,\, 0.1,\, 0.25,\, 0.5,\, 0.75,\, 0.9\}$.}
    \label{fig:fields}
\end{figure*}

\subsubsection{Analysis of the baryonic gas properties}
\label{subsubsec:fields-gas_analysis}

We present our analysis of the gas properties for the TNG100 simulation. Here, we use the temperature, density and metallicity data of the low-resolution simulation run and employ discrete natural neighbours to interpolate the fields over the full simulation volume on a regular grid of spacing $0.2\,h^{-1}\mathrm{Mpc}$ (see \cref{fig:dm_gas_density} for a visualisation). We use the PS-DTFE method to infer the dark matter density at the same resolution, and distinguish the different web environments by evaluating whether a cell is in a megaparsec neighbourhood\footnote{We will comment on this identification in more detail in the context of the galaxy properties in the different caustic environments, see \cref{sec:galaxies}.} of any simplex of the numerically determined caustic skeleton (see \cref{subsec:illustris-caustics}). We define the voids through the absence of the caustics, i.e. as those volumes that are not in such a neighbourhood. 

Our results, summarised in \cref{fig:fields}, clearly demonstrate that the baryonic gas is sensitive to the environmental dark matter density and consequently exhibits pronounced differences across the different cosmic web elements. Firstly, the upper panels show that the density of baryonic gas is approximately linearly proportional to the environmental dark matter density. This proportionality persists near-equally over the full range of dark matter density order of magnitudes, so that the relative amount of baryonic to dark matter is about one in five to six in all web environments, in accordance with the mean baryonic density $\bar{\rho}_{\textrm{bar}} \approx 1.3 \cdot 10^{10}\,h^{2} M_{\odot} \mathrm{Mpc}^{-3}$ and the mean dark matter density $\bar{\rho}_{\textrm{DM}} \approx 7.2 \cdot 10^{10}\,h^{2} M_{\odot} \mathrm{Mpc}^{-3}$. The gas therefore traces the dark matter density, and consequently exhibits the web-like morphology suggested in \cref{fig:dm_gas_density}.
We observe a mild deviation from linearity notably for the walls and the lower-density tail of the swallowtail filaments, suggesting that tidal effects have a more dominant influence on the gas distribution in these environments. Overall, we nevertheless find that in accordance with the dark matter densities, the voids have the lowest gas densities, while the walls, filaments and clusters are significantly richer in baryons, with the highest gas densities found in the most dark-matter-dense clusters. Since baryonic gas constitutes the primary reservoir for galaxy formation, we therefore expect significantly higher concentrations of galaxies in the high-density filamentary and cluster environments. This can already be intuited from \cref{fig:dm_gas_density}, and we will examine the galaxy populations in more detail in \cref{subsec:galaxies-populations}. Finally, we note that the upper panels suggest a pronounced difference between the swallowtail and umbilic filaments. While we leave a detailed investigation of their respective morphologies and dynamical origins to future work, we find that the umbilics constitute the more massive filaments and consequently accommodate higher gas densities tracing the higher environmental dark matter densities.

The middle panels of \cref{fig:fields} similarly reveal a correlation between the dark matter density and the gas temperature, albeit less linear than in the case of the gas density. The underdense voids and moderately dense walls accommodate the coolest gas, with typical temperatures of about $T_{\mathrm{gas}} \sim 10^3\, \mathrm{K}$ and $T_{\mathrm{gas}} \sim 10^3 - 10^5\, \mathrm{K}$ respectively. In contrast, the gas in filaments and clusters is significantly hotter, $T_{\mathrm{gas}} \sim 10^6\, \mathrm{K}$, with the highest temperatures of $T_{\mathrm{gas}} > 10^7\, \mathrm{K}$ reached in the most overdense regions of the cosmic web. Our results are in good agreement with numerous preceding analyses, see e.g. \cite{Martizzi+2019}. The non-linear relationship between the environmental density and the gas temperature suggests a clear role of non-linear hydrodynamical processes that heat up the gas near the higher caustics of the dark matter flow. An apparent transition occurs when the extended and moderately overdense walls fold into the swallowtail filaments, marking the point at which these effects become relevant and the gas temperature rapidly shifts from a cooler ($T \lesssim 10^4 K$) into a hotter phase ($T\gtrsim  10^5 K$). As is evident in the histograms, this gives rise to two distinct modes in the cosmic gas temperature distribution. Moreover, we note that the temperature of the baryonic gas is an important clue to the formation of galaxies across the different web environments: the heating of the gas in the denser environments suppresses the clustering of baryonic clumps and the active formation of stars, thus suggesting a more quiescent population of galaxies in the present-day filaments and clusters. We will elaborate on this and related properties in detail in \cref{sec:galaxies}.

Finally, the lower panels \cref{fig:fields} show an even stronger bimodality in the gas metallicity, pointing to a clear division of the cosmic web into metal-poor and metal-rich environments. In the low-density regions, the gas metallicity is approximately vanishing, with numerical artefact values of about $Z_{\mathrm{gas}} \sim 10^{-9}$ set through the resolution of the TNG100 simulation model. The gas metallicity is only considerable in the dense environments of the present-day cosmic web, with values reaching up to $Z_{\mathrm{gas}} \sim 10^{-3}$ in the clusters and umbilic filaments, as well as the high-density tail of the swallowtail filament family. This reaffirms the qualitative bimodality of the cosmic gas properties suggested above by the gas temperature. Moreover, the gas metallicity is an unambiguous indicator of the cosmic history of galaxy formation across the different web environments. The pronounced gas metallicities in the highly dense filaments and clusters reveal that these have, in the course of their evolution, hosted significant populations of galaxies that burned the surrounding gas into metals. While a detailed treatment of galaxy formation in the evolving cosmic web is beyond the scope of our study, we note again that the gas metallicity hints at pronounced differences in the properties of the galaxy populations that may be observed in the present-day web environments.

Altogether, our results suggest that the baryonic gas is an integral part of the present-day cosmic web: while it is the dark matter that traces out the geometric backbone of the large-scale structure through the caustics in its collisionless flow, it is the hydrodynamic evolution of the gas falling into the same that governs the formation of galaxies. Having identified clear environmental differences, we now move on to the main part of our study and investigate in detail the galaxy properties embedded in the different web environments.


\section{Properties of galaxies in the multiscale cosmic web}
\label{sec:galaxies}

The preceding analysis demonstrates that the baryonic gas exhibits distinct properties across the different web environments; in particular, its temperature and metallicity indicate varying star formation activities and evolutionary histories over cosmic time. At the same time, as highlighted in the introduction, the observational evidence of the morphology-density relation \citep{Dressler1980} and the colour-density relation (e.g. \cite{Cooper+2007}) quantifies the observed bimodality of galaxy populations inferred from cosmic surveys, see \cite{Strateva+2001, Baldry+2004, Balogh+2004, Menanteau+2005} to name but a few. The environmental dependence of the galaxies' colours and star formation rates \citep{Kaufman+2004, Hogg+2004, Balogh+2004, Blanton+2005} further suggests pronounced differences between galaxies across web environments. These trends have recently been explored in the simulated cosmic web of the SIMBA \citep{Metuki+2015}, the Horizon-AGN simulation \citep{Laigle+2018}, the EAGLE simulation\citep{Xu+2020}, the IllustrisTNG \citep{Hasan+2023, Yu+2025, NandiPandeySarkar2026} cosmological simulation suites and, most recently, the Flamingo and Colibre simulations \citep{Schaye:2023, Kugel:2023, Schaye:2025}.

\subsection{Overview of galaxy properties}
\label{subsec:galaxies-overview}

To study the impact of the different cosmic web environments on the embedded galaxies, we study in detail the properties of galaxies in the different caustic environments constituting the geometric backbone of the multiscale cosmic web in the IllustrisTNG simulations. The following properties of the luminous 
galaxies in the IllustrisTNG data catalogues:

\begin{enumerate}
    \item \textbf{Colour band magnitude difference $g-r$} $\quad$ A standard property for spectroscopic analyses, the galaxy rest-frame $g-r$ colour index is the dimensionless difference in the total band magnitudes of the subhalo when passed through the Sloan Digital Sky Survey (SDSS) \citep{SDSS2004} spectroscopic $g$- and $r$-filters. The colour index is computed from the summed emission of all bound stellar particles in a given subhalo, see \cite{Nelson+2017}.
    \item \textbf{Specific star formation rate (sSFR)} $\quad$  The sSFR is defined as the star formation rate (SFR) -- measured in dimensions of (stellar) mass over time -- divided by the mass of the subhalo. The SFR is obtained from the values stored in the subhalo catalogue. We convert the IllustrisTNG internal units to report the sSFR in units of $h\, \mathrm{Gyr}^{-1}$.
    \item \textbf{Stellar metallicity $Z$} $\quad$ The mass-weighted mean stellar metallicity, which we henceforth concisely refer to as the stellar metallicity, is given by $Z = \langle Z_{\star} \rangle / Z_{\odot}$, that is, the mass-weighted mean metallicity of all stars contained in a subhalo, divided by the metallicity of the Sun. Explicitly, the numerator reads  $\langle Z_{\star} \rangle = (\sum_i m_i Z_i) / \sum_i m_i$, where the index $i$ runs over the stellar particles contained in the subhalo and the metallicity $Z_i$ is the mass of all metals (elements heavier than helium) in a star. $Z$ is dimensionless.
    \item \textbf{Baryon fraction $f_{\rm{bar}}$} $\quad$ The baryon fraction of a subhalo is estimated as the total baryonic mass $M_{\rm{bar}} = M_{\star} + M_{\rm{gas}} + M_{\rm{BH}}$ over the total mass $M_{\rm{halo}}$ of the subhalo, $f_{\rm{bar}}  =  M_{\rm{bar}} / M_{\rm{halo}} $. Here, $M_{\star}$ is the stellar mass, $M_{\rm{gas}}$ the mass of all gas particles, and $M_{\rm{BH}}$ the black hole mass contained in a subhalo, with the total mass $M_{\rm{halo}} = M_{\rm{bar}} + M_{\rm{DM}}$ given by the sum of the baryonic and dark matter component. $f_{\rm{bar}}$ is dimensionless.
\end{enumerate}

We extract the properties listed above directly from the halo catalogues accompanying the final-configuration snapshot files of the TNG100 and TNG300 simulations. These list the subhaloes identified by the \verb|Subfind| algorithm. The luminous galaxies are given by the subhaloes with stellar masses $M_{\star} > 0$. To consider only those subhaloes that are likely of cosmological origin, we restrict our analysis to the objects for which $\verb|SubhaloFlag|=1$. While this criterion significantly reduces the number of spurious detections, it is in practice quite conservative, see \cite{Nelson+2019}. To reliably infer the mass-weighted mean stellar metallicities, $g-r$ colour indices and specific star formation rates, we further restrict the selection to those subhaloes that contain at least $N_{\star}=20$ stellar particles. This yields about $10^5$ and $6 \cdot 10^5$ luminous galaxies for the TNG100 and TNG300 simulations, respectively.

\subsection{Assigning galaxies to the caustic skeleton environments}
\label{subsec:galaxies-identification}

We associate the galaxies with the different caustic environments by identifying the subhaloes that lie within the vicinity of the numerically calculated skeleton simplices corresponding to the walls, filaments and cluster nodes, respectively. Conversely, we define the voids through the absence of caustics; the void galaxies are therefore those subhaloes that do not lie near any simplices. We note at this point that the caustics inferred from the Zel'dovich approximation caustic conditions (\cref{tab:caustic_skeleton}) trace the spine of the caustic skeleton. While providing a rigorous understanding of the geometric backbone of structure formation, the association of galaxies with the different web environments requires the specification of a numerical threshold for what constitutes the vicinity of a caustic. Throughout this work, we choose this threshold as $1\,h^{-1}\textrm{Mpc}$, and therefore identify the galaxies in the caustic environments as residing within a megaparsec neighbourhood of the numerically calculated skeleton simplices. Comparing visually with \cref{fig:caustics_scale_space}, we find this to be a suitable choice that appropriately defines the web environments of both the small-$\sigma$ and large-$\sigma$ caustic network.

Although our definition of a caustic neighbourhood is inherently numerical in nature, we have confirmed that variations in the threshold length over a reasonable range have a subdominant effect compared to the systematic variation in galaxy properties across different web environments.

In principle, the caustic skeleton enables sophisticated approaches to the neighbourhood identification, as the exact (rather than Zel'dovich approximation) caustics of the non-perturbative displacement field rigorously delineate the present-day multistream regions constituting the dark matter web surrounding the caustic spine. One could also augment the skeleton-based methods with morphological criteria based on the local overdensity or its gradient. While such an approach would allow for a more refined treatment of both small- and large-scale morphological features in the present-day mass distribution, it would again require the introduction of numerically defined, and ultimately arbitrary, threshold values.
Consequently, we set these considerations aside in the present work and instead focus on the galaxy properties across the different web environments as directly characterised by the phase-space dynamics of the dark matter.

\begin{figure*}
    \centering
    \begin{subfigure}[b]{0.24\textwidth}
        \includegraphics[width=\textwidth]{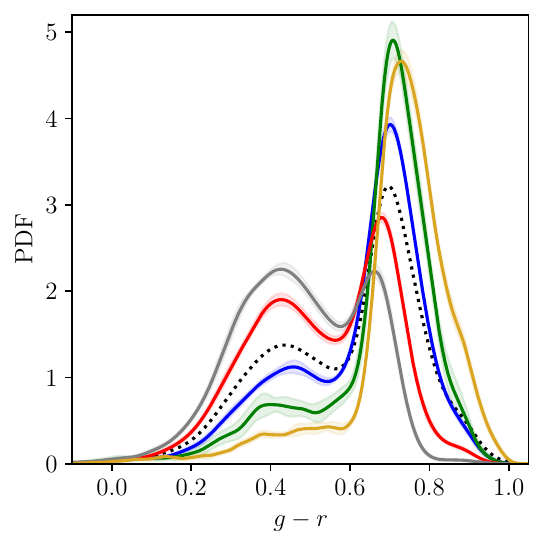}
    \end{subfigure}
    \begin{subfigure}[b]{0.24\textwidth}
    \includegraphics[width=\textwidth]{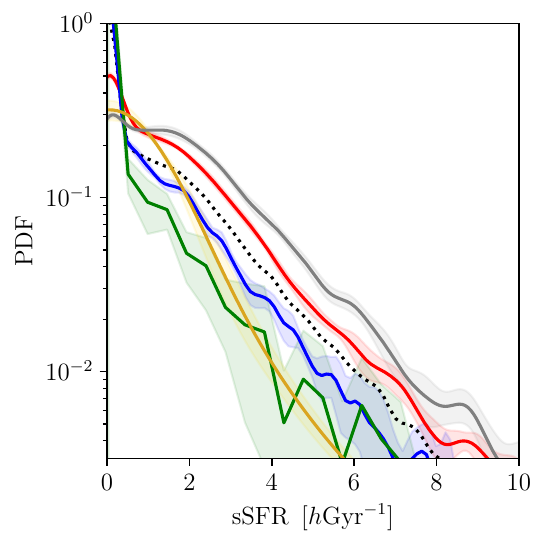}
    \end{subfigure}
    \begin{subfigure}{0.24\textwidth}
        \includegraphics[width=\textwidth]{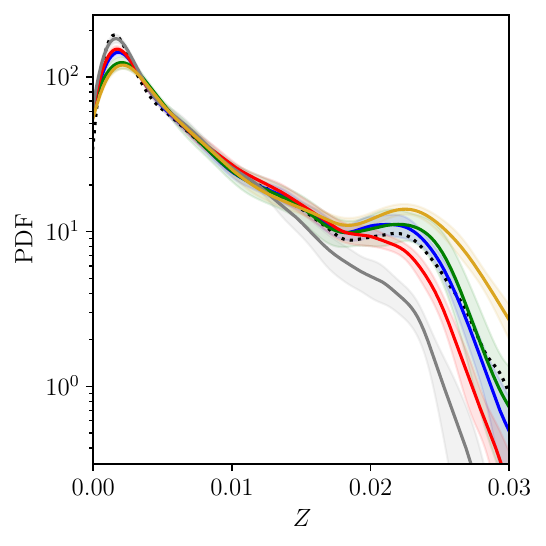}
    \end{subfigure}
    \begin{subfigure}{0.24\textwidth}
        \includegraphics[width=\textwidth]{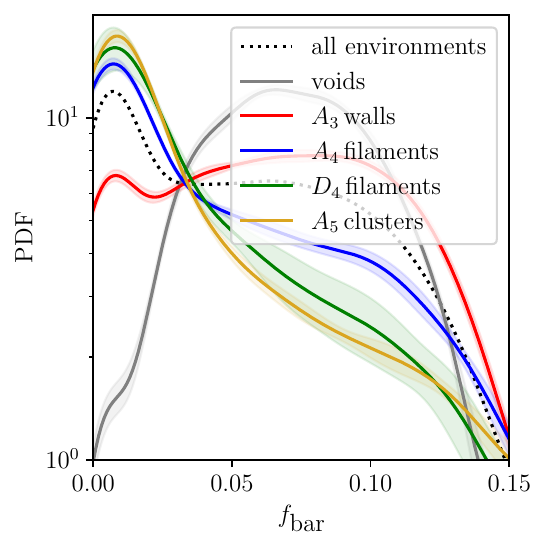}
    \end{subfigure}
    \caption{Galaxy properties across the different web environments in the TNG100 simulation, traced by the caustic skeleton at smoothing scale $\sigma=1.5 \, h^{-1} \textrm{Mpc}$. The PDFs of the galaxy properties are obtained from the histograms of the data of the luminous galaxies contained in the halo catalogue, classified as residing either within void regions or within the different caustic environments, as indicated in the plot legends. We use kernel density estimation to obtain smooth PDFs. Shown from left to right are the PDFs of the $g-r$ colour index, specific star formation rate, metallicity and baryon fraction. The black dotted lines show the PDFs for the total galaxy population in the simulation volume. For each environment, bootstrap sampling is used to divide the population sample into ten subsamples, from which we evaluate the PDF and estimate the $1\sigma$-uncertainty bands (shaded regions) as the mean and standard deviation of the histograms of the subsamples.}
    \label{fig:histograms_tng100}
\end{figure*}

\subsection{Galaxy properties in caustic skeleton environments}
\label{subsec:galaxies-properties}

Having calculated the caustic skeleton according to the discussion of \cref{subsec:illustris-caustics}, we classify the galaxies as described \cref{subsec:galaxies-identification} and evaluate the distributions of their characteristic properties. In the current section, we focus on the intermediate-scale caustic skeleton at smoothing scale  $\sigma = 1.5 \,h^{-1}\rm{Mpc}$, and present our analysis for both the TNG100 and TNG300 data.

\subsubsection{Analysis for the TNG100 data}

\Cref{fig:histograms_tng100} shows the marginal distributions of the colour index $g-r$, the specific star formation rate (sSFR), the metallicity $Z$ and the baryon fraction $f_{\rm{bar}}$ of the TNG100 galaxies residing in the different web environments and the structural components of the caustic skeleton. We find clear differences in the properties of their embedded galaxy populations. The intermediate density environment of filaments, and the high-density environments of clusters and their filamentary extensions, accommodate redder galaxies with higher metallicities, lower baryon fractions and lower specific star formation rates. The lower-density sheets and cosmic voids are characterised by a population of bluer, active galaxies that have lower metallicities, but higher baryon fractions and specific star formation rates. For each of the properties shown in \cref{fig:histograms_tng100}, a clear trend may be identified: the wall galaxies are redder than those in the cosmic voids, the filament galaxies are in turn redder than those in the walls, and finally the cluster galaxies are redder than those in cosmic filaments. Along the same lines, the metallicity increases and the baryon fraction and specific star formation rate decrease proceeding from the galaxy population in voids, walls, filaments and clusters. These results are consistent with the results of preceding studies (e.g. \cite{HaynesGiovanelli:1986,Alpaslan:2014, Metuki+2015,Kraljic+2017, Laigle+2018, Xu+2020, GalarragaEspinosa+2022, Kuchner+2022, Hasan+2023, Hasan+2025,Yu+2025, NandiPandeySarkar2026}), and provide a direct confirmation of the observationally motivated colour-density relation (e.g. \cite{Cooper+2007}). Moreover, the uncertainties on the probability distributions demonstrate that the differences are statistically significant at the level of one up to several standard deviations, implying that it is statistically unlikely that the data are drawn from the same underlying distribution. 

The caustic skeleton formalism and its rigorous identification of the different web environments therefore affirm that the large-scale structure has an impact on the properties of the embedded galaxies, which directly translates into observable differences in the physical cosmic web.

\paragraph*{Nature of filaments:} Within the context of the cosmic web, filaments stand out as the most prominent structural features. They form the backbone of the web-like network, have surface masses and galaxy number densities that render them easily visible, while they are the dynamically dominant component of the cosmic web \citep{Kugel+2026}. Moreover, around half of the dark matter, gas and galaxies in the universe reside in filamentary environments (see e.g. \cite{Cautun+2014,Ganeshaiah+2019}).

Interestingly, within the context of the caustic skeleton theory, we may identify two geometrically distinct families of filaments. These are the swallowtail $A_4$ filaments and the umbilic $D_4$ filaments. 
Although we defer a detailed treatment of their dynamical origin and current-time morphology to the upcoming \cite{HertzschFeldbruggeWeygaert2026}, we note the blue and green curves in \cref{fig:histograms_tng100} reveal differences in the properties of the galaxy populations embedded in the swallowtail and umbilic filaments. The histograms for both filament families are consistent with the existing literature, as the galaxy population in both the swallowtail and umbilic is redder and more quenched than in the walls, but bluer and less quenched than in the clusters.  As for the differences between the filament populations, our analysis shows that there is a mild but systematic shift of the galaxies in the umbilic filaments towards the redder, more quenched tail of the probability distributions. This is because the umbilic filaments, arising from the second eigenvalue $\lambda_2$ acquiring the shell-crossing value $\lambda_2 = \lambda_1=b_c^{-1}$, are generally denser than the swallowtail filaments whose collapse is due to only the first eigenvalue field $\lambda_1$. As we will address in more detail in forthcoming work, we find the umbilic filaments to be shorter and thicker in spatial extent, with higher overdensities; the umbilics are also rarer than the prominent swallowtails that comprise the majority of the cosmic filaments. The paucity of the data points for the umbilic filament galaxies renders the statistical uncertainties larger, notably for the sSFR in the right panel of \cref{fig:histograms_tng100}. 

In summary, our results demonstrate, for the first time, that the umbilic filaments constitute an `intermediate' environment between the swallowtail filaments and clusters. They form the filamentary extensions of and 
around cluster nodes. As such, they tend to host redder and more metal-rich galaxies with lower baryon fractions and star formation activities.

\begin{figure*}
    \centering
    \begin{subfigure}[b]{0.24\textwidth}
        \includegraphics[width=\textwidth]{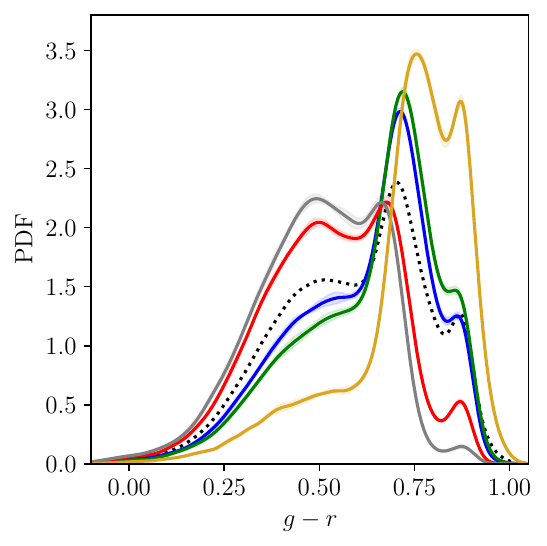}
    \end{subfigure}
    \begin{subfigure}[b]{0.24\textwidth}
    \includegraphics[width=\textwidth]{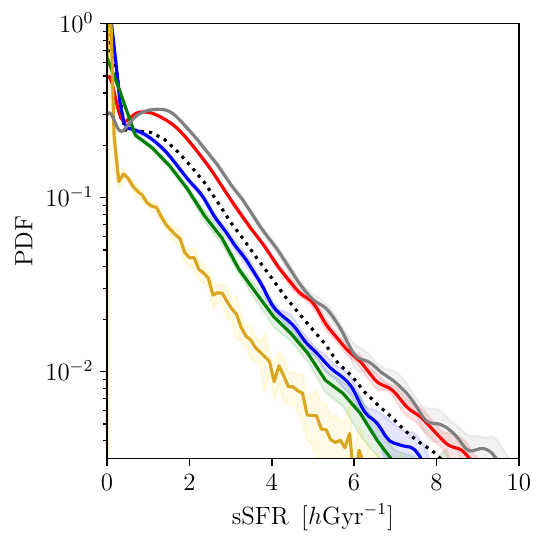}
    \end{subfigure}
    \begin{subfigure}{0.24\textwidth}
        \includegraphics[width=\textwidth]{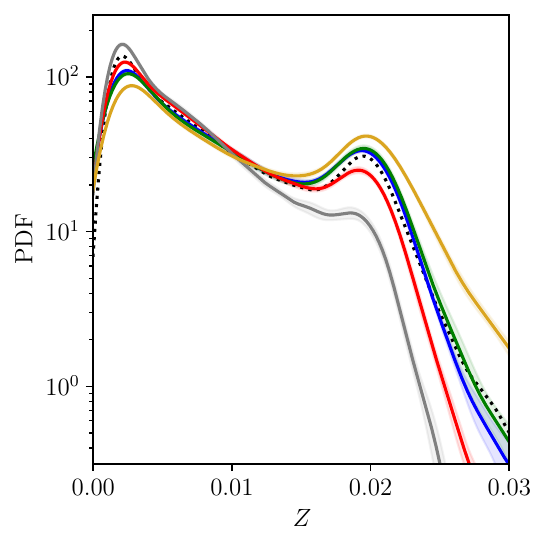}
    \end{subfigure}
    \begin{subfigure}{0.24\textwidth}
        \includegraphics[width=\textwidth]{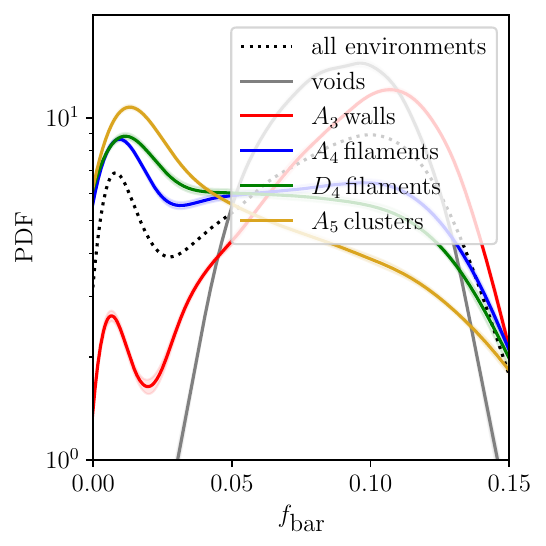}
    \end{subfigure}
    \caption{Galaxy properties across the different web environments in the TNG300 simulation; analogous to \cref{fig:histograms_tng100} for the TNG100 simulation. Again, the web environments are traced by the caustic skeleton at smoothing scale $\sigma=1.5 \, h^{-1} \textrm{Mpc}$, and bootstrap sampling is used to divide the populations into ten subsamples and estimate the uncertainties on the PDFs.}
    \label{fig:histograms_tng300}
\end{figure*}

\subsubsection{Analysis for the TNG300 data}

In \cref{fig:histograms_tng300}, we repeat the analysis for the TNG300 data and again investigate the galaxy populations in the different environments of the caustic skeleton identified at $\sigma = 1.5 \,h^{-1}\rm{Mpc}$. The TNG300 data comprise about 6 times as many galaxies as the TNG100 data, which results in smaller error bars in the histograms. Moreover, the larger cosmological volume contains more large-scale web features, additionally mitigating cosmic variance. However, TNG300 particle masses are greater than the TNG100 particle masses, and the properties of individual galaxies are thus less finely resolved. We have already raised this issue in \cref{subsec:illustris-overview} and follow \cite{Pillepich+2017a, Nelson+2019} in stating that numerical convergence, in particular with regard to galaxy properties, remains an inherent limitation of the IllustrisTNG suite. Nevertheless, we find good qualitative agreement between galaxy properties in the TNG100 and TNG300 simulations. As before, we recover a clear hierarchy of the colour, sSFR, metallicity and baryon fraction across the different web elements, ranging from the blue and actively star-forming void galaxies to the red and quiescent cluster galaxies. Moreover, we again identify a clear, albeit weaker, difference between the swallowtail and umbilic filament populations. The quantitative differences between the TNG100 and TNG300 date are most pronounced in the baryon fraction in the right panel, as the first peak at around zero is suppressed in the TNG300 data. We reason that this may be due to our fixed subhalo selection criterion $N_{\star}\geq20$ (see \cref{subsec:galaxies-overview}), which is more restrictive for the higher-particle-mass TNG300 data. We observe a more surprising numerical discrepancy in the colour index $g-r$, as the TNG300 galaxy colour distribution appears to be trimodal rather than bimodal, with an additional peak at around $g-r\approx 0.9$. While we stress again that we leave the issue of IllustrisTNG's numerical convergence aside, we find that this third and highly red peak is significantly occupied only for the cluster galaxies, residing in the most dominant overdensities.

Overall, our results for both the TNG100 and TNG300 data demonstrate that there exists a clear hierarchy of galaxy properties in the cosmic web traced by the large-scale caustic skeleton. The range of blue and actively star-forming galaxies in the low-density environments (voids and walls) to the red and quiescent galaxies in the intermediate- and high-density environments (filaments and cluster nodes) is consistent with the existing literature and directly confirms the well-established colour-density relation \citep{Kaufman+2004, Hogg+2004, Balogh+2004, Blanton+2005, Cooper+2007}.

\subsection{The multiscale cosmic web and galaxy properties}
\label{subsec:galaxies-scales}

\begin{figure*}

    \centering
    \includegraphics[width=\textwidth]{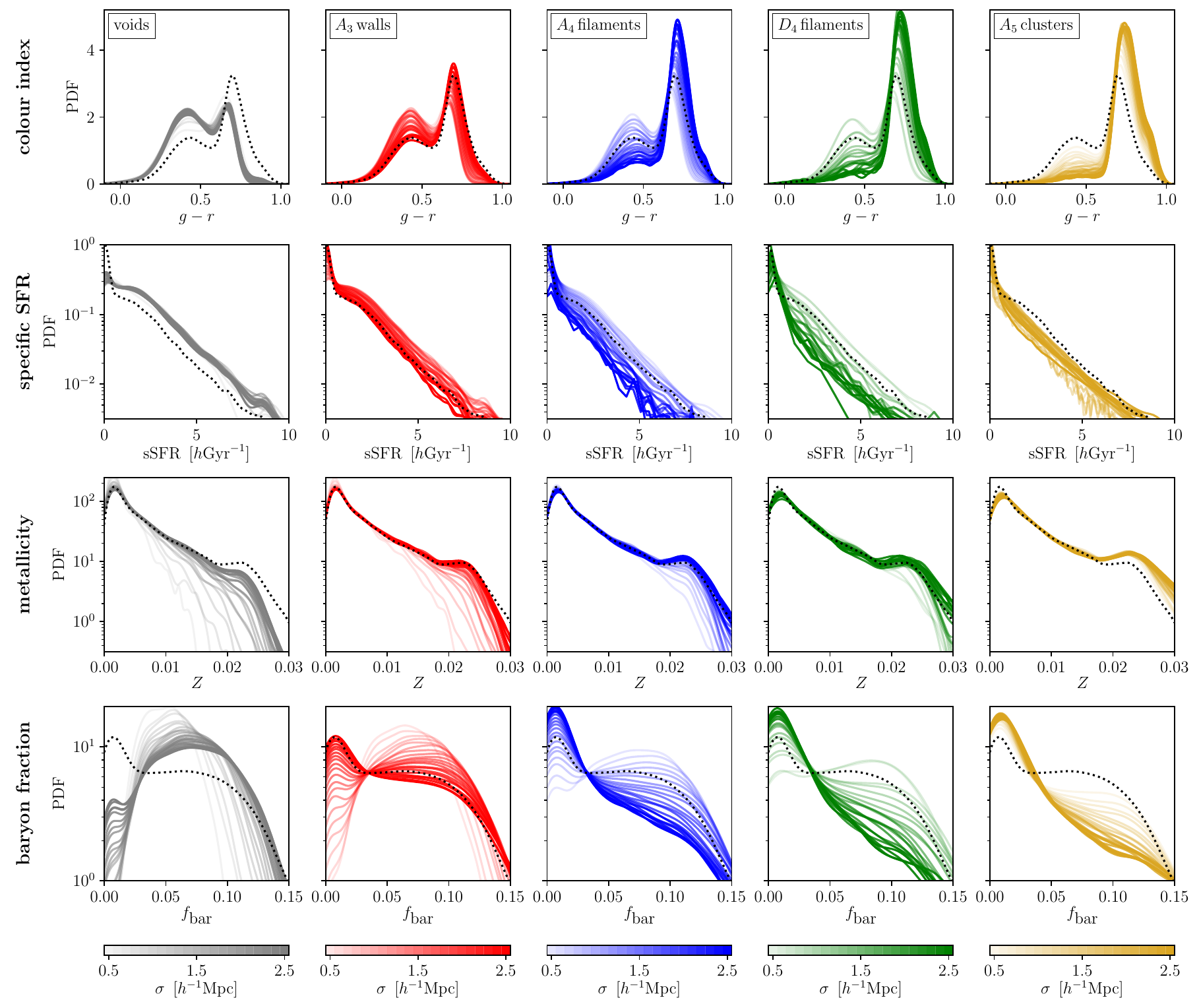}
    
    \caption{Galaxy properties in the scale-space cosmic web of the TNG100 simulation. Shown are the PDFs of the $g-r$ colour index (upper row), specific star formation rate (second row), metallicity (third row) and baryon fraction (lower row) in the different caustic environments (columns) for varying smoothing scales. The caustic skeleton is evaluated from $\sigma=0.5\,h^{-1}\textrm{Mpc}$ to $\sigma=2.5\,h^{-1}\textrm{Mpc}$ in steps of $0.1\,h^{-1}\textrm{Mpc}$. For each smoothing scale, the web elements and galaxy populations are identified, and the PDFs are estimated analogously to \cref{fig:histograms_tng100} and \cref{fig:histograms_tng300}. The respective PDFs are shown in different colours corresponding to the different smoothing scales, see colour bars.}
    \label{fig:histograms_scales}
\end{figure*}

The web dependence of galaxy properties is significantly complicated by the fact that the cosmic large-scale structure is a multiscale network whose filamentary nature may be identified over a range of about five orders of magnitude of the cosmic density field, see \cite{Cautun+2014}. Evidently, one expects the galaxy properties to be sensitive to these environments. Within the caustic skeleton framework, we are able to explicitly address this fundamental aspect of the cosmic web, and assess the influence of the scale of caustic features on the properties of galaxies. To this end, we look at the trends in the distribution of galaxy properties, such as colour and star formation activity, as a function of the scale of the caustic skeleton. Differences between the distribution of these properties may then be ascribed to the presence or absence of filaments, walls and nodes in the branching network of the skeleton at a particular scale. 

The major share of preceding studies on the cosmic web environments has assessed the matter distribution on a single representative scale. In particular,  morphological web identification methods typically deploy a smoothed tracer field, such as the present-day dark matter density field, to classify the cosmic volume into voids, walls, filaments and cluster nodes. As this identification is typically evaluated at a single smoothing scale, simulation-based studies such as \cite{Xu+2020, Yu+2025, NandiPandeySarkar2026} have consequently established the relative content of blue and active to red and quiescent galaxies for the various web features at one characteristic length scale. 

While the \verb|NEXUS(+)| formalism \citep{AragonCalvo+2007, Cautun+2013} does formally take account of the multiscale nature of the web-like cosmic mass distribution, it ultimately yields a single morphologically dominant scale at each location within a simulation or survey volume. In the present study, we take the attention to the multiscale nature of the mass distribution one step further and study the galaxy population trends as a function of an entire continuum of physical length scales. 

As we discussed \cref{subsec:caustics-scale_space}, the caustic skeleton is inherently sensitive to the multiscale nature of the cosmic web due to the identification of the caustics of the dark matter flow from the primordial fluctuations smoothed at varying length scales. \Cref{fig:caustics_scale_space} illustrates how, at each for each individual smoothing scale, the caustics trace the present-day web features down to the length scale under consideration. When taking into account their ensemble, the caustic skeleton provides a theoretically robust and continuous classification of the cosmic web over the full range of cosmologically relevant length scales. We now apply this notion to study, for the first time, the multiscale web dependence of the galaxy properties in the IllustrisTNG suite. We focus our analysis here on the TNG100 simulation, for which we evaluate the caustic skeleton on scales  $\sigma = 0.5, \ldots, 2.5 \,h^{-1} \rm{Mpc}$ in steps of $0.1 \,h^{-1} \rm{Mpc}$. For each scale, we identify the galaxies residing in the different caustic environments and subsequently evaluate the probability density functions of the properties as we did in the preceding section. \Cref{fig:histograms_scales} summarises our results.

At large smoothing scales, the walls, filaments and cluster nodes are identified by the most dominant caustics. The visualisation of \cref{fig:caustics_scale_space} demonstrates how these define and partition the large-scale cosmic mass distribution through multistream regions spanning up to several tens of megaparsecs. Particularly for the filaments and cluster nodes, the large-scale caustics typically come with spatially extended and high density contrasts, making them the most overdense environments in the multiscale cosmic web. Confirming the colour-density relation, it is these caustics that accommodate the reddest and most quenched galaxies, which is directly seen by the upper two rows of \cref{fig:histograms_scales}. Moreover, the large-$\sigma$ filaments and clusters are depleted of their baryons at the current time, and constitute the only environments with considerable numbers of galaxies reaching metallicities of about $Z \approx 0.03$.
In contrast, the cosmic voids are identified by the absence of caustics. At large $\sigma$, as the caustics only identify the most dominant multistream regions and overdensities; the voids may hence contain significant overdense substructures, as seen in the upper right panel of \cref{fig:caustics_scale_space}. Whereas at low $\sigma$, the small-scale voids do not have significant substructures, and therefore constitute the environments of the most pronounced and near-uniform underdensities. Consequently, one expects that it is these environments that accommodate the bluest, most actively star-forming and least metal-rich galaxies. This is confirmed in the left panels of \cref{fig:histograms_scales}.

For all smoothing scales separately, we consistently find the bimodal colour distribution and the clear progression of blue to red galaxies across the voids, walls, filaments and clusters. The latter has been discussed in detail \cref{subsec:galaxies-properties}. However, our analysis reveals that there is a second systemic sequence, namely that of scale space, that affects the galaxy properties in each environment separately. As visualised in the left panels of \cref{fig:caustics_scale_space}, the lower-$\sigma$ caustics encompass also the more tenuous features of the cosmic mass distribution, and therefore feature less extended and milder overdensities. The embedded galaxies in these environments are generally bluer, less metal-rich, contain more baryons and exhibit higher present-day star formation activity. The large-$\sigma$ web elements, in turn, constitute the more dominant shell-crossed regions, such as the major filamentary arteries, the principal large-scale transport channels, and the large flattened supercluster planes \citep[see e.g][]{MEinasto:2025}. The galaxies in these more overdense environments are redder and more quiescent. While the traditional quadripartite web identification assumes characteristic galaxy properties for the different web environments, our analysis instead shows that the typical (i.e. mean or median) properties of the populations depend on the scale at which the multiscale web is measured. Concretely, in our case, this is encapsulated in the value of $\sigma$ at which the caustics are determined. Perhaps surprisingly, the histograms \cref{fig:histograms_scales} suggest that galaxy properties form a continuum in scale space. Is it apparent, for instance, that the probability distributions of all four galaxy properties studied here are near-distinguishable for the large-$\sigma$ walls (solid red) and the small-$\sigma$ swallowtail filaments (light blue). Similar statements can be made for the voids and the walls, and the filaments and clusters, respectively, as well as the distinction between the swallowtail and umbilic filaments. 

For the latter, we note that for each scale $\sigma$ shown in \cref{fig:histograms_scales}, the galaxy populations in the umbilic filaments appear to be redder and less active than the swallowtails comprising the majority of the filamentary caustics. This is consistent with the observation in the preceding section, where we found the large-scale umbilics to constitute an `intermediate' environment between the typical filaments (the swallowtails) and the clusters. With the results of our scale-space analysis, we now conclude that this is true for all smoothing scales.

Despite the continuous nature of the galaxy properties in the multiscale web, we observe some differences in the variations within the different environments. Firstly, while the void galaxies' baryon fractions and metallicities exhibit a clear dependence on the defining smoothing scale, this is less so for their colour and sSFR distributions. We note that the void galaxy sample for the small scale web is small, and the results have to be treated with care. Nevertheless, the histograms in the upper two rows suggest that the galaxies in the single-stream regions (i.e. the cosmic voids) are always bluer and more actively star-forming than the populations in the multistreaming cosmic web, with the characteristic colour and sSFR not depending significantly on the void length scale. Similarly, the scale-dependent variation of the cluster galaxy properties is less pronounced than for the filaments and walls. In particular, we find that the tidal environment of the clusters always accommodates a similar population of metal-rich galaxies with low sSFRs, even at low $\sigma$. Conversely, the scale-dependence is most pronounced for the cosmic filaments. At low $\sigma$, these host a substantial fraction of blue galaxies with high baryon fractions and elevated star formation activity; whereas at the largest scales, the galaxy properties in the filaments resemble those in the clusters, and the populations become nearly exclusively red and quiescent. We conclude that this variation arises as a direct consequence of the wide density range of up five orders of magnitude across which the cosmic filaments persist, see \cite{Cautun+2014}. Our results highlight the need for a careful multiscale treatment, and encourages more detailed investigations of galaxy colours, morphologies and kinematical properties in the scale-space cosmic web.

\begin{figure*}
    \centering
    \begin{subfigure}[b]{0.32\textwidth}
        \includegraphics[width=\textwidth]{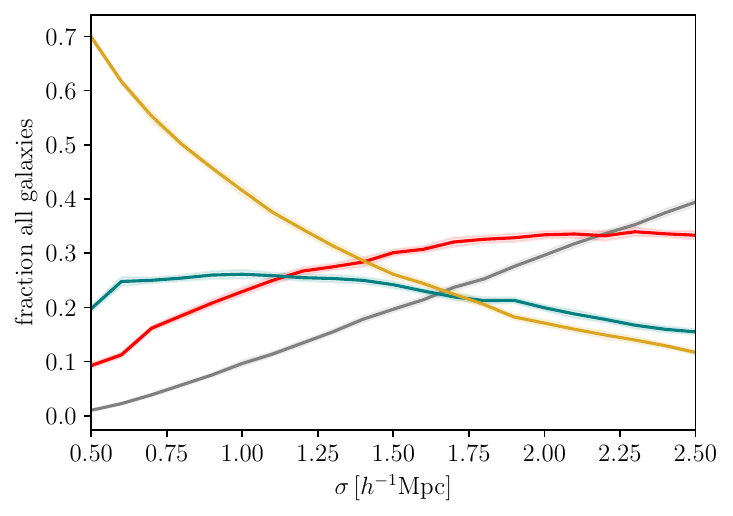}
        \caption{fraction of all galaxies}
    \end{subfigure}
    \begin{subfigure}[b]{0.32\textwidth}
    \includegraphics[width=\textwidth]{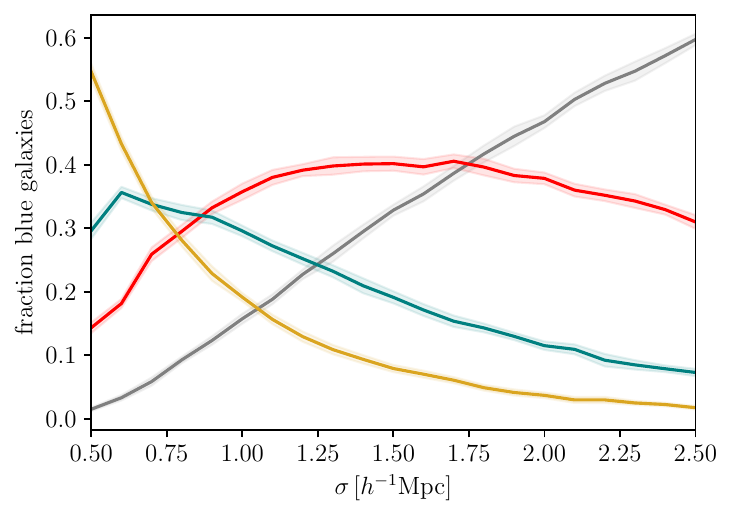}
        \caption{fraction of blue galaxies}
    \end{subfigure}
    \begin{subfigure}[b]{0.32\textwidth}
    \includegraphics[width=\textwidth]{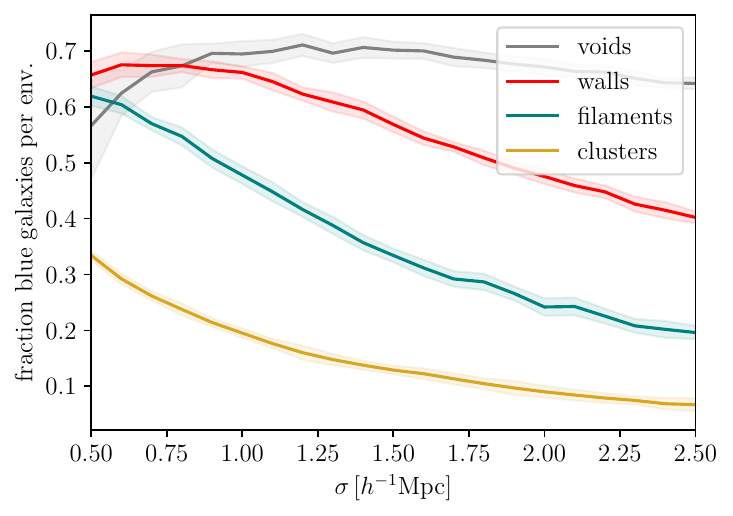}
        \caption{fraction of blue galaxies per environment}
    \end{subfigure}
    \caption{Galaxy populations in the scale-space cosmic web of the TNG100 simulation. The left panel shows the fraction of the simulation's total galaxy population residing in the voids, walls, filaments and cluster nodes, as a function of the defining smoothing scale. For each smoothing scale $\sigma$, the caustic skeleton is evaluated and galaxies in the different environments are counted. The uncertainties are estimated by the  sample standard deviations upon bootstrapping the total population into 25 subsamples.
    Analogously to the left panel, the middle panel shows the fraction of the total blue galaxy population ($g-r < 0.58$) residing in the different environments, and the right panel displays the fraction of blue galaxies within each environment. Again, these are shown as a function of the smoothing scale, and with the uncertainties estimated using bootstrapping into 25 subsamples.}
    \label{fig:populations}
\end{figure*}

\subsection{Galaxy populations in the multiscale cosmic web}
\label{subsec:galaxies-populations}

We conclude this section with a brief discussion of the overall galaxy populations in voids, walls, filaments and clusters. We have seen above that the different environments have a significant influence on the characteristic properties of their embedded galaxies. The galaxies in the less overdense environments (voids and walls) are generally bluer and actively star-forming, whereas those residing in the denser filaments and clusters are typically redder and more metal-rich and quiescent. However, it is at this point not clear how these results translate into the overall population of galaxies observed in the sky.
While the immense cosmic voids constitute about 77 percent of the total cosmic volume, they contain only about 15 percent of the overall dark and baryonic matter. In contrast, the filaments and clusters account for only about 6 and 0.1 percent of the cosmic volume, while at the same time accommodating 50 and 11 percent of the cosmic matter, respectively, see \cite{Cautun+2014} and references therein. Consequently, it is expected that the major share of the overall galaxy population -- both of early- and late-type morphology, or of blue and red colour -- resides in the extended cosmic filaments. Yet, the discussion of \cref{subsec:galaxies-scales} highlights that this statement requires careful treatment in scale space, as the definition of what constitutes a cosmic void, wall, filament and cluster depends on the underlying measurement scale.

We therefore now analyse the populations of the galaxies in the different environments of the caustic skeleton evaluated over a range of cosmologically relevant length scales. As in the preceding sections, we identify the galaxies residing in the different caustic environments, and subsequently evaluate the fraction of (blue) galaxies in the voids, walls, filaments and clusters. For simplicity and clearer comparison with the literature, we now group the more prominent $A_4$ and rarer $D_4$ filaments together, and comment only on the populations in the entirety of the cosmic filament network. We summarise our results in \cref{fig:populations}, which displays, as a function of the smoothing scale $\sigma$, the populations of galaxies in the scale-space cosmic web elements.

The left panel shows the fraction of galaxies contained in the voids, walls, filaments and clusters, respectively. Before identifying quantitative properties, it is evident that the fraction of galaxies in the web elements depends on their defining length scale, as the plots are not constant. At low $\sigma$, the caustic skeleton is sensitive to small-scale shell-crossing events, and therefore covers nearly the entirety of the mild to dominant cosmic overdensities (see \cref{fig:caustics_scale_space}) in the simulation volume. Consequently, nearly all the galaxies are embedded in the ensemble of the walls, filaments and cluster nodes. At these low spatial scales, the cosmic web includes galaxy-sized features and structures of the cosmic web, such as filamentary tendrils and galaxy-sized halos at the nodes. 

are in fact galaxy-sized halos. In contrast, the voids separating the low-$\sigma$ caustic environments are nearly empty and do not contain significant substructure; the fraction of void galaxies is therefore close to vanishing. We observe that at low $\sigma$, a significant fraction of the galaxies reside near the cluster nodes. These correspond to small-scale knots -- including the early-time galactic-scale halos -- in the swallowtail and umbilic filaments, which the small scale caustic skeleton is sensitive to.\footnote{We stress that care must be taken when comparing this result to the literature, as conventional web identification methods typically measure clusters at large scales and additionally require high threshold overdensities. We deliberately set these criteria aside here to maintain an agnostic approach based only on the phase-space dynamics of the dark matter fluid. In doing so, we identify a large number of small-scale galaxy-scale and grouplike clumplike features from the low-$\sigma$ caustic skeleton, and consequently obtain a large fraction of galaxies lying within a megaparsec neighbourhood of the same. The analysis of the preceding section, in particular \cref{fig:histograms_scales}, shows that the low-$\sigma$ nodes are indeed compact overdensities hosting redder galaxy populations. However, due to their small-scale nature, we expect that conventional morphological filters would generally not identify these features as clusters, but rather associate them with overdensities within the cosmic filament network. Reflecting the hierarchical formation of structure, the near-scale-invariant nature of the primordial perturbations seeds a web-like pattern of the dark matter distribution down to galactic scales. With the Zel'dovich approximation being valid at early times even at small scales, the cluster node condition of \cref{tab:caustic_skeleton} traces the dynamical origin of the galaxy-scale overdensities that subsequently undergo non-linear evolution to virialise into groups or individual galaxy-sized (sub-)haloes. Hence, at small smoothing scales, the caustic skeleton naturally traces the `node-like' environments of small individual objects,  below the virial radius of typical large-scale clusters. Consequently, a significant portion of those galaxies that would conventionally be associated with the (small-scale) filaments are identified with small-scale individual galaxy- and group-size halos. It results in a low fraction of the filament galaxy population. For a better comparison with the literature, one could group these classifications together below a certain smoothing scale (e.g. $\sigma \lesssim  1.0 \,h^{-1}\textrm{Mpc}$) or below a certain density contrast (e.g. $\rho / \bar{\rho} \lesssim  100$). However, we stress again that we leave such deliberations aside to maintain an agnostic approach based only on the caustics without introducing further numerical parameters.}
This is a direct manifestation of the hierarchical structure formation, in which the near-scale-invariant nature of the primordial fluctuations seeds a web-like dark mass distribution down to galactic scales. As we increase the smoothing scale and identify the caustics at large $\sigma$, the portion of galaxies in the filaments and clusters decreases. This is because, while constituting the most dominant overdensities, the large-$\sigma$ filaments and clusters cover less and less spatial extent of the web-like mass distribution. In turn, the large-$\sigma$ voids contain significant substructure (see \cref{fig:caustics_scale_space}), such that at $\sigma=2.5 \,h^{-1}\textrm{Mpc}$ we identify about a third of the galaxies as residing in voids. We also find that at large $\sigma$, the fraction of wall galaxies exceeds that of filament galaxies. This is a direct confirmation of the results of \cite{Hertzsch+2026}, where the authors demonstrated that large-scale cosmic walls contain a large number of haloes that trace their filamentary substructure,  identified in turn by the small- and intermediate-$\sigma$ filamentary caustics (and interspersing cluster nodes).

The left panel thus demonstrates that the fractions of the overall galaxy populations in the different web environments depend on the defining smoothing scale. To further investigate the galaxy colour bimodality we have found above, we now discuss the total population of blue galaxies across the different environments. We define these to be the galaxies whose colour index is less than the mean colour index of all galaxies, $g-r < 0.58$; we note that this threshold value corresponds to approximately the local minimum of the bimodal distribution of the total galaxy population, see  \cref{fig:histograms_tng100}. In the middle panel of \cref{fig:populations}, we show the total number of blue galaxies across the different environments as a function of $\sigma$. Again, we find a clear dependence of the number of embedded galaxies on the defining smoothing scale. At low $\sigma$, the walls, filaments, and even the nodes, cover also moderate cosmic overdensities, and therefore accommodate nearly the entirety of the blue galaxy population. Whereas in the large-$\sigma$ regime, their contribution steadily decreases. At $\sigma=2.5\,h^{-1}\textrm{Mpc}$, the clusters correspond to the most dominant knots in the cosmic web, which contain nearly no blue galaxies. Similarly, the large-$\sigma$ filaments account for only about a tenth of the overall blue galaxy population. Conversely, the total number of blue galaxies in the voids near-linearly increases with the smoothing scale. Starting at nearly zero for low $\sigma$, the large-$\sigma$ voids contain about a third of the total number of galaxies (see left panel), but more than half of the total number of blue galaxies. Clearly, this is in accordance with the expectation from \cref{fig:histograms_scales}, considering that the large-$\sigma$ voids make up the major share of the overall cosmic volume. At the largest scales ($\sigma=2.5\,h^{-1}\textrm{Mpc}$), we find that about 90 percent of the blue galaxies reside in walls and voids together, while this fraction reduces to about 70 percent at intermediate scales ($\sigma=1.5\,h^{-1}\textrm{Mpc}$) and to about 15 percent at the smallest scales ($\sigma=0.5\,h^{-1}\textrm{Mpc}$).

Finally, the right panel of \cref{fig:populations} shows the fraction of blue galaxies within the galaxy population of each caustic environment. The shown values correspond to the integrals of the marginal distributions in \cref{fig:histograms_scales}  to the mean colour $g-r = 0.58$, and hence provide a consistency check for the discussion of the total populations shown in the left and middle panel of \cref{fig:populations}. As expected, the small-$\sigma$ walls and filaments, and to a lesser extent the cluster nodes, still accommodate a considerable fraction of blue galaxies. As the smoothing scale increases, the caustics become sensitive only to the more dominant overdensities, and the galaxy populations consequently become redder. Consistent with \cref{fig:histograms_scales}, we can identify a clear progression over a wide range of smoothing scales: the wall galaxies are always bluer than the filament galaxies, which in turn are always bluer than the clusters. These are the most red environments, and their fraction of blue galaxies quickly falls off to about 10 percent at the largest scales, while the fractions remain about 25 and 40 percent for the filaments and walls, respectively. The voids constitute the environments with the bluest populations, as the fraction is nearly constant at about 60-70 percent over a wide range of scales; as for the low-$\sigma$ voids, we reason the inversion of the colour trend may be due to the small number of void galaxies, thus increasing the errors on the determination of the blue fraction.

Altogether, the results of this section show that the cosmic web and its multiscale nature are pivotal to understanding properties of the galaxies that we observe in the sky. Having investigated in detail their dependence on the different caustic environments, we now move to the last part of this paper and study another aspect of the cosmic mass distribution, namely the impact of the web formation time on the properties of the embedded galaxies.


\section{Dependence of galaxy properties on the cosmic web formation times}
\label{sec:formation_time}

Each individual structural feature in the cosmic web has its own formation time. The caustic skeleton model traces the formation history of the entirety of the emerging, merging and evolving mass elements of the cosmic web. In this section, we make use of the formalism's unprecedented identification of the formation times of the cosmic large-scale structure elements to study the properties of galaxies in the aged present-day cosmic web.

\begin{figure*}
    \centering
    \includegraphics[width=\linewidth]{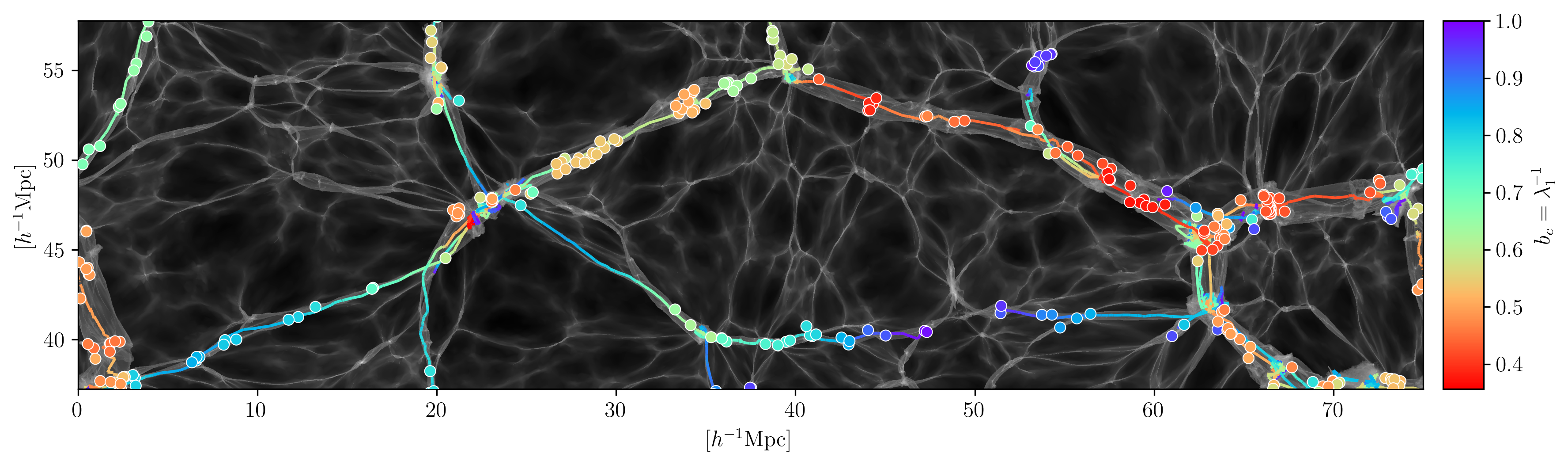}
    \caption{The formation time of the present-day cosmic web in the TNG100 simulation. Shown is a slice of the density field (same slice as \cref{fig:triple_plot} and \cref{fig:caustics_scale_space}) with the slices through the present-day cosmic walls traced by the intermediate-scale cusp sheets at $\sigma=1.5\,h^{-1} \textrm{Mpc}$. The mass elements in the walls are coloured by their formation time $b_c = \lambda_1^{-1}$ as determined by the first eigenvalue $\lambda_1$ of the primordial mass element out of which they emerged (see \cref{tab:caustic_skeleton} and main text). The dots show the dark matter haloes that reside within a slab of thickness $\epsilon = 0.05\,h^{-1}\textrm{Mpc}$ around the shown slice, and within a megaparsec neighbourhood of the identified walls. The haloes are accordingly coloured by the formation time of their nearest mass element in the cusp sheets.}
    \label{fig:caustics_age}
\end{figure*}

\subsection{Cosmic web formation time from simulations}
\label{sec:formation_time-sims}

The formation history of the cosmic web is crucial to understanding its present-day morphology. Emerging through a multiscale and hierarchical network of caustics, the overdensities of the shell-crossed web elements accrete matter as time progresses. Consequently, one expects that older web elements have undergone more phase-mixing and have different morphologies from the youngest, nascent components of the cosmic large-scale structure. Solidifying this insight, \cite{FeldbruggeWeygaert2024} and \cite{Hertzsch+2026} have recently shown in constrained simulations that the dark matter density fields associated with the different cosmic web environments exhibit a systematic dependence on both the length scale and the formation time of the defining caustics. It is natural to expect that this dependence further affects the properties of the web-embedded dark matter haloes and the luminous galaxies we observe today.

In this section, we investigate this dependence by evaluating the formation time of the different elements of the cosmic web from its caustic skeleton. 
Specifically, we identify the formation time with the moment of first shell-crossing, marking the onset of non-linear gravitational collapse and the start of the phase-mixing process. In the Zel'dovich approximation, one can directly infer the formation time $b_c$ from the first eigenvalue field of the deformation tensor,  $b_c=\lambda_1^{-1}$. In the caustic skeleton, this corresponds to the formation of the fold caustic (see \cref{tab:caustic_skeleton}). Reflecting the accuracy of the formalism, we find that this prediction is a good approximation to the formation time not only of the web elements that have formed until the present day, but also those that are yet to form in the cosmic future, see the forthcoming \cite{FeldbruggeHertzschWeygaert2026}. In principle, one may further improve on this estimate by applying higher-order Lagrangian Perturbation Theory (LPT) \citep{Bouchet+1995, Catelan1995, RampfHahn2021}, of which the Zel'dovich approximation is the first-order solution. Remarkably, \cite{RampfHahn2021} have shown that high-order LPT is accurate until first shell-crossing. Along with the ‘perfect' collapse model by \cite{Wislocka+2025}, these methods may offer powerful tools for investigating the formation time of the caustic skeleton delineating the cosmic web. However, we leave such applications to future studies, and proceed here by evaluating the web formation time through the prediction from the Zel'dovich approximation.

\Cref{fig:caustics_age} illustrates the formation time of the network of large-scale cusp sheets corresponding to the cosmic walls. Emerging as isolated components, as time progresses, the sheets merge to form a connected network. Clearly, the formation time is imprinted in the morphology of the present-day structures, as the older walls in \cref{fig:caustics_age}  have acquired more mass and appear thicker and more overdense than their younger counterparts. This figure visually motivates a careful treatment of the galaxy properties with respect to the formation time of their embedding web environment.

It is challenging to construct the formation times of individual structures in the cosmic web using conventional morphology filters. By classifying the configuration of a tracer field, e.g. the dark matter density, at a single instant in time, such methods do not capture the phase-space evolution of the cosmic mass distribution. In principle, it is possible to obtain a dynamical picture of the cosmic web by applying morphology filters to classify the structures in the different snapshot files. This was done in the context of galaxy properties at varying redshifts by \cite{Kraljic+2017, Hasan+2023, Perez+2024b, Hasan+2025, Jego+2026}. However, there does not seem to be rigorous and natural way to track the temporal evolution, merger and morphogenesis of the individual web elements purely on morphological grounds, especially when structures may appear or disappear as numerical artefacts between consecutive snapshot files.

These limitations set the caustic skeleton apart as a cosmic web identification method, which inherently traces the formation history of the entirety of the present-day large-scale structure. As a Lagrangian formalism, the caustic skeleton yields a continuous classification of the cosmic web not only over length scales (see \cref{fig:caustics_scale_space} and \cref{subsec:galaxies-scales}), but also over its formation time. We reiterate that to each mass element in the present-day cosmic web, as evaluated on a scale $\sigma$, the caustic skeleton associates a formation time $b_c$ through the first eigenvalue $b_c = \lambda_1^{-1}$ of the primordial mass element out of which the late-time position originated. We now employ this simple yet powerful notion to investigate, for the first time, the dependence of the galaxy properties in the IllustrisTNG simulations on the age of the embedding web elements.

\subsection{The dependence of galaxy properties on the web formation times}
\label{sec:formation_time-galaxies}

\begin{figure*}
    \centering
    \includegraphics[width=\textwidth]{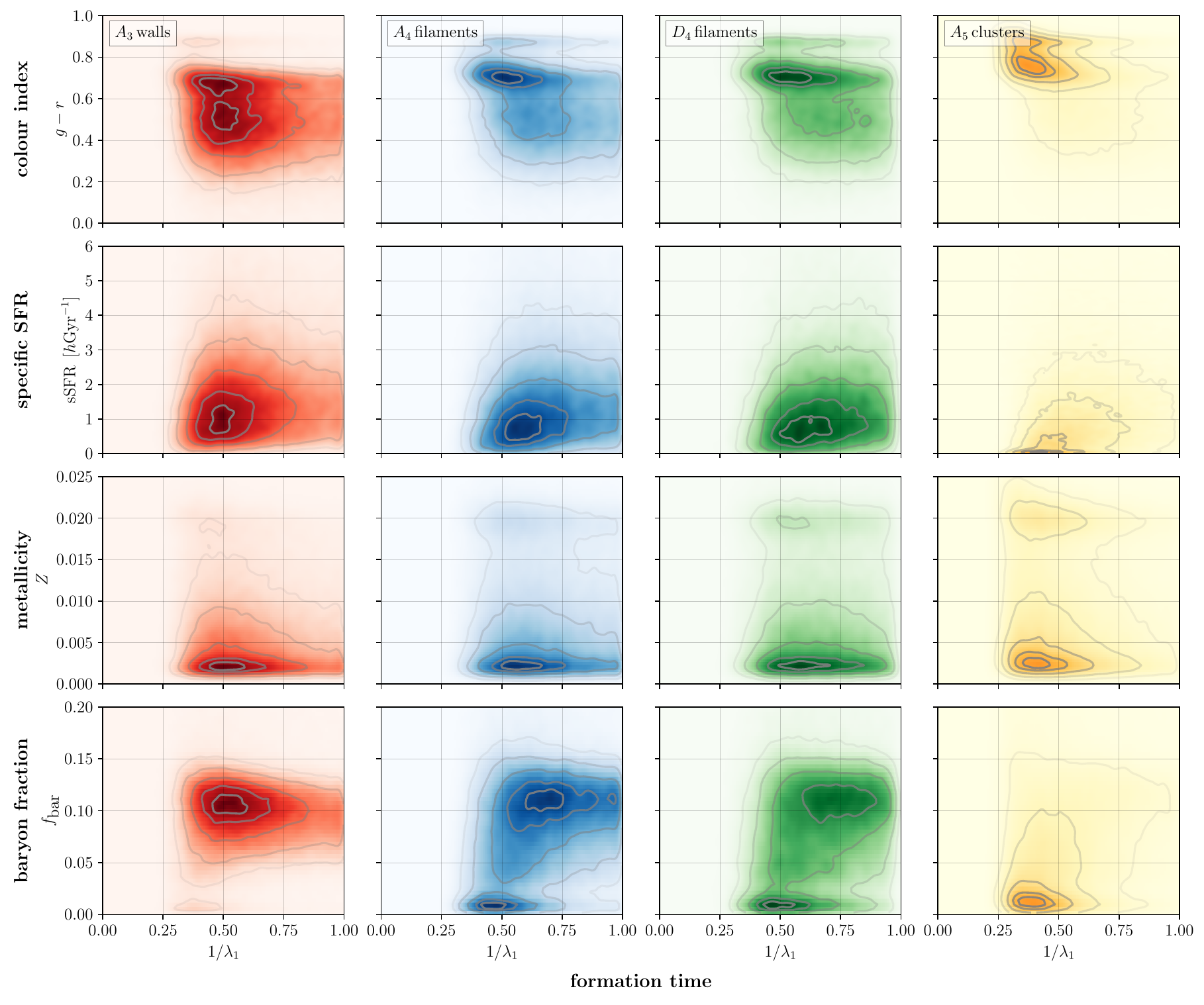}
    \caption{Dependence of galaxy properties on the web formation time in the TNG300 simulation. The web elements are traced by the intermediate-scale caustic skeleton, evaluated at $\sigma=1.0 \,h^{-1} \textrm{Mpc}$. Shown are the bivariate PDFs of the $g-r$ colour index (upper row), specific star formation rate (second row), metallicity (third row) and baryon fraction (fourth row) against the web formation time. The columns correspond to the different caustic environments. The PDFs are given as the kernel density estimations of the histograms obtained 
    from the galaxies near the caustic environments in the full TNG300 simulation volume. For each galaxy, the nearest mass element in the cosmic web is identified, and the formation time $b_c = \lambda_1^{-1}$ of the mass element is evaluated from the first eigenvalue $\lambda_1$ of the corresponding primordial mass element. The contours of the PDFs are displayed at levels $\{0.1, 0.25, 0.5, 0.75, 0.9\}$.}
    \label{fig:formation_time_histograms_1}
\end{figure*}

\begin{figure*}
    \centering
    \includegraphics[width=\textwidth]{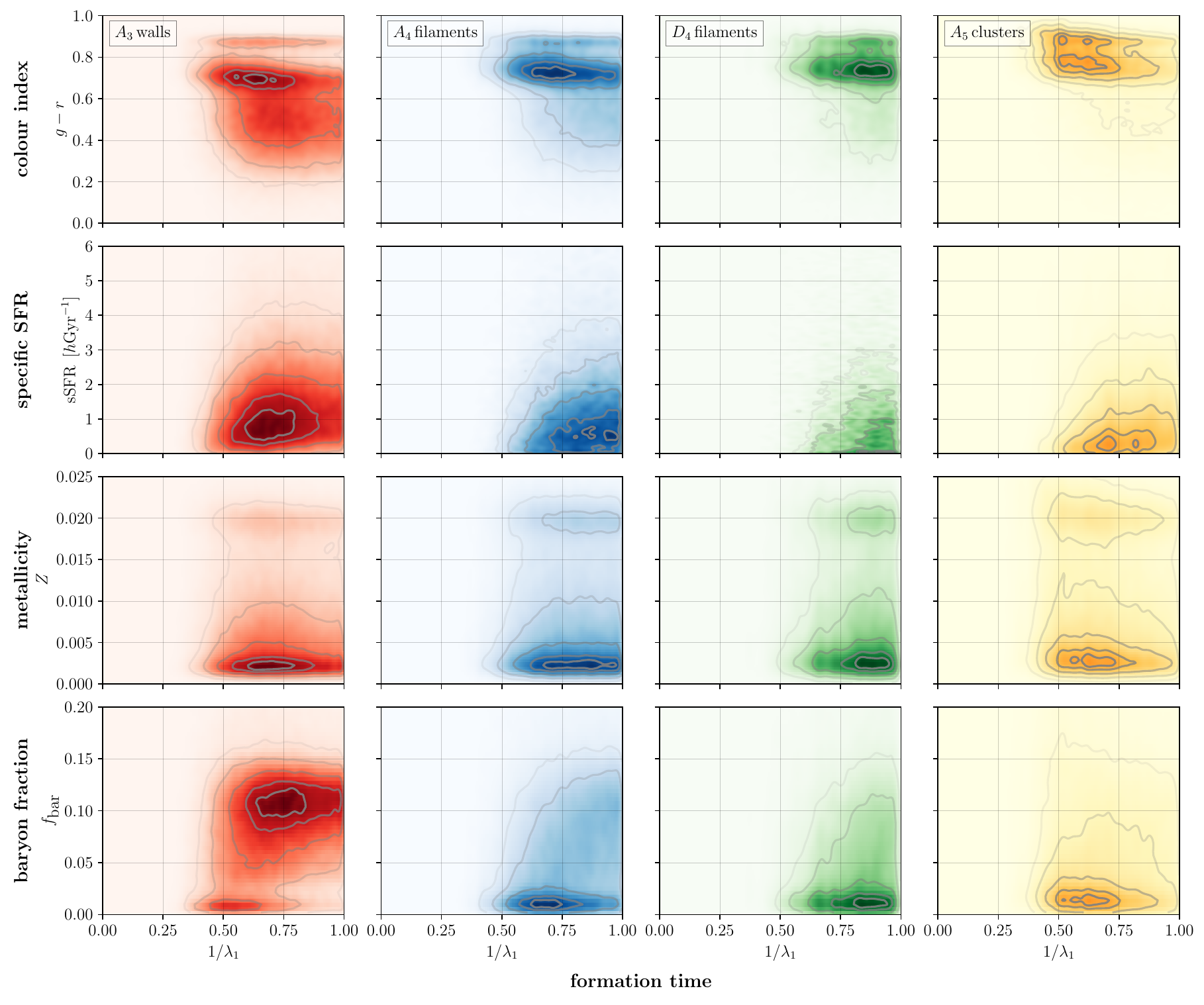}
    \caption{Same as \cref{fig:formation_time_histograms_1}, but with the cosmic web elements identified by the large-scale caustic skeleton, evaluated at smoothing scale $\sigma=2.0 \, h^{-1} \textrm{Mpc}$.}
    \label{fig:formation_time_histograms_2}
\end{figure*}

As in \cref{sec:galaxies}, we use the galaxies' position given in the halo catalogue and associate these with the walls, filaments and clusters identified by the caustic skeleton at a chosen smoothing scale $\sigma$. We then measure, for each galaxy, the formation time  $b_c = \lambda_1^{-1}$ of the nearest mass element in the cosmic web (i.e. of the galaxy's embedding); \cref{fig:caustics_age} illustrates this for the galaxies near the cusp sheets. Having obtained the ages of the respective mass elements in the cosmic web,  we investigate their relations to the characteristic galaxy properties that we studied in \cref{sec:galaxies}. We do not consider the voids here, as we are explicitly interested in the multistreaming web elements that have formed up to the current time. Moreover, to consider a representative sample of large-scale cosmic web elements at different ages, we present our analysis only for the TNG300 simulation, which contains a large number of distinct large-scale walls and filaments. While we have obtained similar results for the TNG100 simulation, its smaller volume and finite number of large-scale web features render the analysis significantly more prone to cosmic variance.

We first study the relation between the web formation time and galaxy properties at smoothing scale $\sigma = 1.0 \,h^{-1}\textrm{Mpc}$. At this length scale, the caustic skeleton traces a significant portion of the overdensities in the present-day cosmic web (cf. \cref{fig:caustics_tng300,fig:caustics_scale_space}), and a substantial portion of the galaxies in the simulation volume are identified as residing within the walls, filaments and clusters. The histograms of \cref{fig:formation_time_histograms_1} show that there are systematic relationships between the $g-r$ colour index, specific star formation rate (sSFR), metallicity and baryon fraction of the galaxies and the formation time of their embedding web environments. Before commenting on these, we note that the marginal distributions of the formation times in \cref{fig:formation_time_histograms_1} are not to be confused with the the actual distributions of the formation times of the mass elements in the walls, filaments and clusters, which are beyond the scope of the present article; we refer the reader to \cite{Hertzsch+2026} and forthcoming work by \cite{HertzschFeldbruggeWeygaert2026}. By measuring the properties of the galaxies and their nearby mass elements, we consider here an inherently biased sample: the data points come only from those regions that have acquired sufficient overdensities for galaxy formation to occur in the first place. One therefore expects the majority of the galaxies to reside in older parts of the cosmic web, while its youngest, nascent components have not yet acquired pronounced overdensities and substantial galaxy populations. This explains the paucity of data points for $\lambda_1^{-1} \approx 1.0$ in the shown histograms.

In \cref{sec:galaxies}, we argued that the marginal distributions of \cref{fig:histograms_tng300} and \cref{fig:histograms_scales} reveal clear differences in the galaxy populations across different web environments. By including the web formation time in the analysis, we will now recover the same observations, but provide a more nuanced view of the galaxies' characteristic properties. Notably, within one family of the web elements, we find that those mass elements that shell-crossed earlier have typically acquired higher overdensities over cosmic time, and therefore accommodate more quenched galaxies at the present day. The most active galaxies, on the other hand, typically reside in those parts of the web that formed more recently. 

\begin{figure*}
    \centering
    \includegraphics[width=\textwidth]{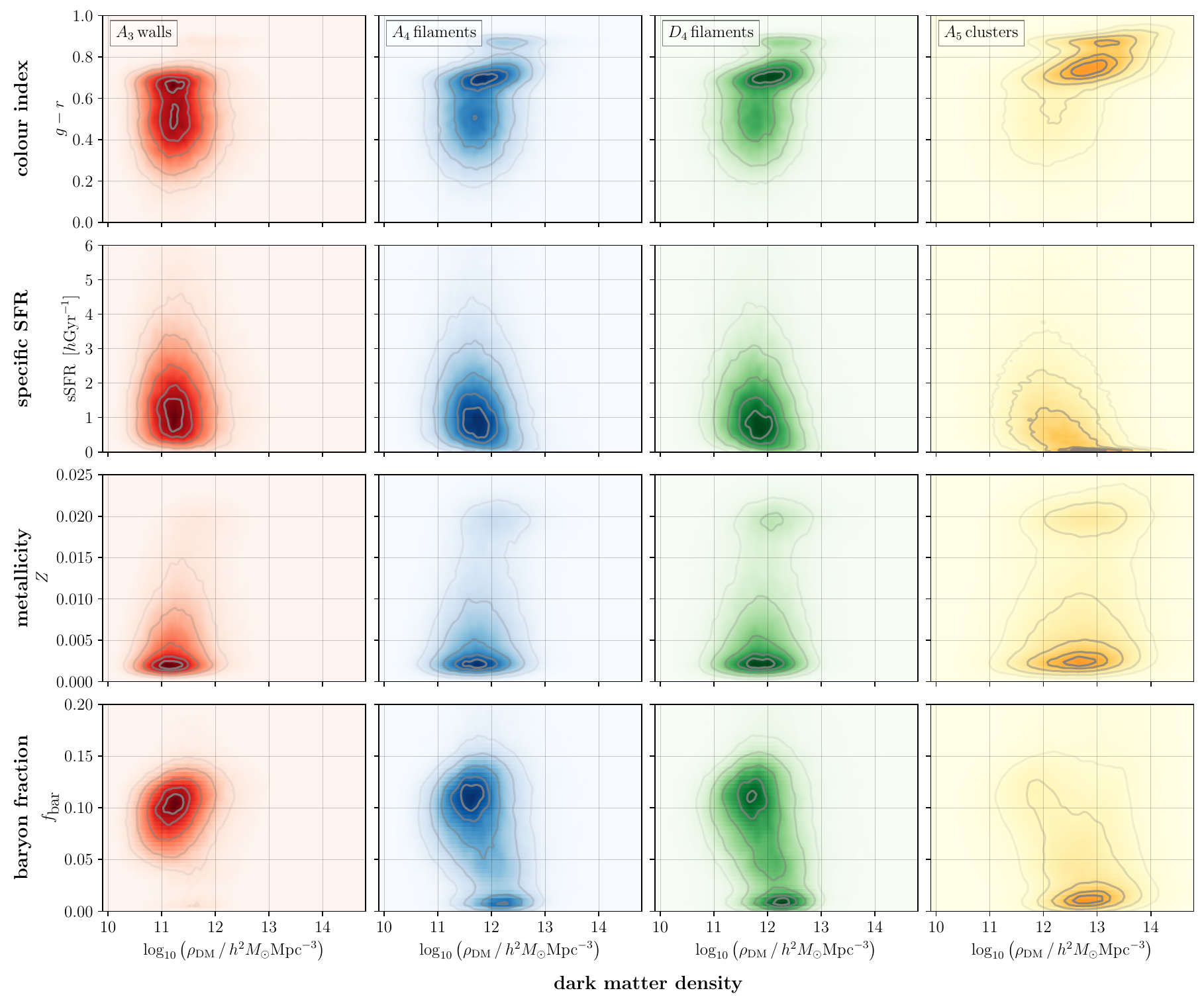}
    \caption{Galaxy properties and environmental dark matter densities in the TNG300 simulation; analogous to \cref{fig:formation_time_histograms_1,fig:formation_time_histograms_2} for the dependence on the web formation time. The web elements are identified by the intermediate-scale caustic skeleton at $\sigma=1.0\,h^{-1}\textrm{Mpc}$. The galaxies are classified as residing in the different caustic environments (corresponding to the different columns), and the bivariate PDFs of the galaxy properties against the environmental dark matter densities are obtained as kernel density estimations of the respective histograms. The contours of the PDFs are displayed at levels $\{0.1, 0.25, 0.5, 0.75, 0.9\}$.}
    \label{fig:density_histogram}
\end{figure*}

This is consistently demonstrated in the different panels of \cref{fig:formation_time_histograms_1}. The upper panel shows that, consistent with \crefrange{fig:histograms_tng100}{fig:histograms_scales}, the characteristic galaxy population colour in the walls is bluer than in the filaments, which in turn host bluer galaxies than the dense clusters. While differing significantly across the different cosmic web environments, the colour index exhibits a clear inverse trend with the formation time of the embedding web elements across all environments: the reddest galaxies in each environment typically reside within the oldest web parts of the cosmic web, while bluer galaxies are found in its younger, nascent features. 
Accordingly, the second row shows that the star formation activity of the galaxies residing in the oldest parts of the cosmic web is suppressed at the present day; the embedded red galaxy populations are typically the most quenched. Conversely, while quenching may also affect galaxies in younger environments, we find a near-linear trend of higher star formation activities with later formation times. This trend is apparent even for the galaxies in the recently formed filaments and clusters, as these host the bluer and active fraction of the overall filament and cluster populations, see \cref{fig:histograms_tng300}. This is particularly the case for the youngest swallowtail filaments.
The age of the cosmic web not only affects the present-day star formation activity, but also the metallicities of the embedded galaxies. The third row of \cref{fig:formation_time_histograms_1} shows that the most metal-rich galaxies reside in the older parts of the cosmic web. We find that, in particular for the clusters, the pronounced second peak in the marginal metallicity distribution (see  \cref{fig:histograms_tng300}) is suppressed for the galaxies in the youngest web components; to a lesser extent, the same can be said for the galaxies in the walls and the filaments. 
Finally, the lower row of \cref{fig:formation_time_histograms_1} shows the corresponding relationships for the galaxies' baryonic content. We find that for the walls, the present-day baryon fraction does not significantly depend on the formation time. For the filaments, on the other hand, it is evident that the galaxies near the oldest mass elements have the lowest baryon fractions, as these environments have been effectively stripped of their baryons over cosmic time. Conversely, filament galaxies only retain considerable baryon fractions in younger environments. A similar statement can be made for the clusters, although the trend is less pronounced here.

We reiterate at this point that the identification of the cosmic web from the caustic skeleton depends on the length scale at which the caustics are measured. We discussed the scale-space nature of the cosmic web and the properties of its embedded galaxies in detail in \cref{subsec:galaxies-scales}. When relating these properties to the formation time of the web elements, the scale-space nature needs to be taken into account. In the analysis above, we considered the caustics at $\sigma=1.0 \,h^{-1} \textrm{Mpc}$ and therefore included moderate overdensities and less pronounced web elements traced by the intermediate-scale caustics. We now consider the large-scale web traced by the caustic skeleton at $\sigma=2.0 \,h^{-1} \textrm{Mpc}$, and show the respective histograms in \cref{fig:formation_time_histograms_2}. By smoothing the primordial potential perturbation with a larger-$\sigma$ window function, the magnitude of the tidal eigenvalues is lowered (see right panel \cref{fig:caustics_scale_space}), and the large-$\sigma$ caustics are therefore found to form at later times $b_c=\lambda_1^{-1}$, in agreement with the hierarchical model of structure formation; for a detailed discussion, see \cite{Hertzsch+2026}. The histograms of \cref{fig:formation_time_histograms_2} confirm this, as the formation of the web elements is now found to set in at around $b_c \approx 0.5$, later than $b_c \approx 0.3$ as in \cref{fig:formation_time_histograms_1}. Moreover, the large-$\sigma$ caustics constitute the more tidally dominant, spatially extended and overdense elements of the multiscale web. As in \cref{fig:histograms_scales}, the galaxy populations are therefore overall redder and more quenched than in the skeleton at $\sigma=1.0\,h^{-1}\textrm{Mpc}$. Yet, within these overall more quenched populations, we find the same trends that we have seen above. Notably, the upper panel of \cref{fig:formation_time_histograms_2} shows that the galaxy populations are again bluer when embedded in younger parts of the cosmic web, consistently observed across all environments. We note that the distribution of the wall galaxies' colours is now shifted more towards the red end than in the $\sigma=1.0 \,h^{-1}\textrm{Mpc}$ case. While galaxies in younger wall components are still on the bluer spectrum, the overall population is now less bimodal, consistent with the respective marginal distributions shown \cref{fig:histograms_scales}. Similarly, the bluer population of filament galaxies resides solely in the young parts of the cosmic web, and the overall fraction of blue galaxies is now reduced compared to the  $\sigma=1.0 \,h^{-1}\textrm{Mpc}$ case. The same is seen for the clusters. Moreover, we find that notably in the filaments and clusters, considerable star formation rates persist only in the youngest web elements, whereas the galaxy populations have been largely quenched in the older parts of the cosmic web, just as was seen above.  The metallicity, on the other hand, does not exhibit as clear a trend as in the previous case. We reason that this is because we here consider the dominant, more overdense large-scale caustics, which typically accommodate metal-rich galaxy populations at the current time, even when having formed more recently. Finally, we find that the galaxies in the oldest large-scale filaments and clusters have been largely depleted of their baryons, while considerable baryon fractions only persist for the population in the youngest walls.

\subsection{Density and formation time}

Altogether, the results confirm that the formation time of the cosmic web elements has a significant influence on the properties of the embedded galaxy populations. In particular, our measurements are consistent with the assumption that older parts of the cosmic web have acquired higher overdensities, and therefore accommodate redder, more quenched galaxies. In \cref{fig:density_histogram}, we show the corresponding histograms of the galaxy properties against the dark matter density of the embedding environment, evaluated for the intermediate-scale caustic skeleton at $\sigma=1.0 \, h^{-1}\textrm{Mpc}$. The marginal distributions of the densities near galaxies are similar to ones shown for the full simulation volume in \cref{fig:fields}, in that the walls constitute moderately overdense environments (compared to the underdense voids) with density contrasts $\rho / \bar{\rho} \approx 1-10$ over the mean cosmic density $\approx 8.6 \cdot 10^{10}\,h^2 M_{\odot} \textrm{Mpc}^{-3}$. In comparison, the filaments at $\sigma=1.0 \, h^{-1}\textrm{Mpc}$ reach density contrasts up to  $\rho / \bar{\rho} \approx 100$, while the clusters host overdensities of up to an order of magnitude more, that is $\rho / \bar{\rho} \approx 1000$. As expected, the upper panel of \cref{fig:density_histogram} shows a clear trend of the galaxy colours with the environmental densities, directly confirming the observationally supported colour-density relation (e.g. \cite{Cooper+2007}). Particularly for the large-density end of the filament and cluster populations, we find a clear and approximately linear relationship between the order of the environmental density contrast and the galaxy colour index. Moreover, we recover the results of \cref{fig:histograms_tng300} in that we can associate the distinct peaks in the marginal colour distribution with the walls, filaments and clusters respectively. The star formation rates across the environments behave accordingly, as they are suppressed with increasing environmental densities. The quenching is particularly pronounced in the densest clusters; here, solely quiescent populations are found beyond densities of about $\rho \sim 10^{13.5} \, h^2 M_{\odot} \rm{Mpc}^{-3}$. We further obtain a clear trend of high-metallicity galaxy populations residing in denser environments, consistent with the trends discerned in the marginal distributions in \cref{fig:histograms_tng300}. For instance, we find that the second peak of the marginal distribution is pronounced only for the dense clusters. Finally, we find an inverse trend of the baryon fraction with the environmental densities. While the galaxies in the low-density walls have retained a significant portion of their baryonic content, the baryon fraction is near-linearly decreasing with the environmental density in both the swallowtail and umbilic filaments. The latter constitute the denser filament family, and the galaxies in the umbilics are therefore generally less baryon-rich than in the swallowtails. This trend clearly continues in the clusters, where baryon fractions are significantly suppressed and eventually near-vanishing in the highest-density environments.

\begin{figure*}
    \centering

    \begin{subfigure}{\textwidth}
        \centering
        \includegraphics[width=0.48\textwidth]{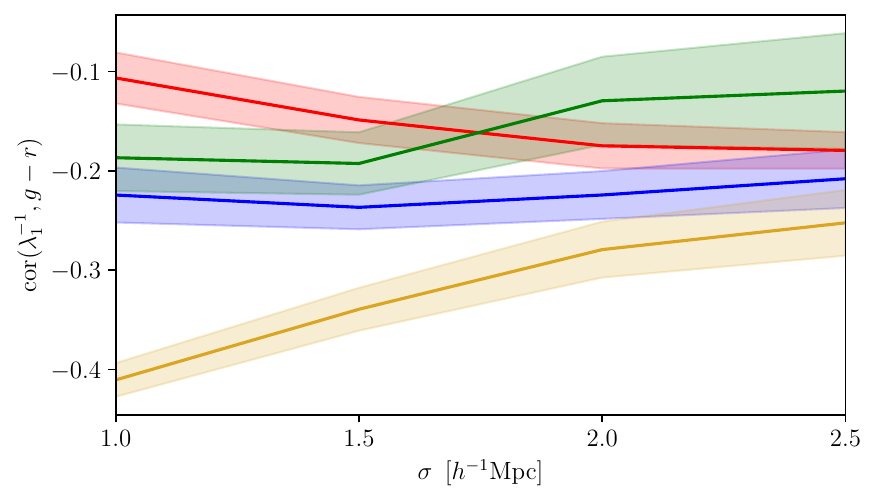}
        \hfill
        \includegraphics[width=0.48\textwidth]{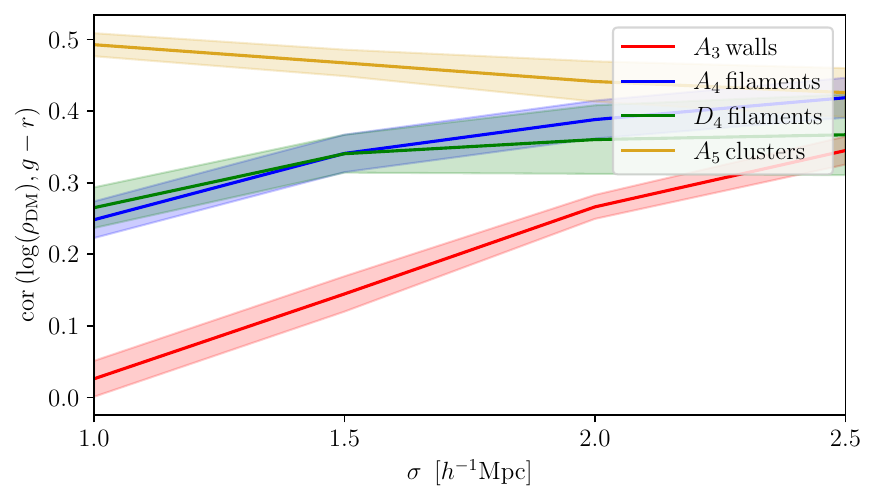}
        \caption{scale-space correlation of the $g-r$ colour index}
    \end{subfigure}

    \vspace{1.5em}

    \begin{subfigure}{\textwidth}
        \centering
        \includegraphics[width=0.48\textwidth]{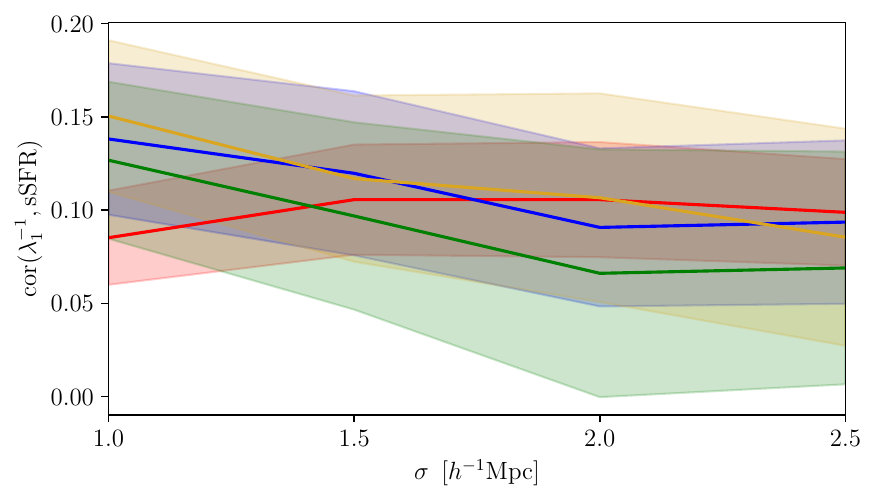}
        \hfill
        \includegraphics[width=0.48\textwidth]{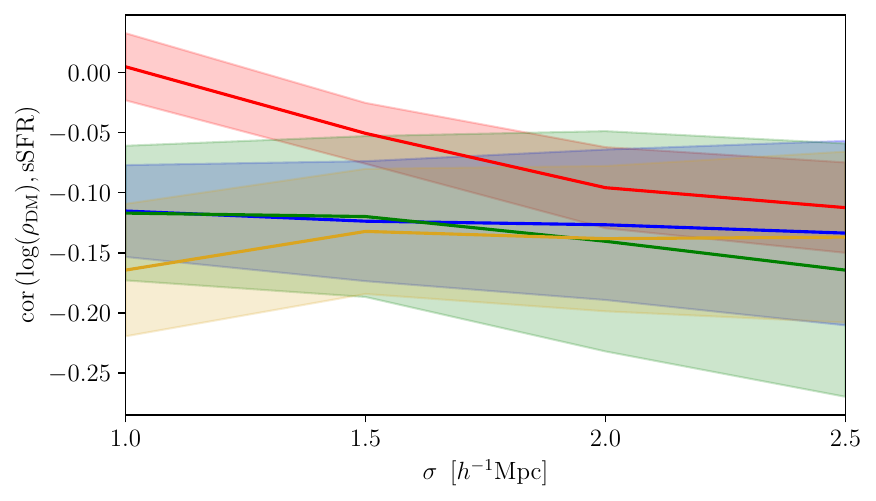}
        \caption{scale-space correlation of the specific star formation rate}
    \end{subfigure}

    \caption{Correlations of the TNG300 galaxy properties with the web formation time (left) and the environmental dark matter density (right) in scale space. Shown are the correlation coefficients for the galaxies' $g-r$ colour index (upper row) and specific star formation rate (lower row) 
    as a function of $\sigma$, across the different caustic environments (see legend). We determine the TNG300 caustic skeleton for smoothing scales $\sigma= 1.0,\, 1.5,\, 2.0,\, 2.5 \, h^{-1} \textrm{Mpc}$. For each scale, the embedded galaxies are identified and the correlation coefficient between the galaxy property and the web formation time or environmental density is computed (cf. \crefrange{fig:formation_time_histograms_1}{fig:density_histogram}). The uncertainties are estimated by drawing 100 subsamples from the different galaxy populations. We evaluate the correlation coefficient for each of these and report the measured correlation and $1\sigma$-uncertainty bands (shaded region) as the mean and standard deviation, respectively.}
    \label{fig:correlations}
\end{figure*}

\subsection{Quantifying the scale-space correlation of galaxy properties and web formation times}
\label{sec:formation_time-correlation}

The results of the previous section demonstrate that both the formation time and the density contrast of the web elements are relevant to understanding their local properties. Of course, both are intricately related, as the large-scale cosmic web elements are shaped in the linear and mildly non-linear regime of the dark matter dynamics, but subsequently undergo non-linear matter accretion to form the connected and highly overdense network we see today. One therefore naturally expects a significant degeneracy between the formation time and the density contrasts, as was found in \cite{Hertzsch+2026} regarding the morphology of walls formed at different times and at varying length scales. Recent studies on galaxy properties, in particular \cite{NandiPandeySarkar2026} on the IllustrisTNG simulations, have proposed numerical quantifications of the relationships between the galaxy properties in the web environments. However, this and related analyses have typically treated the web through morphology filters at a single characteristic length scale; the statistical summaries have therefore left the multiscale of the cosmic mass distribution aside.

We now revisit these investigations in the context of our scale-space analysis of the galaxy properties and the web formation time. We are interested in their statistical relations over the range of cosmologically relevant length scales tracing the cosmic web. We investigate these by evaluating the TNG300 caustic skeleton at scales $\sigma =\{1.0, \,1.5,\, 2.0,\,2.5\,h^{-1} \mathrm{Mpc}\}$. Further, to obtain an interpretable quantification of the relationships, we measure the linear correlation coefficients between the respective distributions. These highlight the existence of pronounced relationships (or inverse relationships) between the galaxy properties and their web embedding. This is particularly relevant for the galaxy colours and the sSFR, which we have found above to exhibit clear and nearly linear trends with the web formation time and the dark matter density, consistent with the well-established colour-density relation (e.g. \cite{Cooper+2007}). We correlate these properties and show the results in the \cref{fig:correlations}. In the upper left panel, we find that the formation time is negatively correlated to the colour of the galaxy populations across all web environments over the full range of length scales. The negative correlation is most pronounced for the small-scale clusters, meaning that the galaxies near the oldest clusters at $\sigma=1.0\,h^{-1}\mathrm{Mpc}$ are significantly redder than the ones near the younger ones. The correlations for the filaments and clusters at larger length scales are weaker, yet persist as statistically significant. While we also measure a negative correlation coefficient for the walls, in this case it is lesser in magnitude at smaller $\sigma$, which we reason is due to the overall low density and hence reduced galaxy production in the low-$\sigma$ walls. Conversely, the lower left panels show that the sSFR has a positive correlation with the web formation time, meaning that galaxies in environments that formed later are more active. The correlation coefficients are statistically significant, yet are lesser in magnitude and more variable than for the colours, implying that the relationships are less linear. Moreover, the sSFR correlations are less significant at large $\sigma$, where the large-scale skeleton traces only the most dominant web structures. Again, we attribute the surprising behaviour of the wall correlation coefficient to suppressed galaxy formation in low-$\sigma$ walls. We analogously calculated the scale-space correlation coefficients for the baryon fraction and the metallicity, but omit these here, as we find that the distributions are not well approximated by linear relations, with the correlation coefficients being consistent with zero.

In the right panels of \cref{fig:correlations}, we show the scale-scale space correlations of the colour index and the sSFR against the dark matter density contrast. We find that the colour has a positive correlation with the environmental density, with the correlation coefficient being most significant for the largest-scale filaments and clusters, which accommodate the highest overdensities, as we argued for \cref{fig:density_histogram}. Again, this quantitatively confirms the observationally supported colour-density relation. Accordingly, the sSFR is negative correlated with the dark matter density,  implying that the more active galaxies reside in less dense environments. The correlation coefficients here are fairly stable over the chosen range of length scales, and most pronounced at large-$\sigma$.

Overall, \cref{fig:correlations} along with the preceding figures suggests that the galaxy properties exhibit significant correlations with both the formation times of the cosmic web elements and their characteristic density contrasts. The linear correlations with the latter are more pronounced, and directly interpretable with the characteristic densities of the walls ($\rho/\bar{\rho}\approx 1-10$), filaments ($\rho/\bar{\rho}\approx 10-100$) and clusters ($\rho/\bar{\rho}\approx 100-1000$) respectively. Our study demonstrates that the formation history is relevant to understanding the assembly of galaxy populations in the cosmic web, and implicitly influences their present-day properties. Moreover, while we have focused on stellar-population-related properties in this article, we expect that the formation history of the cosmic web has a significant impact on the dynamical properties of the present-day galaxy population. In particular, we expect that the shapes and spin alignments of galaxies exhibit strong dependencies on the formation histories of the various cosmic web components. By applying the identification methods of caustic skeleton formalism, this may shed more light on a multitude of preceding works, such as the recent study by \cite{Hasan+2025}. We leave such investigations to future work.

\section{Conclusion}

The cosmic web is the largest known structural pattern in the Universe. Its intricate and multiscale filamentary morphology is woven by the non-linear gravitational dynamics of the collisionless dark matter and the infall and hydrodynamic evolution of the baryonic gas. Constituting the embedding of all galaxy formation, it has been known since the seminal work by \cite{Dressler1980} that the cosmic web influences the properties of the embedded galaxies. Numerous observational analyses, \cite{Strateva+2001, Baldry+2004, Balogh+2004, Menanteau+2005} to name but a few, revealed that the cosmic galaxy population is bimodal, composed on the one hand of blue, actively star-forming galaxies and on the other hand of red, quiescent galaxies. \cite{Kaufman+2004, Hogg+2004, Balogh+2004, Blanton+2005} and many subsequent works demonstrated that these properties can be directly attributed to the environmental densities, thus establishing the famous \textit{colour-density relation} (e.g. \cite{Cooper+2007}). In recent years, the advent of large-scale simulations has enabled detailed analyses of this relation across different web environments, namely the voids, walls, filaments and clusters, see e.g. \cite{Metuki+2015, Laigle+2018, Xu+2020, Hasan+2023, Yu+2025, NandiPandeySarkar2026}. These preceding analyses have used morphology filters such as the \verb|DisPerSE| method \citep{Sousbie2011, Sousbie2011b}, the Bisous scheme \citep{StoicaGregoriMateu2005, Tempel+2016}, variations of the T-web formalism \citep{Hahn+2007, Forero-Romero+2009, Hoffman+2012} or other web identification schemes (see \cite{Libeskind+2018}) to classify the present-day mass distribution into different environments. While providing a consistent picture of galaxy properties in the cosmic web, these identifications are typically done at a single characteristic length scale only, thus discarding the multiscale nature of the cosmic mass distribution. Moreover, morphology filters do not take into account the phase-space dynamics shaping the cosmic web, and therefore lack a rigorous understanding of what constitutes the structural components of the cosmic web.
In this article, we revisited these analyses and, for the first time, applied the caustic skeleton formalism to trace the multiscale geometric backbone of structure formation and the embedded galaxies in the IllustrisTNG simulation suite \citep{Nelson+2019}.

The caustic skeleton model \citep{ArnoldShandarinZeldovich1982, Feldbrugge+2018} is a mathematically rigorous and parameter-free formalism that traces the formation history of the cosmic web from the singularities of the dark matter flow. A series of recent studies \citep{FeldbruggeWeygaert2023, FeldbruggeYanWeygaert2023, FeldbruggeWeygaert2024, Hertzsch+2026} has demonstrated its unprecedented power in identifying and statistically analysing the multiscale nature of the present-day cosmic mass distribution from its initial conditions. The formalism unambiguously associates the multistream regions, walls, filaments and clusters characterising the present-day matter configuration with an exhaustive list of caustics determined from the tidal configuration of the primordial potential perturbation. A particularly interesting prediction of the caustic skeleton model is the existence of two distinct filament families, of which we have presented first evidence in this article, and which will be further analysed in \cite{HertzschFeldbruggeWeygaert2026}. Employing its rigorous identification of the caustic environments, we applied the formalism to trace the scale-space cosmic web of the TNG100 and TNG300 simulations.

After a review of the caustic skeleton formalism in \cref{sec:caustics} and an overview of the IllustrisTNG suite in \cref{subsec:illustris-overview}, we commenced in \cref{subsec:illustris-caustics} by providing a detailed explanation of the construction of the caustic skeleton in the simulations. Unfortunately, the suite's glass pre-initial conditions have up to this point prohibited a naive Lagrangian treatment of the dark matter dynamics from the primordial conditions. This has constituted a major challenge in analysing the formation history of the cosmic web in the IllustrisTNG suite. To overcome this issue, we implemented an accurate and efficient numerical interpolation scheme that allows us to reconstruct the primordial potential out of which the simulations originated. This enables, for the first time, a Lagrangian treatment of the simulations and is expected to prove useful beyond the caustic skeleton formalism. Equipped with the ability, we identified the caustics and associated the present-day mass distribution and embedded galaxies with the different web environments.

In \cref{sec:field}, we used this identification to analyse the continuum fields characterising the cosmic web. Focussing first on the dark matter, we applied the phase-space DTFE method \citep{Feldbrugge2024, FeldbruggeHertzsch2025} to present, for the first time, the multistreaming dark matter density field. We demonstrated how the log-normal nature of the density field arises due to a summation of higher-streaming configurations, which will be analysed in more detail by \cite{Alferink+2026}. With the higher-streaming configurations being at the heart of the caustic skeleton formalism, we then moved on to study properties of the baryonic gas in the different caustic environments delineating the cosmic web. We find that across all structural components, namely the voids, walls, filaments and clusters, the baryonic gas density traces dark matter density near-linearly, so that the density ratio of baryonic gas to dark matter is about $5.3$ in all environments. However, significant variations arise for the gas temperature and metallicity. We find that the former transitions from a cool phase ($T \lesssim 10^4 \mathrm{K}$) to a hot phase ($T \gtrsim 10^5 \mathrm{K}$) from the walls to the filaments. This hints at significant differences in star formation activity across the web environments, as the heating of the gas suppresses baryonic clumps in the higher-density environments. Moreover, we demonstrated that the gas metallicity is strongly bimodal, with most of the cosmic volume exhibiting negligible metal fractions, while only the clusters and highest-density filaments accommodate considerable portions of elements heavier than helium. This is an unambiguous clue to differences in star formation histories across the web environments, as significant stellar activity over cosmic time has occurred only in the densest regions of the present-day mass distribution.

With the baryonic gas properties hinting at differences of the galaxy populations across the cosmic web, we moved in \cref{sec:galaxies} to the main part of our study, namely a detailed analysis of the galaxy properties in the different caustic environments constituting the present-day cosmic web. We focused in this article on the $g-r$ colour indices, specific star formation rates, metallicities and baryon fractions of the luminous galaxies in the TNG100 and TNG300 data catalogues. We first found that the caustic identification recovers the well-established progression of galaxy properties across the traditional quadripartite web environments from morphology filters. We find the void galaxies are bluest, with the lowest metallicities, retaining the highest baryon fractions and being actively star-forming. Whereas the galaxies in the densest environments, namely the clusters, are quenched; that is, these galaxies are red, with high metallicities, low baryon fractions and low star formation rates. The galaxy quenching is hierarchical in that the galaxies in the moderately dense walls are typically redder than those in voids, while still being on the bluer and more active end. The filament galaxies, in turn, are redder and more quenched than those in the walls, but less so than the galaxies in the clusters. Our results directly confirm preceding IllustrisTNG analyses, such as \cite{Yu+2025, NandiPandeySarkar2026}, and agree with the well-established colour-density relation (e.g. \cite{Cooper+2007}). We also present the first evidence of the two distinct filament families made up by the swallowtail and umbilic caustics, as the latter are typically more overdense and consequently host a mildly more red and quenched galaxy population. A detailed investigation of the filament morphologies will be addressed in \cite{HertzschFeldbruggeWeygaert2026}.

The caustic skeleton is sensitive to the multiscale nature of the cosmic mass distribution and provides a natural framework for tracing it from the scale-space network of dark matter singularities. By smoothing the primordial potential on a range of cosmologically relevant length scales, this article provided, for the first time, a detailed scale-space classification of the IllustrisTNG voids, walls, filaments and clusters. While previous studies on galaxy properties have largely ignored the impact of the multiscale nature, in \cref{subsec:galaxies-scales}, we went beyond the single-scale web identification and discussed the galaxy properties across the different scale-space web environments. In doing so, we obtained clear evidence of a second systemic progression that extends the traditional quadripartite web identification: the galaxies in the large-scale, more dominant overdensities, traced by the large-scale caustics, are typically redder, whereas those residing in smaller-scale caustics tend to be bluer. As such, the characteristic galaxy properties are highly dependent on the scale at which the web is identified. Succinctly summarised in \cref{fig:histograms_scales}, our results suggest that the galaxy properties form a continuum over the scale space of the different environments. For instance, we find that the galaxy population in small-scale swallowtail filaments is rather similar to that residing in large-scale walls. The scale-space nature of the cosmic web therefore blurs the traditional quadripartite galaxy populations, instead highlighting the continuous nature of the cosmic mass distribution and the existence of caustics (and consequently galaxy embeddings) from small to large length scales.

In \cref{subsec:galaxies-populations}, we further elaborated on this notion by evaluating the overall populations of galaxies in the different scale-space web environments. We find that the defining length is crucial to determining the number of galaxies residing in the voids, walls, filaments and clusters. For instance, the small-scale voids contain very little of the overall cosmic matter, and their galaxy content is close to vanishing. Whereas when measured at the largest scales (which we take as $\sigma=2.5\,h^{-1}\textrm{Mpc}$ in this study), the voids contain significant substructure (see \cref{fig:caustics_scale_space}) and account for up to about 40 percent of all galaxies. Similarly, we find that the fraction of blue galaxies across the different environments requires careful treatment in scale space. For instance, small-scale filaments may contain up to about 35 percent of all blue galaxies; whereas when measuring at the largest scales, it is the walls and voids that make up 30 and 60 percent of the blue galaxies, respectively, thus accounting for a total of 90 percent of the overall blue galaxy population. To our knowledge, these results constitute the first quantification of the galaxy content across the scale-space web environments. A unique feature of our analysis is that we highlight the existence of the web-like dark mass distribution down to galactic scales: in particular, we find that the butterfly cluster condition at the smallest scales naturally traces the dynamical origin of those overdensities that later virialise into groups or individual (sub-)haloes. These galaxy-scale `cluster-like' environments further illustrate the subtle nature of the multiscale cosmic web and its embedded galaxy population. 

In the last part of this article, \cref{sec:formation_time}, we applied the caustic skeleton formalism to study another aspect of the cosmological large-scale structure: for the time, we investigated the impact of the web formation time on the properties of the present-day embedded galaxies. While the formation time of the present-day structural features is not directly accessible to traditional identification methods based on morphology filters, the caustic skeleton -- being a phase-space formalism -- offers an unprecedented view into the formation history of the cosmic web. We used this notion to identify for each galaxy near a present-day wall, filament or cluster the formation time of its embedding web element, as we illustrated in \cref{fig:caustics_age}. Revisiting the analyses of \cref{sec:galaxies}, we subsequently evaluated, across the different web environments, the dependence of the galaxies' colours, star formation activities, metallicities and baryon fractions on the formation time. Here, we find clear trends in particular for the colours and star formation rates: galaxies in older web environments tend to be redder and more quenched, while those in only recently formed web environments are typically bluer and more active. We compared this dependence to the traditionally considered colour-density relation. To do so, we evaluated, over the cosmologically relevant range of length scales, the linear correlation of the colour and star formation rates with the formation time and environmental dark matter density. We find that, although the correlation with the density contrast is generally more pronounced, the dependence on the web formation time may be of similar order. Our results highlight that the dynamics of the web formation play a significant role in the assembly of the galaxy population, and encourage further applications of Lagrangian web identification methods.

There are several promising avenues for further simulation-based investigations of galaxy properties from the caustic skeleton. Firstly, while we have ignored galaxy morphologies throughout this article, \cite{Dressler1980} and numerous subsequent works demonstrated that these are highly dependent on the environmental densities. The caustic skeleton may provide a more nuanced view on how the early- and late-type galaxies are embedded in the different environments of the scale-space cosmic web. Another promising avenue in this context is a study of galaxy properties in the evolving cosmic web from its infancy up to the present day. In this article, we have focused on the cosmic web in its current state; yet the web elements and the embedded galaxies are the results of cosmic evolution over the course of billions of years. With the rigorous and continuous classification of the structural components over time and scale-space, the caustic skeleton may offer a fruitful framework for systematically investigating the history of galaxy formation and may yield new insights into current observational tensions concerning galaxy morphologies (e.g. \cite{Ferreira+2022}). In particular, the morphogenesis of dark matter caustics may shed further light on the timescales and dynamical processes that give rise to atypical galaxy populations, namely the red spirals and blue ellipticals, see \cite{Bamford+2009}. An extension of the present work may systematically study their dynamical origin and address the discrepancies between the morphology-density and the colour-density relation. 

Related to the galaxy morphologies, the continuous classification of the web over cosmic time may also be key in understanding the kinematical evolution of dark matter haloes and luminous galaxies. Directly applicable to the IllustrisTNG suite, this enables detailed investigations of halo shape and spin alignments with the prominent cosmic filaments. Moreover, systematic constrained simulation suites of the different caustic environments \citep{FeldbruggeWeygaert2023, FeldbruggeWeygaert2024, Hertzsch+2026, HertzschFeldbruggeWeygaert2026} offer a controlled environment to investigate the success and limitations of tidal torque theory (see e.g. \cite{Lopez+2025}). Together, these tools may prove useful in revisiting recent kinematical analyses such as \cite{Hasan+2025}.

Finally, our results demonstrate the key dependencies of the galaxy properties on cosmic environments as simulated by the IllustrisTNG suite. An intriguing continuation of our work may concern the galaxy properties in the physical Cosmic Web in our local neighbourhood. While survey data of the Universe provide detailed knowledge notably about the colour of nearby galaxies, recently developed reconstruction techniques  (see e.g. \cite{JascheWandelt2013, JascheLavaux2019, Valade+2022, Valade+2024, McAlpine+2025}) allow for elaborate simulations of the initial conditions and dynamical evolution out of which the present-day web emerged. By evaluating the caustic skeleton of such constrained simulations, one can directly compare the galaxy properties in the physical Cosmic Web to those measured in modern simulations. In light of the wealth of current and forthcoming data from state-of-the-art cosmological surveys — notably DESI, Euclid, LSST, 4MOST, and Roman — our study of galaxy properties within the cosmic web is both timely and of primary physical importance.